\RequirePackage[hyphens]{url}
\RequirePackage{snapshot}
\documentclass{vldb}

\usepackage{graphicx}
\usepackage{balance}
\usepackage{bm}

\usepackage{color}
\usepackage{listings}
\usepackage{colortbl}
\usepackage{verbatim}

\usepackage{amsmath}
\usepackage{amssymb}
\usepackage{xspace}

\usepackage{algorithm}
\usepackage{algorithmicx}
\usepackage[noend]{algpseudocode}
\usepackage{tikz}
\usepackage{tabularx}
\usepackage{multirow}
\usetikzlibrary{shapes,backgrounds,arrows}
\usepackage[textfont=bf,labelfont=bf,subrefformat=parens,position=below]{subfig}
\usepackage{listings}
\usepackage{caption}

\usepackage{stmaryrd}

\usepackage{paralist}
\usepackage[colorlinks=true,
           urlcolor=blue,
           citecolor=blue,
           linkcolor=blue,
           ]{hyperref}
\usepackage{mathtools}
\usepackage{rotating}
\usepackage{adjustbox}
\usepackage{dsfont}
\usepackage{enumitem}

\usepackage{cleveref}

\usetikzlibrary{decorations,arrows,shapes,shadows}
\usepackage{environ}
\usepackage{etoolbox}

\usepackage{graphicx}
\usepackage{balance}
\usepackage[disable]{todonotes}

\newbool{CompileTechReport}
\newcommand{\detailedproof}[2]{\ifbool{CompileTechReport}{
\begin{proof}
#2
\end{proof}
}{\noindent\textit{Proof Sketch:}#1\qed}\smallskip}
\newcommand{\ifnottechreport}[1]{\ifbool{CompileTechReport}{}{#1}}
\newcommand{\iftechreport}[1]{\ifbool{CompileTechReport}{#1}{}}

\booltrue{CompileTechReport}

\newcommand{\mypara}[1]{\smallskip\noindent\textbf{#1.}}

\DeclareRobustCommand{\BG}[1]{{\todo[color=red!40,inline]{\textbf{Boris says:}{#1}}}}
\DeclareRobustCommand{\BGDel}[2]{{\todo[inline,color=red!40]{\textbf{Boris deleted:}{#1}\textbf{because:} {#2}}}}
\DeclareRobustCommand{\BGcom}[1]{}
\DeclareRobustCommand{\SL}[1]{{\todo[inline,color=blue!50]{\textbf{Seokki says:}{#1}}}}

\DeclareRobustCommand{\RevDel}[1]{\todo[inline,color=green!60]{\textbf{Deleted for revision:}{#1}}}

\newcommand{\card}[1]{\left|{#1}\right|}

\newtheorem{theo}{Theorem}
\newtheorem{defi}{Definition}

\newtheorem{exam}{Example}

\definecolor{black}{rgb}{0,0,0}
\definecolor{grey}{rgb}{0.8,0.8,0.8}
\definecolor{red}{rgb}{1,0,0}
\definecolor{green}{rgb}{0,1,0}
\definecolor{darkgreen}{rgb}{0,0.5,0}
\definecolor{darkpurple}{rgb}{0.5,0,0.5}
\definecolor{darkdarkpurple}{rgb}{0.3,0,0.3}
\definecolor{blue}{rgb}{0,0,1}
\definecolor{shadegreen}{rgb}{0.95,1,0.95}
\definecolor{shadeblue}{rgb}{0.95,0.95,1}
\definecolor{shadered}{rgb}{1,0.85,0.85}
\definecolor{oddRowGrey}{rgb}{0.95,0.95,0.95}
\definecolor{evenRowGrey}{rgb}{0.85,0.85,0.85}

\newcommand{\mathtext}[1]{\thickspace\text{#1}\thickspace}
\newcommand{\mathtab}{\thickspace\thickspace\thickspace}
\newcommand{\projection}{\Pi}
\newcommand{\selection}{\sigma}
\newcommand{\aggregation}{\gamma}
\newcommand{\union}{\cup}

\newcommand{\join}{\bowtie}
\newcommand{\antijoin}{\vartriangleright}
\newcommand{\duprem}{\delta}

\newcommand{\rename}{\rho}

\newcommand{\rschema}[1]{\mathbf{#1}}

\newcommand{\iftels}[3]{\text{\bf if}\thickspace (#1) \thickspace \text{\bf then} \thickspace #2 \thickspace \text{\bf else} \thickspace #3}
\newcommand{\isnull}[1]{\mathbf{isnull}(#1)}
\def\ojoin{\setbox0=\hbox{$\bowtie$}
  \rule[-.02ex]{.25em}{.4pt}\llap{\rule[\ht0]{.25em}{.4pt}}}
\def\leftouterjoin{\mathbin{\ojoin\mkern-5.8mu\bowtie}}

\newcommand{\defas}{\coloneqq}
\DeclareMathOperator*{\argmax}{argmax}

\newcommand{\thead}[1]{{\cellcolor{black}{\textcolor{white}{\textbf{#1}}}}}

\newcommand{\hlrow}{\rowcolor{shadered}}

\newcommand{\query}{Q}
\newcommand{\dlImp}[0]{\,\ensuremath{\mathtt{{:}-}}\,}
\newcommand{\dlNeg}{\neg\,}
\newcommand{\varvec}[1]{\overline{{#1}}}
\newcommand{\bodyOf}[1]{\mathsf{body}(#1)}
\newcommand{\headOf}[1]{\mathsf{head}(#1)}
\newcommand{\varsOf}[1]{\mathsf{vars}(#1)}
\newcommand{\attrsOf}[1]{\mathsf{attrs}(#1)}

\newcommand{\dlcomp}{\psi}

\newcommand{\matches}{\curlyeqprec}
\newcommand{\allMatches}{\mathcal{M}}

\newcommand{\rel}[1]{\ensuremath{\mathtt{#1}}}
\newcommand{\cnst}[1]{\textsf{#1}}

\newcommand{\adom}{adom}

\newcommand{\allAnnotRules}{\ensuremath{\mathcal{A}}\xspace}
\newcommand{\db}{D}

\newcommand{\dataDomain}{\ensuremath\mathbb{D}}

\newcommand{\placeholders}{\ensuremath\mathbb{P}}
\newcommand{\val}{\nu}

\newcommand{\deri}{\ensuremath{d}}
\newcommand{\fderi}{\ensuremath{d}}
\newcommand{\sderi}{\ensuremath{d_{s}}}

\newcommand{\proofWN}{\mathcal{P}}
\newcommand{\wnSize}{n_P}
\newcommand{\proofA}{\mathcal{A}}
\newcommand{\aSize}{n_A}
\newcommand{\succDeri}{\samp_{np}}
\newcommand{\sSize}{n_{np}}

\newcommand{\arityOf}{arity}
\newcommand{\consOf}{C}
\newcommand{\gconsOf}{C}

\newcommand{\completeness}{cp}
\newcommand{\informativeness}{info}
\newcommand{\score}{sc}
\newcommand{\lbc}{\underline{\completeness}}
\newcommand{\ubc}{\overline{\completeness}}
\newcommand{\lbsc}{\underline{\score}}
\newcommand{\ubsc}{\overline{\score}}

\newcommand{\lca}{\textsc{LCA}\xspace}
\newcommand{\lcasm}{lca}

\newcommand{\overSamp}{\ensuremath{\textsc{OS}}\xspace}
\newcommand{\samp}{\ensuremath{\textsc{S}\xspace}}
\newcommand{\sizeOf}[1]{n_{#1}}
\newcommand{\sampSize}{\ensuremath{\sizeOf{\samp}}\xspace}
\newcommand{\osSize}{\ensuremath{\sizeOf{\overSamp}}\xspace}
\newcommand{\sampleOp}{\ensuremath{\textsc{Sample}}}
\newcommand{\rownumOp}{\#}
\newcommand{\osThresh}{\ensuremath{P_{success}}\xspace}
\newcommand{\prob}{p}
\newcommand{\probProv}{\ensuremath{\prob_{prov}}\xspace}
\newcommand{\probNotProv}{\ensuremath{\prob_{notProv}}\xspace}
\newcommand{\sprov}{\ensuremath{\mathcal{S}}\xspace}
\newcommand{\pslb}{\sprov_{lb}}
\newcommand{\psub}{\sprov_{ub}}

\newcommand{\pat}{\ensuremath{\textsc{Pat}}}

\newcommand{\plOf}[1]{\mathsf{placeh}(#1)}
\newcommand{\pleq}{\preceq_{p}}

\newcommand{\pind}{\bot_{p}}
\newcommand{\parg}[2]{{#1}[{#2}]}

\newcommand{\qVar}[1]{\query_{#1}}
\newcommand{\qBind}{\query_{bind}}
\newcommand{\qDer}{\query_{der}}
\newcommand{\qGoalJoin}{\query_{goals}}
\newcommand{\qSamp}{\query_{sample}}
\newcommand{\qMatch}{\query_{match}}
\newcommand{\checkExistsCond}{\theta_{der}}
\newcommand{\checkPtCond}{\theta_{\pt}}
\newcommand{\checkVarPredCond}[1]{\theta_{#1}}
\newcommand{\checkJoinPreds}{\theta_{join}}
\newcommand{\checkMatchPred}{\theta_{match}}

\newcommand{\qLCA}{\query_{lca}}
\newcommand{\condLCA}{\theta_{lca}}
\newcommand{\projLCA}{A_{lca}}

\newcommand{\explainq}{\textsc{Prov}}

\newcommand{\ProvPoly}{\ensuremath{\mathbb{N}[X]}\xspace}

\newcommand{\greenT}{\textcolor{DarkGreen}{T}}
\newcommand{\redF}{\textcolor{DarkRed}{F}}

\definecolor{DarkGreen}{rgb}{0,0.45,0}
\definecolor{DarkRed}{rgb}{0.8,0,0}
\definecolor{DarkYellow}{rgb}{0.6,0.6,0}
\definecolor{DarkGray}{rgb}{0.2,0.2,0.2}

\newcommand{\listconcat}{\,{\tt ::}\,}

\newcommand{\qType}{\textsf{type}}

\newcommand{\why}{\textsc{Why}}
\newcommand{\whynot}{\textsc{Whynot}}
\newcommand{\aProvQ}{\Phi}

\newcommand{\bobsQ}{\aProvQ_{bob}}
\newcommand{\tWithCons}{\ensuremath{t}}
\newcommand{\tWithConsVars}{\ensuremath{\boldsymbol{t}}}

\newcommand{\pt}{\tWithConsVars}
\newcommand{\run}{r_{\pt}}

\newcommand{\tBobFull}{\rel{AL}(N, \cnst{shared})}
\newcommand{\tBob}{\tWithConsVars_{bob}}

\newcommand{\tExFull}{\rel{\qEx}(X,\cnst{4})}
\newcommand{\tEx}{\tWithConsVars_{ex}}
\newcommand{\qEx}{\query_{ex}}
\newcommand{\rEx}{r_{ex}}
\newcommand{\rExun}{r_{\tEx}}
\newcommand{\pqEx}{\aProvQ_{ex}}

\newcommand{\anAnnotRule}{r(\varvec{c}) - (\varvec{g})}

\newcommand{\allDer}{all-derivations\xspace}
\newcommand{\AllDer}{All-derivations\xspace}
\newcommand{\singleDer}{single-derivation\xspace}
\newcommand{\SingleDer}{Single-derivation\xspace}
\newcommand{\probSucc}{\ensuremath{P_{success}}\xspace}

\newcommand{\ucqNegOrd}{\ensuremath{\mathsf{UCQ}^{\neg<}}\xspace}

\newcommand{\oNotation}[1]{O({#1})}

\definecolor{crred}{rgb}{0.8,0.0,0.0}

\usepackage{etoolbox}
\definecolor{revgreen}{rgb}{0,0.5,0}
\newrobustcmd{\reva}[1]{#1}
\newrobustcmd{\revb}[1]{#1}
\newrobustcmd{\revc}[1]{#1}
\newrobustcmd{\revm}[1]{#1}

\vldbTitle{Approximate Summaries for Why and Why-not Provenance}
\vldbAuthors{Seokki Lee, Bertram Lud\"ascher, Boris Glavic}
\vldbDOI{https://doi.org/10.14778/3380750.3380760}
\vldbVolume{13}
\vldbNumber{6}
\vldbYear{2020}

\Crefname{exam}{Ex.}{Ex.}
\Crefname{figure}{Fig.}{Fig.}
\Crefname{section}{Sec.}{Sec.}
\Crefname{defi}{Def.}{Def.}
\Crefname{theo}{Thm.}{Thm.}

\renewenvironment{itemize}[1]{\begin{compactitem}#1}{\end{compactitem}}

\begin{document}

\definecolor{lstpurple}{rgb}{0.5,0,0.5}
\definecolor{lstred}{rgb}{1,0,0}
\definecolor{lstreddark}{rgb}{0.7,0,0}
\definecolor{lstredl}{rgb}{0.64,0.08,0.08}
\definecolor{lstmildblue}{rgb}{0.66,0.72,0.78}
\definecolor{lstblue}{rgb}{0,0,1}
\definecolor{lstmildgreen}{rgb}{0.42,0.53,0.39}
\definecolor{lstgreen}{rgb}{0,0.5,0}
\definecolor{lstorangedark}{rgb}{0.6,0.3,0}
\definecolor{lstorange}{rgb}{0.75,0.52,0.005}
\definecolor{lstorangelight}{rgb}{0.89,0.81,0.67}
\definecolor{lstbeige}{rgb}{0.90,0.86,0.45}
\definecolor{Comments}{rgb}{0.00,0.50,0.00}
\definecolor{KeyWords}{rgb}{0.00,0.00,0.13}
\definecolor{Strings}{rgb}{0.60,0.00,0.00}

\lstdefinestyle{psql}
{
  tabsize=3,
  basicstyle=\small\upshape\ttfamily,
  language=SQL,
  morekeywords={PROVENANCE,BASERELATION,INFLUENCE,COPY,ON,TRANSPROV,TRANSSQL,TRANSXML,CONTRIBUTION,COMPLETE,TRANSITIVE,NONTRANSITIVE,EXPLAIN,SQLTEXT,GRAPH,IS,ANNOT,THIS,XSLT,MAPPROV,cxpath,OF,TRANSACTION,SERIALIZABLE,COMMITTED,INSERT,INTO,WITH,SCN,PROV,IMPORT,FOR,JSON,JSON_TABLE,XMLTABLE},
  extendedchars=false,
  keywordstyle=\color{blue},
  mathescape=true,
  escapechar=@,
  sensitive=true,
  stringstyle=\color{Strings},
  string=[b]'
}

\lstdefinestyle{psqlcolor}
{
tabsize=2,
basicstyle=\footnotesize\upshape\ttfamily,
language=SQL,
morekeywords={PROVENANCE,BASERELATION,INFLUENCE,COPY,ON,TRANSPROV,TRANSSQL,TRANSXML,CONTRIBUTION,COMPLETE,TRANSITIVE,NONTRANSITIVE,EXPLAIN,SQLTEXT,GRAPH,IS,ANNOT,THIS,XSLT,MAPPROV,cxpath,OF,TRANSACTION,SERIALIZABLE,COMMITTED,INSERT,INTO,WITH,SCN,UPDATED,FOLLOWING,RANGE,UNBOUNDED,PRECEDING,OVER,PARTITION,WINDOW},
extendedchars=false,
keywordstyle=\bfseries\color{lstpurple},
deletekeywords={count,min,max,avg,sum,lag,first_value,last_value},
keywords=[2]{count,min,max,avg,sum,lag,first_value,last_value,lead,row_number},
keywordstyle=[2]\color{lstblue},
stringstyle=\color{lstreddark},
commentstyle=\color{lstgreen},
mathescape=true,
escapechar=@,
sensitive=true
}

\lstdefinestyle{rsl}
{
tabsize=3,
basicstyle=\small\upshape\ttfamily,
language=C,
morekeywords={RULE,LET,CONDITION,RETURN,AND,FOR,INTO,REWRITE,MATCH,WHERE},
extendedchars=false,
keywordstyle=\color{blue},
mathescape=true,
escapechar=@,
sensitive=true
}

\lstdefinestyle{pseudocode}
{
  tabsize=3,
  basicstyle=\small,
  language=c,
  morekeywords={if,else,foreach,case,return,in,or},
  extendedchars=true,
  mathescape=true,
  literate={:=}{{$\gets$}}1 {<=}{{$\leq$}}1 {!=}{{$\neq$}}1 {append}{{$\listconcat$}}1 {calP}{{$\cal P$}}{2},
  keywordstyle=\color{blue},
  escapechar=&,
  numbers=left,
  numberstyle={\color{green}\small\bf}, 
  stepnumber=1, 
  numbersep=5pt,
}

\lstdefinestyle{xmlstyle}
{
  tabsize=3,
  basicstyle=\small,
  language=xml,
  extendedchars=true,
  mathescape=true,
  escapechar=£,
  tagstyle={\color{blue}},
  usekeywordsintag=true,
  morekeywords={alias,name,id},
  keywordstyle={\color{red}}
}

\lstdefinelanguage{json}{
    basicstyle=\normalfont\ttfamily,
    numbers=left,
    numberstyle=\scriptsize,
    stepnumber=1,
    numbersep=8pt,
    stringstyle=\color{Strings},
    showstringspaces=false,
    breaklines=true,
    frame=lines,
    string=[b]"
}

 \lstset{style=psql}

\title{Approximate Summaries for Why and Why-not Provenance\\
(Extended Version)}

\numberofauthors{3}

\author{
\begin{minipage}{.3\linewidth}
\centering
Seokki Lee\\
      \affaddr{Illinois Institute of Technology}\\
      \email{slee195@hawk.iit.edu}
\end{minipage}
\begin{minipage}{.4\linewidth}
\centering
Bertram Lud\"ascher\\
      \affaddr{University of Illinois, Urbana-Champaign}\\
      \email{ludaesch@illinois.edu}
\end{minipage}
\begin{minipage}{.3\linewidth}
\centering
Boris Glavic\\
      \affaddr{Illinois Institute of Technology}\\
      \email{bglavic@iit.edu}
\end{minipage}
}

\maketitle
\graphicspath{ {figs/} }

\begin{abstract}
  Why and why-not provenance have been studied extensively
  in recent years.
 However,  why-not provenance and --- to a lesser degree --- why
provenance can be very large, resulting in severe scalability and
usability challenges.
We introduce a novel \emph{approximate summarization} technique
for provenance to address these challenges.
Our approach uses patterns
to encode why and why-not provenance concisely.
We develop techniques for efficiently computing provenance summaries
that balance informativeness, conciseness, and completeness.
To achieve scalability,
we integrate sampling techniques into  provenance capture and summarization.
Our approach is the first to both scale to large datasets
and generate comprehensive and meaningful summaries.
\end{abstract}

 \section{Introduction}
\label{sec:introduction}

Provenance for relational
queries~\cite{green2007provenance} 
explains how results of a query are derived from the query's inputs.  In
contrast, why-not provenance explains why a query result is missing.
Specifically, \textit{instance-based}~\cite{DBLP:journals/vldb/HerschelDL17}
why-not provenance techniques determine which existing and missing data from a
query's input is responsible for the failure to derive a missing answer of
interest.
In prior work, we have shown how why and why-not provenance can be
treated uniformly  for first-order queries 
using non-recursive Datalog with negation
~\cite{KL13} and have implemented this idea in the \textit{PUG} system
~\cite{LS17,LL18}.
Instance-based why-not provenance techniques 
either (i) 
enumerate all potential ways of deriving a result (\textit{\allDer} approach) 
or 
(ii) return only one possible, but failed, derivation or parts thereof (\textit{\singleDer{}} approach). 
For instance, Artemis~\cite{HH10}, 
Huang et al.~\cite{huang2008provenance}, and 
PUG~\cite{LL18j,LS17,LL18} are \allDer approaches while 
the Y! system~\cite{WZ14a,WH13} is a  \singleDer{} approach.

\newcolumntype{P}[1]{>{\centering\arraybackslash}p{#1}}
\definecolor{LightCyan}{rgb}{0.88,1,1}
\definecolor{TeaGreen}{rgb}{0.82,0.94,0.75}
\newcolumntype{C}[1]{>{\centering\arraybackslash}p{#1}}
\begin{figure}[t]
  \centering 
   \begin{minipage}{1\linewidth}
    \centering
    \begin{minipage}{1\linewidth}
      \centering \small
      \begin{align*}
        & r_1: \rel{AL}(N,R) \dlImp \rel{L}(I,N,T,R,\textsf{queen anne},E), \rel{A}(I,\textsf{2016-11-09},P)\\
      \end{align*}
    \end{minipage}\\[-2mm]
    \begin{minipage}{1\linewidth}
      \centering \hspace{-5mm}
      \scriptsize
      \begin{minipage}{0.92\linewidth}
        \centering $\,$\\[-2mm]
        \setlength\tabcolsep{2.8pt}
        \begin{tabular}{|cccccc|}
         \multicolumn{2}{l}{\textbf{L}isting (input)}  \\[0.5mm]\cline{1-6}
          \thead{Id} & \thead {Name} & \thead{Ptype} & \thead{Rtype} & \thead{NGroup} & \thead{Neighbor} \\
          \rowcolor{TeaGreen} 8403 & central place & apt & shared & queen anne & east\\
          9211 & plum & apt & entire & ballard & adams \\
          2445 & cozy homebase & house & private & queen anne & west\\
          \rowcolor{TeaGreen} 8575 & near SpaceNeedle &	apt & shared & queen anne & lower\\
          4947 & seattle couch & condo & shared & downtown & first hill \\
          2332 & modern view & house & entire &	queen anne & west \\
          \cline{1-6}
        \end{tabular}
      \end{minipage}
     \end{minipage}
   \end{minipage}

    \begin{minipage}{1\linewidth}
     \centering \hspace{-2mm}
      \begin{minipage}{1\linewidth}
       \scriptsize \centering
	\begin{minipage}{.45\linewidth}
          \centering $\,$\\[0.5mm]
          \begin{tabular}{|ccc|}
            \multicolumn{2}{l}{\textbf{A}vailability (input)} \\[.5mm]\cline{1-3}
            \thead{Id} & \thead{Date} & \thead{Price}\\
            9211&	2016-11-09&	130\\
            2445&	2016-11-09&	45\\
            2332&      2016-11-09 & 350 \\
            4947&	2016-11-10&	40\\
            \cline{1-3}
          \end{tabular}
	\end{minipage}
        \hspace{4mm}
	\begin{minipage}{.4\linewidth}
          \centering$\,$\\[-4mm]
           \begin{tabular}{|c|c|}
            \multicolumn{2}{l}{\textbf{A}vailable\textbf{L}istings (output)} \\[.5mm]\cline{1-2}
             \thead{Name} & \thead{Rtype}\\
             cozy homebase & private \\
             modern view & entire \\
            \cline{1-2}
          \end{tabular}
      \end{minipage}

    \end{minipage}\\
  \end{minipage}

  \resizebox{1\linewidth}{!}{
    \begin{minipage}{1.4\linewidth}
      \centering$\,$\\[1mm]
      \setlength\tabcolsep{3.5pt}
      \begin{tabular}{c|c|c|c|c|c|c|c|c}
        \textbf{Attribute} & Id & Name & Ptype & Rtype & NGroup & Neighbor & Date & Price\\ \hline
        \textbf{\#Distinct Values} & 6 & 6 & 3 & 3 & 3 & 5 & 2 & 4 \\
      \end{tabular}
  \end{minipage}
}
$\,$\\[0mm]
  \caption{\revm{Example Airbnb database and query}}
  \label{fig:running-example-db}
\end{figure}

\begin{exam}\label{ex:intro-db-scen}
\Cref {fig:running-example-db} shows a sample of a real-world data\-set
recording Airbnb (bed and breakfast) listings and their availability.
 Each \rel{Listing} has an \emph{id}, \emph{name}, \emph{property type (\texttt{Ptype})}, \emph{room type (\texttt{Rtype})},
   \emph{neighborhood (\texttt{Neighbor})}, and \emph{neighborhood group (\texttt{NGroup})}. The neighborhood groups are larger areas including multiple neighborhoods.
  \rel{Availability} stores \emph{ids} of listings with available \emph{dates} and a \emph{price} for each date.
We refer to this sample dataset as \emph{S-Airbnb} 
and the full dataset as \emph{F-Airbnb} ({\url{https://www.kaggle.com/airbnb/seattle}}).
Bob, an analyst at Airbnb,  investigates a customer complaint about the lack of availability of shared rooms
on \textsf{2016-11-09} in Queen Anne (\texttt{NGroup} = \textsf{queen anne}).
He first uses Datalog rule $r_1$ 
from \Cref{fig:running-example-db} to return all listings (names and room types) available on that date in Queen Anne.
The query result confirms the customer's complaint, since none of the available listings are \textsf{shared} rooms.
Bob now needs to investigate what led to this missing result.
\end{exam}

We refer to
such questions as \textit{provenance questions}. A provenance question is a tuple with constants and placeholders (upper-case letters) which specify a set of (missing) answers the user is interested in (all answers that agree with the provenance question on constant values).
For example, \revm{Bob's question can be written as $\rel{AL}(N,\textsf{shared})$}. 
\AllDer approaches like PUG explain the absence of \textsf{shared} rooms by enumerating all derivations of missing answers that \emph{match} Bob's question. 
That is, all possible bindings of the variables of the rule $r_1$ to
values from the active domain (the values that exist in the database) 
such that \revm{$R$ is bound to \textsf{shared}} and the tuple produced by the grounded rule is missing.
While this explains why shared rooms are unavailable (any tuple with $R=\,$ \textsf{shared}), 
the number of possible bindings can be prohibitively large. 
Consider our toy example \emph{S-Airbnb} dataset. Let us assume that only values from the active domain of each attribute are considered for a variable bound to this attribute to avoid nonsensical derivations, \revm{e.g., binding prices to names}. 
The number of distinct values per attribute are shown on the bottom of \Cref{fig:running-example-db}.
Under this assumption, there are 
$6 \cdot 6 \cdot 3 \cdot 5 \cdot 4 = 2160$ possible ways to derive missing results matching $\rel{AL}(N,\textsf{shared})$. 
For the full dataset \emph{F-Airbnb}, 
there are 
$\sim 15 \cdot 10^{12}$ 
possible derivations. 

\begin{exam}\label{ex:expl-whynot-pug}
Continuing with \Cref{ex:intro-db-scen}, assume that Bob uses
PUG~\cite{LS17} 
to compute an explanation for 
the missing result $\rel{AL}(N,\textsf{shared})$.
A provenance graph fragment is shown in \Cref{fig:part-whynot-lic}. 
This type of provenance graph connects rule derivations (box nodes) with the tuples (ovals) they are deriving, rule derivations to the goals in their body 
(rounded boxes), and goals to the tuples that justify their success or
failure. Nodes are colored red/green to indicate failure/success (goal and rule nodes) or absence/existence (tuple nodes). 
For S-Airbnb, the graph produced by PUG  
consists of all $2160$ failed derivations of missing answers that match $\rel{AL}(N,\textsf{shared})$.
The fragment shown in \Cref{fig:part-whynot-lic} encodes  one of these derivations:
The shared room of the existing listing \textit{Central Place} (\texttt{Id} \textsf{8403}) is not available on \textsf{2016-11-09} at a price of $\$130$,
explaining that this derivation fails because
the tuple $(\textsf{8403},\textsf{2016-11-09},\textsf{130})$ does not exist in the relation $\rel{Availability}$
(the second goal failed).
\end{exam}

\SingleDer approaches address the scalability issue of why-not provenance by only returning a single derivation (or parts thereof).
However, this comes at the cost of 
incompleteness. 
For instance, 
a \singleDer{} approach may return the derivation shown in \Cref{fig:part-whynot-lic}. 
However, such an explanation is not sufficient for Bob's investigation.
What about other prices for the same listing? 
Do other listings from this area have shared rooms that are not available for this date or do they simply not have shared rooms? A single derivation approach cannot answer such questions since it only provides one out of a vast number of failed derivations (or even only a sufficient reason for a derivation to fail as in~\cite{WZ14a,WH13}). \BGDel{, but misses many equally valid explanations.}{redundant}
For 
S-Airbnb, 
no \textsf{shared} rooms are available in Queen Anne on Nov 9th, 2016
because:
 (i) all the existing \textsf{shared} rooms of apartments (listings \textsf{8403} and \textsf{8575}) in 
 Queen Anne are not available on the requested date and
 (ii) no listings in the West Queen Anne neighborhood (listings \textsf{2445} and \textsf{2332}) have \textsf{shared} rooms. 
\textbf{Thus, returning only one derivation is insufficient for justifying the missing answer as only the collective failure of all possible derivations explains the missing answer.} 
\iftechreport{Suppose that Bob has to explain the result of his investigation to his manager and is expected to propose possible ways of how to improve the availability of rooms in Queen Anne. His manager is unlikely to accept an explanation of the form \textit{``There are no shared rooms available on this date, because listing 8403 is not available for \$130 on this day.''}}

\mypara{Summarizing Provenance}
In this paper, we present a novel approach that overcomes the drawbacks of both approaches.
Specifically, we efficiently create summaries that
compactly represent large amounts of provenance information. 
We focus on the algorithmic and formal foundation of this method as well as its experimental evaluation
(we demonstrated a GUI frontend in~\cite{LL18} and our vision in~\cite{SN17}).

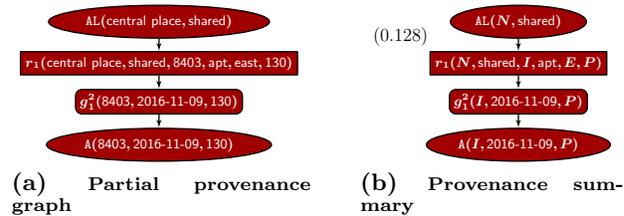
\begin{figure}[t]
  \begin{minipage}{.45\linewidth}
  \centering
  \subfloat[\scriptsize Partial provenance graph]{
    \scalebox{0.58}{\begin{tikzpicture}[>=latex',line join=bevel,line width=.3mm]
  \definecolor{fillcolor}{rgb}{0.63,0,0}
  \definecolor{wcolor}{rgb}{0.83,1.0,0.8};

  \node (REL_Q_WON_NORTH_PARK_assault_) at (293bp,180bp) [draw=black,fill=fillcolor,ellipse,text=white] {$\boldsymbol{\rel{AL}(\cnst{central place},\cnst{shared})}$};

  \node (RULE_0_WON_NO_CRIME_) at (293bp,153bp) [draw=black,fill=fillcolor,rectangle,text=white] {$\boldsymbol{r_1(\cnst{central place},\cnst{shared},\cnst{8403},\cnst{apt},
      \cnst{east},
      \cnst{130})}$};

  \node (GOAL_0_0_WON_NO_CRIME_) at (293bp,128bp) [draw=black,fill=fillcolor,rounded corners=.15cm,inner sep=3pt,text=white] {$\boldsymbol{g_{1}^{2}(\cnst{8403} ,\cnst{2016-11-09}
      ,\cnst{130})}$};

  \node (REL_CRIME_WON_NO_CRIME_) at (293bp,100bp) [draw=black,fill=fillcolor,ellipse,text=white] {$\boldsymbol{\rel{A}(\cnst{8403}, \cnst{2016-11-09},
      \cnst{130})}$};

  \draw [->] (REL_Q_WON_NORTH_PARK_assault_) -> (RULE_0_WON_NO_CRIME_);
  \draw [->] (RULE_0_WON_NO_CRIME_) -> (GOAL_0_0_WON_NO_CRIME_);
  \draw [->] (GOAL_0_0_WON_NO_CRIME_) -> (REL_CRIME_WON_NO_CRIME_);

\end{tikzpicture}

 }
    \label{fig:part-whynot-lic}
  }
\end{minipage}
\hspace{5mm}
\begin{minipage}{.45\linewidth}
  \centering
  \subfloat[\scriptsize Provenance summary]{
    \scalebox{0.58}{\begin{tikzpicture}[>=latex',line join=bevel,line width=.3mm]
  \definecolor{fillcolor}{rgb}{0.63,0,0}
  \definecolor{wcolor}{rgb}{0.83,1.0,0.8};

  \node (REL_Q_WON_NORTH_PARK_assault_) at (293bp,180bp) [draw=black,fill=fillcolor,ellipse,text=white] {$\boldsymbol{\rel{AL}(N,\cnst{shared})}$};

  \node (RULE_0_WON_NO_CRIME_) at (293bp,153bp) [draw=black,fill=fillcolor,rectangle,text=white,label={[xshift=-2.7cm, yshift=0.0cm]\large ($0.128$)}] {$\boldsymbol{r_1(N,\cnst{shared},I,\cnst{apt},
      E,
      P)}$};

  \node (GOAL_0_0_WON_NO_CRIME_) at (293bp,128bp) [draw=black,fill=fillcolor,rounded corners=.15cm,inner sep=3pt,text=white] {$\boldsymbol{g_{1}^{2}(I, \cnst{2016-11-09},
      P)}$};

  \node (REL_CRIME_WON_NO_CRIME_) at (293bp,100bp) [draw=black,fill=fillcolor,ellipse,text=white] {$\boldsymbol{\rel{A}(I ,\cnst{2016-11-09}
      ,P)}$};

  \draw [->] (REL_Q_WON_NORTH_PARK_assault_) -> (RULE_0_WON_NO_CRIME_);
  \draw [->] (RULE_0_WON_NO_CRIME_) -> (GOAL_0_0_WON_NO_CRIME_);
  \draw [->] (GOAL_0_0_WON_NO_CRIME_) -> (REL_CRIME_WON_NO_CRIME_);

\end{tikzpicture}

 }
    \label{fig:summ-whynot-lic}
  }
\end{minipage}
$\,$\\[0mm]
\caption{\revm{Explanations for the missing results 
  $\small\text{AL}(N,\text{shared})$
  }}
\label{fig:summ-expl-whynot}
\end{figure}

\begin{exam}\label{ex:expl-whynot-pug-summ}
Our summarization approach encodes sets of nodes from a provenance graph using \emph{``pattern nodes"}, i.e., nodes with placeholders.\footnote{We deliberately use the term placeholder and not variable to avoid confusion with the variables of a rule.} 
A possible summary 
for 
$\rel{AL}(N,$ $\textsf{shared})$ 
is shown in \Cref{fig:summ-whynot-lic}.
The graph contains a rule pattern node
  $r_1(N,\textsf{shared},I,\textsf{apt},E,P)$. $N$, $I$, $E$, and $P$ are placeholders. 
For each such node, our approach reports the amount of provenance covered by the pattern (shown to the left of nodes).
This summary provides useful information to Bob:
  all shared rooms of apartments in Queen Anne are not available at any price on Nov 9th, 2016 (their ids are not in relation \rel{Availability}).
  Over F-Airbnb, 
  $\sim 12.8\%$ of derivations for 
  $\rel{AL}(N,\textsf{shared})$ match this pattern.

\end{exam}

The type of patterns we are using here can also be modeled as selection queries and has been used to summarize provenance~\cite{WM13,RS14} and for explanations in general~\cite{EA14,GF18}.

\mypara{Selecting Meaningful Summaries}\label{sec:sel-mean-summ}
The provenance of a (missing) answer can be summarized in many possible ways.
Ideally, we want \textit{provenance summaries}
to be \emph{concise} (small provenance graphs),
\emph{complete} (
covering all provenance), and \emph{informative} (
providing new insights). We define informativeness as the number of constants in a pattern that are not enforced by the user's question. The intuition behind this definition is that patterns with more constants provide more detailed information about the data accessed by derivations.
Finding a solution that optimizes all three metrics is typically not possible. Consider two extreme cases: (i) any provenance graph is 
a provenance summary (one without placeholders). Provenance graphs are complete and informative, but not concise;
(ii) at the other end of the spectrum, an arbitrary number of derivations of a rule $r$ can be represented as a single pattern with only placeholders resulting in a maximally concise summary.
However, 
such a summary is not informative since it only contains  placeholders. 
We 
design a summarization algorithm that 
 returns a set of up to $k$ patterns (guaranteeing conciseness) optimizing for a combination of completeness and informativeness.
The rationale behind this approach is
  to ensure that summaries are covering a sufficient fraction of the provenance and at the same time provide sufficiently detailed  information. 
\iftechreport{Most of our results, however, are independent of the choice of ranking metric.}
\BGDel{We argue that this 
  is fulfilled as long as the returned provenance summary is not larger than a certain threshold. This observation motivated our choice to guarantee an upper bound on the size by returning an explanation that consists of the top-$k$ patterns and within solutions that are not larger than this upper bound to optimize for completeness and informativeness.}{redundant}

\mypara{Efficient Summarization} While 
 summarization of provenance has been studied in previous work, e.g.,~\cite{AB15a,XE},
 for why-not provenance we face the challenge that it is infeasible to generate full provenance as input for summarization. 
 For instance, 
 there are $\sim 15 \cdot 10^{12}$ derivations of missing answers matching Bob's question if we use 
   the F-Airbnb dataset.
We overcome this problem 
by (i) integrating summarization with provenance capture 
and (ii) developing a method for sampling rule derivations from the why-not provenance 
without 
materializing it first. 
Our sampling technique is based on the observation that the number of missing answers is typically significantly larger than the number of existing answers. Thus, to create a  sample of the why-not provenance of missing answers matching a provenance question, we can randomly generate derivations that match the provenance question. We, then, filter out derivations for existing answers.
\RevDel{This is possible, because we can check efficiently whether a derivation computes an existing answer.}
This approach is effective, because a randomly generated derivation is much more likely to derive a missing than an existing answer.
While sampling is necessary for performance, it is not sufficient. Even for relatively small sample sizes, enumerating all possible sets of candidate patterns and evaluating their scores to find the set of size up to $k$ with the highest score is not feasible. We introduce several heuristics and optimizations that together enable us to achieve good performance. Specifically, we limit the number of candidate patterns, approximate the completeness of sets of patterns over our sample, and exploit provable upper and lower bounds for the score of candidate pattern sets when ranking such sets.
\BG{
We observe that the union of derivations for existing and missing results (denoted as $\sderi$ and $\fderi$, respectively) 
produced by a rule is the cross-product of the domains of the attributes accessed by the rule. 
Typically, 
$|\fderi| \gg |\sderi|$ 
and, thus, it is reasonable to assume that the probability  $\probNotProv$ to sample a derivation of a missing answer is higher than the probability of sampling derivations for existing tuples $\probProv (= 1-\probNotProv)$ such that
$\probNotProv \gg \probProv$. 
Based on this observation, 
(i) we create a sample of size $\osSize > \sampSize$ (the target sample size) from the cross-product 
by sampling each domain of attributes individually and ``zipping'' the individual samples into a sample of the derivations of a rule
and, then, 
(ii) remove from the sample all derivations that belong to $\sderi$ (can be checked efficiently by accessing the database).
Note that 
(ii) may lead to the sample of size $\sampSize' < \sampSize$ which necessitates a sample of size $\osSize$.
We demonstrate how to choose $\osSize$ such that the resulting sample contains with high probability at least $\sampSize$ derivations of missing answers 
for the provenance question. 
We further demonstrate experimentally 
that the samples produced in this way lead to summaries of quality that is very close to the quality 
over full provenance. 
\BG{The paragraph above is good, but I think we can cut it if we really need space}
}

\mypara{Contributions}
To the best of our knowledge, we are the first to address both the usability and scalability (computational) challenges of why-not provenance through summaries. 
\RevDel{PUT BACK IN CR MAYBE: For example, our approach only needs seconds to generate the provenance summary 
shown in \Cref{fig:summ-whynot-lic}.}
Specifically, we make the following contributions: 

\begin{itemize}
\item Using patterns, we 
  generate meaningful summaries
  for 
  the why and why-not provenance
   of unions of conjunctive queries with negation  \revm{and inequalities (\ucqNegOrd)}. 
\item We develop a summarization algorithm that applies sampling during provenance capture and 
  avoids 
  enumerating full why-not provenance.
  Our approach out-sources most computation to 
  a database system.
\item We experimentally 
  \revm{compare our approach with a single-derivation approach and Artemis~\cite{HH10} and demonstrate that it efficiently produces high-quality summaries.}\BG{We also compare against a single derivation approach.} 
\end{itemize}

The remainder of this paper is organized as follows.
We cover preliminaries 
in \Cref{sec:prelim} and define 
the provenance summarization problem 
in \Cref{sec:prob-def}. We present an overview of our approach
in \Cref{sec:overview} and, then, discuss sampling, pattern candidate generation, 
and top-$k$ summary construction  (\Cref{sec:sampling-provenance,sec:generating-pattern-c,sec:estimating-pattern-c,sec:computing-top-k-summ}).
We 
present experiments 
in \Cref{sec:experi}, discuss related work in \Cref{sec:rel-work}, and conclude in \Cref{sec:conclusions}.

 \section{Background}\label{sec:prelim}

\subsection{Datalog}
\label{sec:dl}

A Datalog program $\query$ consists of a finite set of rules $r: \rel{R}(\varvec{X}) \dlImp \rel{g_1}(\varvec{X_1}),$ $\ldots,$ $\rel{g_m}(\varvec{X_n})$ 
where $\varvec{X_j}$ denotes a tuple of variables and/or constants. 
$\rel{R}(\varvec{X})$ is the \emph{head} of
the rule, denoted as $\headOf{r}$, and $\rel{g_1}(\varvec{X_1}), \ldots,
\rel{g_m}(\varvec{X_n})$ 
is the \emph{body} (each $\rel{g_j}(\varvec{X_j})$
is a \emph{goal}).
We use $\varsOf{r}$ to denote 
the set of variables in $r$. 
The set of relations in the schema
 over which $\query$ is defined is referred to as the extensional database
 (EDB), while relations defined through rules in $\query$ form the
 intensional database (IDB), i.e., 
  the heads of rules.
All rules $r$ of $\query$ 
have to be \emph{safe}, i.e., every
variable in $r$ must occur in a positive literal in $r$'s body.
\revm{Here, we consider union of conjunctive queries with negation 
  and comparison predicates 
  (\ucqNegOrd).
Thus, all rules of a query have the same head predicate and
goals 
in the body are either \emph{literals}, i.e.,
 atoms $\rel{L}(\varvec{X_j})$  or their negation $\neg
 \rel{L}(\varvec{X_j})$, or
 comparisons of the form $a \diamond b$ where $a$ and $b$ are either constants or variables and $\diamond \in \{< , \leq, \neq, \geq, >\}$}.
For example, considering the Datalog rule $r_1$ from \Cref{fig:running-example-db},
$\headOf{r_1}$ is $\rel{AL}(N,R)$ and 
$\varsOf{r_1}$ is $\{I,N,T,R,E,P\}$. The rule is safe since all head variables occur in the body and all goals are positive. 
\RevDel{As mentioned in \Cref{sec:introduction}, our summarization approach supports \ucqNegOrd (unions of conjunctive queries with negation and inequalities). 
That is, a restriction of FO queries where all literals in the body of a rule are EDB relations and all rules have the same head predicate.}

The active domain $\adom(D)$ of a database $D$ (an instance of EDB relations) is the set of all constants that appear in $D$.
We assume the existence of a universal domain of values $\dataDomain$ which is a superset of the active domain
of every database.
The result of evaluating 
$\query$ over 
$D$, denoted as $\query(D)$, contains all IDB tuples $Q(t)$ for which there exists a successful rule derivation with head $Q(t)$.
A derivation of $r$ is the result of applying a valuation $\val: \varsOf{r} \to \dataDomain$ which maps the variables of 
$r$ to constants \revm{such that all comparisons of the rule hold, i.e., for each comparison $\dlcomp(\varvec{Y})$ the expression  $\dlcomp(\val(\varvec{Y}))$ evaluates to true.}
Note that the set of all derivations of $r$ is independent of $D$ since the constants of a derivation are from $\dataDomain$.
Let $\varvec{c}$ be a list of constants from $\dataDomain$, one for each variable of $r$.
We use $r(\varvec{c})$ to denote the rule derivation that assigns constant $c_i$ to variable $X_i$ in $r$.
Note that variables are ordered by the position of their first occurrence in $r$, 
e.g., the variable order for $r_1$ (\Cref{fig:running-example-db}) is $(N,R,I,T,E,P)$.
A rule derivation is successful (failed) if all (at least one of) the goals 
in its body are successful (failed). 
A positive/negative literal goal is successful if the corresponding tuple exists/does not exist. 
A missing answer for 
$\query$ and 
$D$ is an IDB tuple $Q(t)$ for which all derivations failed. 
For a given 
$D$ and 
$r$, 
we use $\db \models r(\varvec{c})$ to denote that 
$r(\varvec{c})$ is successful over $\db$.
Typically, as mentioned in \Cref{sec:introduction}, not all 
failed derivations constructed in this way are sensible, e.g., a derivation may assign an integer to an attribute of type string. 
We allow users to control which values to consider for which attribute (see~\cite{LS17,LL18j}).
For simplicity, however, 
we often assume a single universal domain $\dataDomain$.

\subsection{Provenance Model}
\label{sec:rule-deri}

We now explain the provenance model 
introduced in \Cref{ex:expl-whynot-pug}. 
As demonstrated in~\cite{LL18j}, this provenance 
model is equivalent to the provenance semiring model for positive queries~\cite{green2007provenance} and to its extension for first-order (FO) formula~\cite{T17}.
In our 
model, existing IDB tuples are connected to the successful rule derivations that derive them while missing tuples are connected to all failed derivations that could have derived them.
Successful derivations are connected to successful goals. 
Failed  derivations 
are only connected to failed goals (which justify the failure).
Nodes in provenance graphs  carry two types of labels: (i) a label that determines the node type (tuple, rule, or goal) and additional information, e.g., the arguments and rule identifier of a derivation, and (ii) a label indicating success/failure. 
We encode 
(ii) as colors in visualizations of such graphs.  
As shown in~\cite{LS17}, provenance in this model can equivalently be represented as sets of successful and failed rule derivations as long as the success/failure state of goals are known. 
\BGDel{That is, the translation between provenance graphs and such ``extended'' rule derivations is lossless.}{repeats the point made in the previous sentence}
\BGDel{Thus, alternatively, we can define provenance as a set of rule derivations for which we record the success of their goals.}{}
\BGDel{For that purpose, we introduce \emph{annotated rule derivations} which are rule derivations paired with a list of boolean values  recording the state of goals.}{}

\begin{defi}\label{def:annot-rule-deri}
  Let 
  $r$ be a Datalog rule $\rel{Q}(\varvec{X}) \dlImp \rel{R_1}(\varvec{X_1}), \ldots$
  $,\rel{R_{l}}(\varvec{X_l}), \dlNeg \rel{R_{l+1}}(\varvec{X_{l+1}}),$ $\ldots, \dlNeg \rel{R_{m}}(\varvec{X_n}), \dlcomp(\varvec{Y_1}), \ldots, \dlcomp(\varvec{Y_k})$
  where each
    $\dlcomp_i$ is a comparison. Let  $D$ be a database. 
  An \emph{annotated derivation}  $d = \anAnnotRule$ of 
  $r$ consists of a list of constants $\varvec{c}$  
  and a list of goal annotations
  $\varvec{g} = (g_1, \ldots, g_m)$ 
  such that (i) $r(\varvec{c})$ is a rule derivation, and (ii)
 $g_i = \greenT \mathtext{if } i  \leq l \wedge D \models R_i(\varvec{c_i}) \mathtext{or } i  > l \wedge D \not\models R_i(\varvec{c_i})$ and $g_i = \redF$ otherwise.
\end{defi}

An example failed annotated  derivation  of 
rule $r_1$ (\Cref{fig:running-example-db}) is $d_1 = r_1(\cnst{central place}, \cnst{shared}, \cnst{8403}, \cnst{apt},  \cnst{east},  \cnst{130}) - (\greenT, \redF)$
from \Cref{fig:part-whynot-lic}.
That is, while 
  $\rel{A}(\cnst{8403}, \cnst{2016-11-09}, \cnst{130})$ failed, 
  $\rel{L}(\cnst{8403}, \cnst{central place},	\cnst{apt}, \cnst{shared}, \cnst{queen anne}, \cnst{east})$ is successful.
 Using annotated derivations, we can explain the existence or absence of a (set of) query result tuple(s).
 We use $\allAnnotRules(\query,\db, r)$ to denote all annotated derivations of rule $r$ from $\query$ according to $\db$, $\allAnnotRules(\query,\db)$ to denote $\bigcup_{r \in \query} \allAnnotRules(\query, \db, r)$, and $\allAnnotRules(\query, \db, t)$ to denote the subset of $\allAnnotRules(\query, \db)$ with head $Q(t)$. Note that by definition,
 valuations that violate any comparison of a rule are not considered to be rule derivations.
\BG{Note that constants from a rule, \cnst{2016-11-09} and \cnst{queen anne} are  $r_1$ are not listed in rule derivation.
  The constants \cnst{2016-11-09} and \cnst{queen anne} are  $r_1$ Only assignments to variables are listed in rule derivations.}

We now define  \emph{provenance questions (PQ)}. 
Through the type of 
a PQ ($\why$ or $\whynot$), the user specifies whether she is interested in missing or existing results. 
In addition, the user provides a tuple $\tWithConsVars$ of constants (from $\dataDomain$) and 
placeholders to indicate what tuples she is interested in.   We refer to such tuples as pattern tuples 
(\textit{p-tuples} for short) and use bold font to distinguish them from tuples with constants only.
We use capital letters to denote placeholders and variables, and lowercase to denote constants.
We say a tuple $\tWithCons$ \emph{matches} a p-tuple $\tWithConsVars$,
written as $\tWithCons \matches \tWithConsVars$, if we can 
unify $\tWithCons$ with $\tWithConsVars$ by applying a valuation $\val$ 
that substitutes placeholders 
in $\tWithConsVars$ with constants from $\dataDomain$ such that $\val(\tWithConsVars) = \tWithCons$, e.g.,
$\rel{AL}(\cnst{plum},\cnst{shared}) \matches \rel{AL}(N, \cnst{shared})$ using $\val \defas N \to \cnst{plum}$.
The provenance of all existing (missing) tuples  matching $\tWithConsVars$  constitutes the answer of a $\why$ ($\whynot$) PQ.

\begin{defi}[Provenance Question]\label{def:question}
   Let $\query$ be a query.
  A provenance question $\aProvQ$ over $\query$ is a pair $(\tWithConsVars, \qType)$ where $\tWithConsVars$ is a p-tuple 
   and $\qType \in \{\why, \whynot\}$.
\end{defi}

 Bob's 
 question from \Cref{ex:intro-db-scen} can be written as $\bobsQ = (\tBob,$ $\whynot)$ where \revm{$\tBob = \rel{AL}(N, \cnst{shared})$}, i.e.,
 Bob wants an explanation for \textit{all} missing answers \revm{where $R = \cnst{shared}$.} 
The graph shown in \Cref{fig:part-whynot-lic} is part of the provenance for $\bobsQ$.
\BGDel{$\explainq$ implicitly explaining PQs.}{}

\begin{defi}[Provenance]\label{def:provenance-model}
  Let $\db$ be a database, $\query$ an $n$-nary \revm{\ucqNegOrd} query,  
  and $\tWithConsVars$ an $n$-nary p-tuple. 
  We define the
  why and why-not provenance of $\tWithConsVars$ over $\query$ and $\db$ as:\\[-5mm]
  \begin{align*}
    \why(\query,\db,\tWithConsVars) &= \bigcup_{\tWithCons \matches \tWithConsVars \wedge \tWithCons \in \query(\db)} \why(\query,\db,\tWithCons)\\
  \why(\query,\db,\tWithCons) &= \{ d \mid 
                                 d \in \allAnnotRules(\query, \db, \tWithCons) \wedge \db \models d 
                                \}\\[1mm]
  \whynot(\query,\db,\tWithConsVars) &= \bigcup_{\tWithCons \matches \tWithConsVars \wedge \tWithCons \not\in \query(\db)} \whynot(\query,\db,\tWithCons)\\
  \whynot(\query,\db,\tWithCons) &= \{ d \mid 
  d \in \allAnnotRules(\query, \db, \tWithCons) \wedge \db \not \models d 
                                   \}
  \end{align*}
The provenance $\explainq(\aProvQ)$ of a provenance question $\aProvQ$ is:
  \begin{align*}
    \explainq (\aProvQ) =
    \begin{cases}
      \why(\query, \db, \tWithConsVars) & \mathtext{if} \aProvQ = (\tWithConsVars, \why)\\
      \whynot(\query, \db, \tWithConsVars) & \mathtext{if} \aProvQ = (\tWithConsVars, \whynot)
    \end{cases}
  \end{align*}
\end{defi}

\BGDel{To capture the provenance, we use a query instrumentation technique called emph{firing rules}
which has introduced in~cite{kohler2012declarative} and extended for negation and why-not by PUG~cite{LS17}
(with the efficient rewriting algorithm).}{seems disconnected from the rest of this section. Where should this live? Intro? or Sec 4 or 5?}

\section{Problem Definition}
\label{sec:prob-def}

We now formally define the problem addressed in this work: how to summarize 
the provenance $\explainq(\aProvQ)$ 
of a 
provenance question $\aProvQ$. 
For that, we introduce derivation patterns that concisely describe provenance 
and, then, define provenance summaries as sets of such patterns. We also develop quality metrics for such summaries that model completeness and informativeness as introduced 
in \Cref{sec:introduction}.
\subsection{Derivation pattern}
\label{sec:d-pattern}

A \emph{derivation pattern} is an annotated rule derivation whose arguments can be both constants and placeholders.

\begin{defi}[Derivation Pattern]\label{def:deri-patt}
  Let $r$ be a rule 
  with $n$ variables and $m$ goals and 
  $\placeholders$ an infinite set of placeholders.
  A \emph{derivation pattern} 
  $p = r(\bar{e}) - (\bar{g})$ 
  consists of a list $\bar{e}$ of length $n$ where $e_i \in \dataDomain \cup \placeholders$ 
  and $\bar{g}$, a list of $m$ booleans. 
\end{defi}

Consider
 pattern $p_1 = r_1(N,\cnst{shared},I,\cnst{apt},E,P) - (\greenT, \redF)$
 for rule $r_1$ (\Cref{fig:running-example-db}) shown in \Cref{fig:summ-whynot-lic}. 
 Pattern $p_1$ represents the set of failed derivations 
 matching $\rel{AL}(N,\cnst{shared})$
where 
the listing is an apartment (\cnst{apt})  
and for which the $1^{st}$ goal succeeded (the listing exists in Queen Anne) and the $2^{nd}$ goal failed (the listing is not available on Nov 9th, 2016).
We use $\parg{p}{i}$ to denote the $i$th argument of pattern $p$ and omit goal annotations if they are irrelevant to the discussion.

\subsection{Pattern Matches}
\label{sec:patt-mat}
A derivation pattern $p$ 
represents the set of derivations 
 that ``match'' the pattern.
We define \emph{pattern matches} as valuations that replace the placeholders in a pattern with constants from $\dataDomain$. 
In the following, we use $\plOf{p}$ to denote  the set of placeholders of a pattern $p$.

\begin{defi}[Pattern Matches]\label{def:p-match}
A derivation pattern $p = r(\bar{e})-(\bar{g_1})$ matches an annotated rule derivation $\deri = r(\bar{c}) - (\bar{g_2})$, 
written as $p \matches \deri$, if there exists a valuation $\val:\plOf{p} \to \dataDomain$ such that $\val(p) = \deri$ and $\bar{g_1} = \bar{g_2}$. 
\end{defi}

Consider 
\revm{$p_1 = r_1(N,\cnst{shared},I,\cnst{apt},E,P) - (\greenT, \redF)$}
and 
\revm{$\deri_1 = r_1(\cnst{central place},\cnst{shared},\cnst{8403},\cnst{apt},\cnst{east},\cnst{130}) - (\greenT, \redF)$} 
(from Fig.
 \ref{fig:summ-expl-whynot}). 
We have $p_1 \matches \deri_1$ since the valuation \revm{$N \to \cnst{central place}$, $I \to \cnst{8403}$,  $E \to \cnst{east}$, and $P \to \cnst{130}$} maps $p_1$ to $\deri_1$ 
and the goal indicators ($\greenT,\redF$) are same for  $p_1$ and $\deri_1$.

\subsection{Provenance Summary}
\label{sec:prov-summ}
We call $p$ a pattern for 
a p-tuple $\tWithConsVars$ if $p$ and $\tWithConsVars$ agree on constants, e.g., $p_1$ is a pattern for 
$\tBob = \revm{\tBobFull}$ since $\parg{p}{2} = \parg{\tBob}{2} = \textsf{shared}$.
We use $\pat(\query,\tWithConsVars)$ to denote the set of all 
patterns for $\tWithConsVars$ and $\query$.

\begin{defi}[Provenance Summary]\label{def:summ-deri}
  Let 
  $\query$ be a \\$\ucqNegOrd$ query  
  and $\aProvQ = (\tWithConsVars, \qType)$ a
provenance question. A \emph{provenance summary} $\sprov$ for $\aProvQ$ is a subset of $\pat(\query,\tWithConsVars)$.
\end{defi}

Based on the \Cref{def:summ-deri}, any subset of $\pat(\query, \tWithConsVars)$ is 
a  summary. However, summaries do differ in conciseness, informativeness, and completeness.
Consider a 
summary 
for $\bobsQ$ 
consisting of 
\revm{$p_2 = r_1(N,\cnst{shared},I,T,E,P) - (\greenT,\redF)$ and $p_2' = (N,\cnst{shared},I,T,E,P) - (\redF,\redF)$.} This summary covers 
$\explainq(\bobsQ)$.\footnote{Pattern $p_2'' = r_1(N,\cnst{shared},I,T,E,P) - (\redF,\greenT)$ has no matches, because non-existing listings cannot be available.} 
However, the pattern only consists of placeholders and constants from $\bobsQ$ --- no new information is conveyed.
Pattern
$p_3 = r_1(\cnst{plum},\cnst{shared},$ $\cnst{9211},\cnst{apt},
      \cnst{east},
      \cnst{130}) \\- (\greenT,\redF)$ 
 consists only of constants. It provides detailed information 
but covers only one derivation.

\subsection{Quality Metrics}
\label{sec:qual-met}

We now introduce a quality metric that combines \emph{completeness} and \emph{informativeness}.
We define \emph{completeness} as the fraction of  $\explainq(\aProvQ)$ 
matched by a pattern. 
For a 
question $\aProvQ$, 
query $\query$, and database $\db$,
we use $\allMatches(\query,\db,p,\aProvQ)$ to denote all derivations in $\explainq(\aProvQ)$ that match a pattern $p$:
\begin{align*}
  \label{eq:match-in-prov}
  \allMatches(\query,\db,p,\aProvQ) \defas \{ d \mid d \in \explainq(\aProvQ) \wedge d \matches p \}
\end{align*}

Considering the pattern $p_1$ 
from \Cref{fig:summ-whynot-lic} and the derivation $\deri_1$ 
from \Cref{fig:part-whynot-lic},
we have \revm{$\deri_1 \in \allMatches(r_1,\db, p_1,\bobsQ)$}. 
\begin{defi}[completeness]\label{def:recall}
  Let 
  $\query$ be a query, $\db$ a database, $p$ a pattern, and $\aProvQ$ a provenance
  question. 
  The \emph{completeness} 
  of $p$ is defined as
  $\completeness(p) =
  \frac{|\allMatches(\query,\db,p,\aProvQ)|}{|\explainq(\aProvQ)|} $.
\end{defi}

We 
also define  \textit{informativeness} which measures how much new information is conveyed by a pattern.

\begin{defi}[Informativeness]\label{def:info}
  \hspace{-0.5mm}For a
  pattern $p$ and 
  question $\aProvQ$ with p-tuple $\tWithConsVars$, let 
  $\consOf(p)$ 
  and $\gconsOf(\pt)$ 
denote the number of constants in $p$ 
and $\tWithConsVars$, respectively. 
The \emph{informativeness} $\informativeness(p)$ of $p$ is defined as
$\informativeness(p) = \frac{\consOf(p) - \gconsOf(\tWithConsVars)}{\arityOf(p) - \gconsOf(\tWithConsVars)}$.
\end{defi}

For Bob's question $\bobsQ$ and pattern \revm{$p_1 =r_1(N,\cnst{shared},I,$ $\cnst{apt},
      E,
      P)- (\greenT,\redF)$}, 
we have $\informativeness(p_1) = 0.2$ 
because
$\consOf(p_1)$ is 2 (\textsf{shared} and \textsf{apt}), 
$\gconsOf(\tBob)$ is 1 (\textsf{shared}), and 
$\arityOf(p_1)$ is 6 (all placeholders and constants).
\BGDel{In \Cref{sec:measure-patt-qual}, we explain why measuring the quality of patterns are challenging wrt. pattern matches (\Cref{def:p-match}) 
because finding matches over $\explainq$ to compute completeness is not efficient, even not feasible (e.g., $\whynot(\query,\db,t)$  is typically unavailable as mentioned in \Cref{ex:expl-whynot-summ}).
In \Cref{sec:approx-summ}, we introduce the approach that measures the quality over a sample we create and compute real approximate qualities based on the ratio between the size of sample and $\explainq$.}{Do we need to know this here?} 
\revb{We generalize completeness and informativeness to sets of patterns (summaries) as follows. The completeness of a summary $\sprov$ is the fraction of $\explainq(\aProvQ)$ covered by at least one pattern from $\sprov$.
  For patterns $p_2$ and $p_2'$ 
from \Cref{sec:prov-summ}, 
we have $\completeness(\{p_2,p_2'\}) = \completeness(p_2) + \completeness(p_2') = 1$.
  Note that $\completeness(\sprov)$ may not be equal to the sum of $\completeness(p)$ for $p \in \sprov$ since the set of matches for two patterns may overlap.
  We will revisit overlap in \Cref{sec:computing-top-k-summ}.
  We define informativeness as the average informativeness of
  the patterns in $\sprov$. 
  \begin{align*}
  \completeness(\sprov) &=
    \frac{|\bigcup_{p \in \sprov}\allMatches(\query,\db,p,\aProvQ)|}{|\explainq(\aProvQ)|}
    &\informativeness(\sprov) &= \frac{\sum_{p \in \sprov} \informativeness(p)}{\card{\sprov}}
  \end{align*}
 }
The score of a summary $\sprov$ is then defined as the harmonic mean of completeness and informativeness, i.e., 
\revb{$\score(\sprov) = 2 \cdot \frac{\completeness(\sprov) \cdot \informativeness(\sprov)}{\completeness(\sprov) + \informativeness(\sprov)}$}. \BG{define these for sets of patterns}
We are now ready to define the \textit{top-k provenance summarization problem} which, 
given a provenance question $\aProvQ$, returns the top-$k$ patterns for $\aProvQ$ wrt. \revb{$\score(\sprov)$}.
\vspace{1mm}
\begin{itemize} 
\item \textbf{Input:} 
  A query $\query$, database $\db$, provenance question $\aProvQ = (\tWithConsVars, \qType)$, 
   $k \in \mathbb{N} \geq 1$. \vspace{-2mm}
 \item \textbf{Output:} 
   \begin{minipage}{0.7\linewidth}
 \begin{align*}
    \sprov(\query,\db,\aProvQ,k) &= \revb{\argmax_{\sprov \subset \pat(\query, \tWithConsVars) \wedge \card{\sprov} \leq k} \score(\sprov)}
  \end{align*}
\end{minipage}
\end{itemize}

\SL{USED IN SEC8 FOR TOPK
\revb{When computing a solution to this problem in addition to finding candidate
  patterns, we have to calculate the score of a combinatorial number of sets of
  patterns. 
  Calculating the informativeness of a set of patterns
  is efficient since it only requires access to the patterns while 
  computing completeness 
  is not sufficient. Specifically, the completeness of a set
  of patterns cannot be computed based on the completeness of the patterns of
  the set since the sets of derivations matched by two patterns may overlap. The
  naive approach to address this problem is to compute the matches for each
  candidate set of patterns independently by finding all matches in the
  data. This requires us to run a number of queries that equal to the number of
  candidate patterns sets which is not feasible. However, we identify two
  special cases where the overlap between the matches of two patterns can be
  computed without accessing the data. We will check for these cases in our
  algorithm solving the top-k provenance summarization problem to compute bounds
  on the score of a candidate pattern set. We say pattern $p_2$
  \emph{generalizes} pattern $p_1$ written as $p_1 \pleq p_2$ if $p_2$ and $p_1$
  agree on all arguments that are constant in $p_2$ and all other arguments of
  $p_2$ are placeholders. For instance, $(X,Y,\cnst{a})$ generalizes
  $(X,\cnst{b},\cnst{a})$.  From the definition for matching immediately follows
  that if $p_1 \pleq p_2$ then
  $\allMatches(\query,\db,p_1,\aProvQ) \subseteq
  \allMatches(\query,\db,p_2,\aProvQ)$ since any derivation matching $p_1$ also
  matches $p_2$. That is, if $p_1 \pleq p_2$ then we know that
  $\completeness(\{p_1, p_2\}) = \completeness(p_2)$. We say pattern $p_1$ and
  $p_2$ are \textit{conflicting} written as $p_1 \pind p_2$ if there exists an
  $i$ such that $p_1[i] = c_1 \neq c_2 = p_2[i]$, i.e., the patterns have a
  different constant at the same position $i$. If $p_1 \pind p_2$ then $\allMatches(\query,\db,p_1,\aProvQ) \cap
  \allMatches(\query,\db,p_2,\aProvQ) = \emptyset$ and we have $\completeness(\{p_1,p_2\}) = \completeness(p_1) + \completeness(p_2)$.}
\BG{Move this somewhere later maybe?}
}

 \begin{figure}[t]
\begin{minipage}{1\linewidth}
 \centering 
 \begin{minipage}{0.99\linewidth}
   \centering
     \includegraphics[width=1\columnwidth]{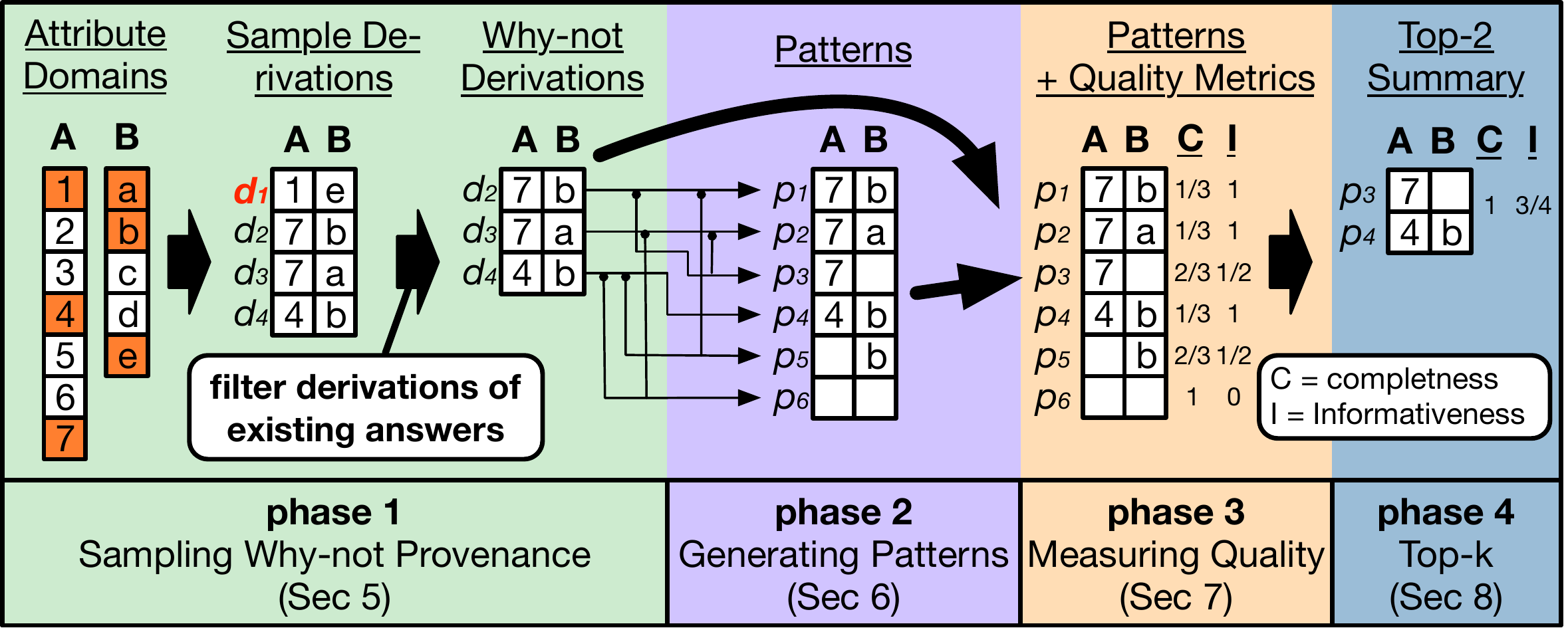}
  \end{minipage}
\\[0mm]
 \caption{Overview of why-not provenance summarization}
\label{fig:pug-summ-process}
\end{minipage}
\end{figure}

\section{Overview}\label{sec:overview}

Before describing our approach in detail in the following sections, we first give a brief overview. To 
  compute a top-$k$ provenance summary  $\sprov(\query, \db, \aProvQ, k)$ for a provenance question $\aProvQ$, 
  we have to (i) compute 
   $\explainq(\aProvQ)$, 
  (ii) enumerate all patterns that could be used in summaries, (iii) calculate matches between derivations and the patterns to calculate the completeness of sets of patterns, and (iv) find a set of up to $k$ patterns for $\aProvQ$ that has the highest score among all such sets of patterns. 
  To compute the exact solution to this problem, we would need to enumerate all derivations from $\explainq(\aProvQ)$. 
  However, this is not feasible for why-not provenance questions 
  since, as we will discuss in the following, the size of why-not provenance 
   $\whynot(\query,\db,\pt)$ is in $\oNotation{\card{\dataDomain}^n}$, i.e., linear in the size of the 
  data domain $\dataDomain$, but exponential in $n$, the maximal number of variables of a rule 
  from query $\query$ that is not bound to constants by 
  $\pt \in \aProvQ$.
  Instead, we present a heuristic approach that uses sampling and outsources most of the computation to a database for scalability.
  \Cref{fig:pug-summ-process} shows an overview of this approach.

\mypara{Sampling Provenance (Phase 1, \Cref{sec:sampling-provenance})}
  As shown in \Cref{fig:pug-summ-process} (phase 1), we develop a technique to compute a sample $\samp$ of $\sampSize$ derivations  from $\whynot(\query,\db,\pt)$ that is unbiased with high probability. 
  We create $\samp$ by (i) randomly sampling a number of $\osSize > \sampSize$ values from the domain of each attribute (e.g., A and B in \Cref{fig:pug-summ-process}) individually, (ii) zip these samples 
  to create derivations,
  and (iii) remove derivations for existing results (e.g., the derivation $d_1$ highlighted in red) to compute a sample of  $\whynot(\query,\db,\pt)$ 
  that with high probability is at least of size $\sampSize$ (in \Cref{fig:pug-summ-process} we assumed $\sampSize =  3$). 
  For why-provenance, 
  we  sample directly from the full provenance for $\aProvQ$ computed using our  query instrumentation
  technique from~\cite{LL18j,LS17}.

\mypara{Enumerating Pattern Candidates (Phase 2, \Cref{sec:generating-pattern-c})}
The number of patterns for a rule 
  with $m$ goals and $n$ variables 
  is in $O(\card{\dataDomain + n}^n \cdot 2^m)$. 
  Even if we only consider patterns that match at least one derivation from 
  $\samp$, 
  the number of patterns may still be a factor of $2^n$ larger than $\samp$. 
  We adopt a heuristic from~\cite{EA14} that, in the worst case, generates  quadratically many patterns (in the size of 
  $\samp$).
As shown in \Cref{fig:pug-summ-process},
  we 
   generate a pattern  $p$ for each pair of derivations $d$ and $d'$ from $\samp$. If $d[i] = d'[i]$ then $p[i] = d[i]$. Otherwise $p[i]$ is a fresh placeholder (shown as an empty box in \Cref{fig:pug-summ-process}).

  \mypara{Estimating Pattern Coverage (Phase 3, \Cref{sec:estimating-pattern-c})}
  To be able to compute the completeness metric
  of a pattern set which is required for scoring  pattern sets in the last step, we need to determine what derivations are covered by which pattern and which of these 
  belong to $\whynot(\query,\db,\pt)$. 
  We estimate completeness based on 
  $\samp$. 
  The informativeness of a pattern can be directly computed from the pattern.

\mypara{Computing the Top-$k$ Summary (Phase 4, \Cref{sec:computing-top-k-summ})}
In the last step 
  (phase 4 in \Cref{fig:pug-summ-process}), we generate sets of up to $k$  patterns from the set of patterns produced in the previous step, rank them based on their scores, and return the set with the highest score as the top-$k$ summary.
We apply a heuristic best-first search method that utilizes efficiently computable bounds for the completeness of sets of patterns to prune the search space.

\begin{figure}[t]
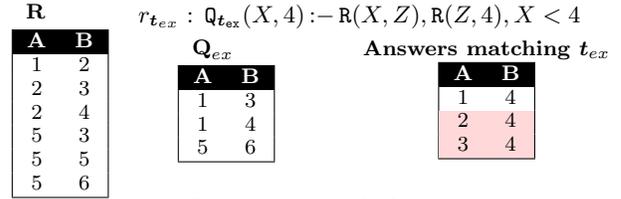


  \textbf{Query: }
  $\rEx:\, \rel{\qEx}(X,Y) \dlImp \rel{R}(X,Z), \rel{R}(Z,Y), X < Y$
  \vspace{1mm}

\textbf{PQ: }
$\pqEx = (\tEx,\whynot)$ where $\tEx = \tExFull$
\vspace{1mm}

\textbf{Query Unified With P-Tuple $\tEx$: }
$$\,\,\,\,\,\,\,\,\,\,\,\,\,\,\,\rExun:\, \rel{\query_{\tEx}}(X,4) \dlImp \rel{R}(X,Z), \rel{R}(Z,4), X < 4$$
\begin{minipage}{1\linewidth}
\vspace{-6mm}
  {\small
\begin{minipage}{0.25\linewidth}
  \begin{tabular}{|cc|}
    \multicolumn{2}{l}{\textbf{R}} \\[.5mm] \hline
    \thead{A}   & \thead{B}        \\
    1           & 2                \\
    2           & 3                \\
    2           & 4                \\
    5           & 3                \\
    5           & 5                \\
    5           & 6                \\
    \hline
  \end{tabular}
\end{minipage}
\begin{minipage}{0.25\linewidth}
  \begin{tabular}{|cc|}
    \multicolumn{2}{l}{$\textbf{Q}_{ex}$} \\[.5mm] \hline
    \thead{A}   & \thead{B}        \\
    1           & 3                \\
    1           & 4                \\
    5           & 6                \\
    \hline
  \end{tabular}
\end{minipage}
\begin{minipage}{0.45\linewidth}
  \centering
  \textbf{Answers matching $\tEx$}\\
  \begin{tabular}{|cc|}
 \hline
    \thead{A}   & \thead{B}        \\
    1           & 4                \\
    \hlrow    2 & 4                \\
    \hlrow    3 & 4                \\
    \hline
  \end{tabular}
\end{minipage}
}
\end{minipage}
\caption{\label{fig:technical-running-example} \revm{Running example for summarization}}
\end{figure}

\section{Sampling Why-not Provenance}\label{sec:sampling-provenance}

In this section, we first discuss how to efficiently generate a sample $\samp$ of annotated derivations of a given size $\sampSize$ from the why-not provenance $\whynot(\query, \db, \tWithConsVars)$ for a provenance question (PQ) $\aProvQ$ (phase 1 in \Cref{fig:pug-summ-process}). This sample will then be used in the following phases of our summarization algorithm.
We introduce a running example in \Cref{fig:technical-running-example} and use it through-out \Cref{sec:sampling-provenance} to \Cref{sec:computing-top-k-summ}.
  Consider the example query $\rEx$ shown on the top of
\Cref{fig:technical-running-example} which returns start- and end-points of paths of length
2 in a graph with integer node labels such that the end-point is labeled with a
lareger number than the start-point. 
Evaluating 
$\rEx$ over the example instance $\rel{R}$ from the same figure yields three results: $\rel{\qEx}(1,3)$, $\rel{\qEx}(1,4)$, and $\rel{\qEx}(5,6)$.
In this example, we want to explain missing answers of the form $\rel{\qEx}(X, 4)$, i.e., 
answering the 
PQ $\pqEx$ 
from \Cref{fig:technical-running-example}.
Recall that,
$\whynot(\query,\db,\tWithConsVars)$ for p-tuple $\pt$ consists of all derivations of tuples $\tWithCons \not\in \query(\db)$ where $\tWithCons \matches \tWithConsVars$.
Assuming $\dataDomain = \{1,2,3,4,5,6\}$, on the bottom right of \Cref{fig:technical-running-example} we show all missing and existing answers matching $\tEx$ (missing answers are shown with red background).

\subsection{Naive Unbiased Sampling}\label{sec:naive-sampl-appr}
  To generate all derivations for 
  missing answers, we can bind the variables of each rule
$r$ of a query $\query$ to the constants from $\tWithConsVars$ to ensure that only
derivations of results which match the PQ's p-tuple $\pt$ are generated. We refer to this process as unifying $\query$ with $\tWithConsVars$. 
For our running example, this yields the rule $\rExun$ shown in \Cref{fig:technical-running-example}.
The naive way to 
create a sample of derivations from $\whynot(\query, \db, \tWithConsVars)$ using this rule is to repeatably sample a value from $\dataDomain$ for each variable, then check whether (i) the predicates of the rule are fulfilled and (ii) the resulting rule derivation computes a missing answer. 
For example, for $\rExun$, we may choose $X = 2$ and $Z = 2$ and get a derivation $d_1 = \rExun(2,2)$. The derivation $d_1$ fulfills the predicate $X < 4$ and its head $\rel{\qEx}(2,4)$ is a missing answer. Thus, $d_1$ belongs to the why-not provenance of $\tEx$. 
Then, to get an annotated rule derivation, we determine its goal annotations by checking whether the tuples corresponding to the grounded goals of the rule exists in the database instance. For this example, $d_1 = \rExun(2,2)-(\redF,\greenT)$ since the first goal $\rel{R}(2,2)$ fails, but the second goal $\rel{R}(2,4)$ succeeds.
There are two ways of how this process can fail  to produce a derivation of $\whynot(\query, \db, \tWithConsVars)$: 
(i) a predicate of the rule may be violated by the bindings generated in this way (e.g., if we would have chosen $X = 5$, then $X < 4$ would not have held) and (ii) the derivation may derive an existing answer, 
e.g., if $X = 1$ and $Z = 3$, we get the failed derivation $\rExun(1,3)$ of the existing answer $\rel{\qEx}(1,4)$.

\mypara{Analysis of Naive Sampling}
  If we repeat the process described above until it has returned $\sampSize$ 
  failed derivations, then this produces an unbiased sample of $\whynot(\query, \db, \tWithConsVars)$. 
  Note that, technically, there is no guarantee that the process will ever terminate since it may repeatedly produce derivations that do not fulfill a predicate or derive existing answers.
Observe that, typically the amount of missing answers is significantly larger than the number of answers, i.e., $\card{\whynot(\query, \db, \pt)} \gg \card{\allAnnotRules(\query,\db,\pt)-\whynot(\query, \db, \pt)}$. As a result, any randomly generated derivation 
is with high probability in $\whynot(\query, \db, \pt)$.
We will explain how to deal with derivations that fail to fulfill predicates in \Cref{sec:samp-prov}.

\mypara{Batch Sampling} 
  A major shortcoming of the naive sampling approach is that it requires us to evaluate queries 
  to test for every produced derivation $d$ whether it derives a missing answer (
  $\headOf{d} \not\in \query(\db)$) and to determine its goal annotations  by checking for each grounded goal $R(\vec{c})$ or $\neg R(\vec{c})$  whether $R(\vec{c}) \in \db$. It would be more efficient to model sampling as  a single batch computation that we can outsource to a database system and that can be fused into a single query with the other phases of the summarization process to avoid unnecessary round-trips between our system and the database.
  However, for batch sampling, we have to choose upfront how many samples to create, but not all such samples will end up being why-not provenance or fulfill the rule's predicates. To ensure with high probability that the batch computation returns at least $\sampSize$ derivations  from $\whynot(\query, \db, \pt)$, we use a larger sample size $\osSize \geq \sampSize$ such that the probability that the resulting sample contains at least $\sampSize$ derivations from $\whynot(\query, \db, \pt)$ is higher than a configurable threshold $\osThresh$ (e.g., 99.9\%).
We refer to this part of the process as \textit{over-sampling}.
We discuss how to generate a query that computes a sample of size $\osSize$
  in \Cref{sec:samp-prov} and, then, discuss how to determine $\osSize$ in \Cref{par:samp-agg}.

\RevDel{
  we introduced an approach which instruments an input Datalog program to
out-source provenance computation to a database system.
\mypara{Challenge 1}\label{par:chal1}
As mentioned in Section\,\Cref{sec:introduction},
the number of derivations in $\whynot(\query,\db,\tWithConsVars)$ is in the order of $\oNotation{\card{\dataDomain}^n}$ where $n$ is the maximal number of variables in a rule of $\query$ excluding variables bound to constants by $\tWithConsVars$ (we call these unbound variables).
Thus, it is not
feasible to
enumerate all derivations in  $\whynot(\query,\db,\pt)$ during summarization. To develop a batch sampling method we have to overcome several challenges: (i) how to express sampling using queries and (ii) how to determine $\osSize$, the number of derivations that the process should produce such that the chance of getting a sample with at least $\sampSize$ derivation from the why-not provenance is larger than a configurable success probability $\osThresh$.
We present our solution for the first challenge in \Cref{sec:samp-prov} and discuss how to determine $\osSize$ in \Cref{par:samp-agg}.
}

\subsection{Batch Sampling Using Queries}
\label{sec:samp-prov}

\RevDel{Our batch sampling technique is based on the following observations: (i) we can partition the set of derivations for tuples matching $\pt$ from $\aProvQ$ into derivations for existing
and missing tuples. The second partition is
$\whynot(\query, \db, \pt)$. Ignoring the goal annotations for now, for any derivation $d \in \allAnnotRules(\query, \db, \pt)$ that derives a tuple $t \matches \pt$ we can decide its membership in $\whynot(\query, \db, \pt)$ by checking whether $t \not\in \query(\db)$
(using the original query result only); (ii) typically $\card{\whynot(\query, \db, \pt)} \gg \card{\allAnnotRules(\query,\db,\pt)-\whynot(\query, \db, \pt)}$ and, thus, any randomly generated derivation for some $t \matches \pt$ is with high probability in $\whynot(\query, \db, \pt)$; and (iii) given a derivation $d$ without goal annotations, we can determine the goal annotations by checking for each grounded goal $R(\vec{c})$ or $\neg R(\vec{c})$  whether  $R(\vec{c}) \in \db$.
Based on these observations, we devise the algorithm
that generates
$\samp$ of size $\sampSize$.}

For simplicity, we limit the discussion to 
queries with a single rule, e.g., \revm{the query $\rEx$ 
  from \Cref{fig:technical-running-example}.
We discuss queries with multiple rules at the end of this section.
The query we generate to produce a sample of size $\osSize$
consists of three steps: generating derivations, filtering derivations of existing answers, determining goal annotations.}

\mypara{1. Generating Derivations}\label{par:samp-gen-derivations}
\revm{We first generate a query that creates a random sample $\overSamp$ of $\osSize$ derivations (not annotated) for which there exists an annotated version in $\allAnnotRules(\query,\db,\pt)$ (all annotated derivations 
  that match head $\query(\pt)$).}
Consider a single rule $r$ with $m$ literal goals, $n$ variables, and $h$ head variables:
$r: \rel{Q}(\varvec{X}) \dlImp \rel{g_1}(\varvec{X_1}),\ldots,\rel{g_m}(\varvec{X_m}),$ \revm{$\dlcomp_1(\varvec{Y_1}),$ $\ldots, \dlcomp_k(\varvec{Y_l})$}.
Let $R_i$ be the relation accessed by goal $\rel{g_i}$, i.e., $\rel{g_i}(\varvec{X_i})$ is either $R_i(\varvec{X_i})$ or $\neg R_i(\varvec{X_i})$.
\BG{For now, we assume that the rule does not contain predicates (how we deal with predicates will be discussed later).}
Let $k$ be the number of
head variables bound by the p-tuple $\pt$ from the 
question $\aProvQ$ for which we are sampling. We use
$\varvec{Z} = Z_{1}, \ldots, Z_{u}$ to denote the $u = n - k$ variables of $r$ that are not bound by $\pt$.
Recall that, to only consider derivations matching $\pt$, we unify the rule with $\pt$ by binding variables in the rule to the corresponding constants from $\pt$. 
We use $\run$ to denote the resulting unified rule.
Note that we will describe our summarization techniques using derivations and patterns for $\run$. Patterns for $r$ can be trivially reconstructed from the results of summarization by plugging in constants from $\pt$.
To generate derivations for such a query, 
we sample $\osSize$ values for each unbound variable independently with replacement, and, then, combine the resulting samples into a sample of $\allAnnotRules(\query, \db, \pt)$ modulo goal annotations.
\revm{Predicates comparing constants with variables, e.g., $X < 4$ in $\rExun$, are applied before sampling to remove values from the domain of a variable that cannot result in derivations fulfilling the predicates.}
Similar to~\cite{LL18j,LS17}, we assume that the user specifies the domain
$\dataDomain_A$ for each attribute $A$ as a unary query that returns
$\dataDomain_A$ (we provide reasonable defaults to avoid overloading the user).
We extend the relational algebra with two operators to be able to express sampling.
Operator $\sampleOp_{n}$ returns $n$ samples which are
chosen uniformly random with replacement from the input of the operator.
We use $\rownumOp_{A}$ to denote an operator that creates an
integer identifier for each input row that is stored in a column $A$
appended to the schema of the operator's input.
For each variable $X \in \varvec{Z}$
with $\attrsOf{X} = \{A_1, \ldots, A_j\}$ ($\attrsOf{X}$ denotes the set of attributes that variable $X$ is bound to by the
rule containing $X$), we create \revm{a query $\qVar{X}$ that unions the domains of these attributes, then applies  predicates
that compare $X$ with constants, and then samples $\osSize$ values.}\\[-4.5mm]
$$\qVar{X} \defas \rownumOp_{id}(\sampleOp_{\osSize}(\revm{\selection_{\checkVarPredCond{X}}(\rename_{X}}((\dataDomain_{A_1} \union \ldots \union \dataDomain_{A_j}))))$$

Here, $\checkVarPredCond{X}$ is a conjunction of all the predicates from $\run$ that compare $X$ with a constant.
The purpose of $\rownumOp$ is to allow us to use natural join to ``zip'' the samples for the individual variables into bindings for all variables of $\run$:\\[-4.5mm]
$$\qBind \defas \revm{\selection_{\checkJoinPreds}}(\qVar{Z_1} \join \ldots \join \qVar{Z_u})$$

Here, $\checkJoinPreds$ is a conjunction of all predicates from $\run$ that compare two variables. Note that the selectivity of $\checkJoinPreds$ has to be taken into account when computing $\osSize$ (discussed in \Cref{par:samp-agg}).
Each tuple in the result of 
$\qBind$ encodes the bindings for one derivation $d$ of a tuple
$t \matches \tWithConsVars$.
\begin{exam}\label{ex:qbind}
  Consider unified rule $\rExun$ from \Cref{fig:technical-running-example}. Assume that 
  $\dataDomain_{A} = \dataDomain_{B} = \projection_{A}(R) \union \projection_B(R)$ and $\osSize = 3$. Variable $X$ is bound to attribute $A$ and $Z$ is bound to both $A$ and $B$. Thus, we generate the following queries:\\[-4mm]
  \begin{align*}
  \qVar{X}   & \defas \rownumOp_{id}(\sampleOp_{\osSize}(\selection_{X < 4}(\rename_{X}(\dataDomain_A))))                             \\
    \qVar{Z} & \defas \rownumOp_{id}(\sampleOp_{\osSize}(\rename_Z(\dataDomain_A \union \dataDomain_B))) \\
\qBind       & \defas \selection_{true}(\qVar{X} \join \qVar{Z})
  \end{align*}
  Evaluated over the example instance, this query may return:\\[-4mm] 
\begin{center}
\revm{
{\small
  \begin{minipage}{0.25\linewidth}
    \textbf{$\qVar{X}$}
    \begin{tabular}{|cc|} \hline
      \thead{id} & \thead{X}               \\ \hline
      1            & 1                         \\
      2            & 2                         \\
      3            & 2                         \\
      \hline
    \end{tabular}
  \end{minipage}
  \begin{minipage}{0.25\linewidth}
    $\qVar{Z}$
    \begin{tabular}{|cc|} \hline
      \thead{id} & \thead{Z}               \\ \hline
      1            & 4                         \\
      2            & 2                         \\
      3            & 4                         \\
      \hline
    \end{tabular}
  \end{minipage}
  \begin{minipage}{0.4\linewidth}
    $\qBind$
    \begin{tabular}{|ccc|} \hline
      \thead{id} & \thead{X} & \thead{Z} \\ \hline
      1            & 1           & 4           \\
      2            & 2           & 2         \\
      3            & 2           & 4           \\
      \hline
    \end{tabular}
  \end{minipage}
}
}
\end{center}
\end{exam}

\mypara{2. Filtering Derivations of Existing Answers}\label{para:samp-filter-existing}
We now construct a query $\qDer$, which checks for each derivation $d \in \overSamp$ for a tuple $t \matches \pt$ whether $t \not\in \query(\db)$ and only retain derivations
  passing this check.
  This is achieved by anti-joining ($\antijoin$) $\qBind$
  with $\query$ which we restricted to tuples matching $\pt$ since only such tuples can be derived by $\qBind$.
  \\[-4mm]
$$\qDer \defas \qBind \antijoin_{\checkExistsCond} \selection_{\checkPtCond}(\query)$$

The query $\qDer$
uses condition $\checkExistsCond$ which equates attributes from $\qBind$ that correspond to head variables of $\run$  with the corresponding attribute from $\query$ \revm{and condition $\checkPtCond$ that filters out derivations not matching $\pt$ by equating attributes with constants from $\tWithConsVars$.} 
\begin{exam}
  \revm{
    From $\qEx$ in  \Cref{fig:technical-running-example}, we remove answers 
    where $Y \neq 4$ (since $\tEx$ binds $Y=4$)
  and
  anti-join on $X$, the only head variable of $\rExun$, with $\qBind$ from \Cref{ex:qbind}.
  The resulting query and its result are shown below. Note that tuple $(1,1,4)$ was removed 
  since it corresponds to a (failed) derivation of the existing answer $\rel{\qEx}(1,4)$.} \\[-6mm]

\revm{
  \begin{minipage}{0.62\linewidth}
  $$\qDer \defas \qBind \antijoin_{X=X} \selection_{Y=4}(\qEx)$$
\end{minipage}\hspace{0.5cm}
\begin{minipage}{0.37\linewidth}
  \vspace{2mm}
    \begin{tabular}{|ccc|}
      \hline
      \thead{id} & \thead{X} & \thead{Z} \\ \hline
      2            & 2           & 2           \\
      3            & 2           & 4           \\
      \hline
    \end{tabular}
  \end{minipage}
}
\end{exam}

\mypara{3. Computing Goal Annotations}
Next, we determine goal annotations for each derivation to create a set of annotated derivations from $\whynot(\query, \db, \pt)$.
Recall that a positive (negative) grounded goal is successful if the corresponding tuple exists (is missing). We can check this by outer-joining the derivations with the relations from the rule's body.
Based on the existence of a join partner, we create boolean attributes storing $g_i$ for $1 \leq i \leq |\bodyOf{r}|$   ($\redF$ is encoded as false). 
For a negated goal, we negate the result of the conditional expression such that $\redF$ is used if a join partner exists. We construct query $\qSamp$ shown below to associates derivations in $\qDer$ with the goal annotations $\bar{g}$.
\begin{align*}
\qSamp &\defas \duprem(\projection_{Z_1,\cdots,Z_u,e_1 \to g_1, \cdots, e_n \to g_m}(\qGoalJoin))\\
\qGoalJoin \defas &\qDer \leftouterjoin_{\theta_1} \projection_{\rschema{R_1}, 1 \to h_1}(\rel{R_1}) \ldots \leftouterjoin_{\theta_n} \projection_{\rschema{R_1}, 1 \to h_m}(\rel{R_m})
\end{align*}
Note that we use duplicate elimination to preserve set semantics. In projection expressions, we use $e \to a$ to denote projection on a scalar expression $e$ whose result is stored in attribute $a$.
Here, the join condition $\theta_i$ equates the attributes storing the values from $\varvec{X_i}$ in $\run$ with the corresponding attributes from $R_i$. \revm{Attributes at positions that are bound to constants in $\run$ are equated with the constant.}
The net effect is that a tuple from $\qDer$ corresponding to a rule derivation $d$ has a join partner in $R_i$ iff the tuple corresponding to the $i^{th}$ goal of $d$ exists in $D$. The expression $e_i$ used in the projection of $\qSamp$ then computes the boolean indicator for goal $g_i$ as follows:
$$
e_i \defas \begin{cases}
  \iftels{\isnull{h_i}}{\redF}{\greenT} & \text{if}\, g_i\, \text{is positive}\\
  \iftels{\isnull{h_i}}{\greenT}{\redF} & \text{otherwise}
\end{cases}
$$\\[-8mm]
\revm{
\begin{exam}
  For our running example, we generate:\\[-4mm]
\begin{align*}
  \qSamp &\defas \duprem(\projection_{\substack{X,Z, \iftels{\isnull{h_1}}{\redF}{\greenT} \to g_1, \\ \iftels{\isnull{h_2}}{\redF}{\greenT} \to g_2}} (\qGoalJoin))\\
  \qGoalJoin &\defas \qDer \leftouterjoin_{X=A \wedge Z=B} \projection_{A,B, 1 \to h_1}(R) \\
         & \hspace{5mm} \leftouterjoin_{Z= A \wedge B = 4} \projection_{A,B, 1 \to h_2}(R)
\end{align*}
Evaluating this query, we get the result shown below.
\begin{center}
  \begin{minipage}{0.6\linewidth}
    $\qSamp$
    \begin{tabular}{|ccccc|}\hline
      \thead{id} & \thead{X} & \thead{Z} & \thead{$h_1$} & \thead{$h_2$} \\ \hline
      2          & 2         & 2         & \redF         & \greenT       \\
      3          & 2         & 4         & \greenT       & \redF         \\
      \hline
    \end{tabular}
  \end{minipage}
\end{center}
The first tuple corresponds to the derivation $\rExun(2,2) - (\redF, \greenT)$ for which the first goal fails since $\rel{R}(2,2)$ does not exist in $\rel{R}$ 
while the second goal succeeds because $\rel{R}(2,4)$ exists. Similarly, the second tuple corresponds to $\rExun(2,4) - (\greenT, \redF)$ for which the first goal succeeds since $\rel{R}(2,4)$ exists while the second goal fails because $\rel{R}(4,4)$ does not exist.
\end{exam}
}

\BG{remove overlapping parts, and move what remains to Sec 5.2}

\mypara{Queries With Multiple Rules}
For queries with multiple rules, we determine $\osSize$ separately for each rule
  (recall that we consider \ucqNegOrd queries where every rule has the same head predicate) and create a separate query for each rule as described above. Patterns are also generated separately for each rule.
In the final step, we then select the top-$k$ summary from the union of the sets of patterns.  
\RevDel{For simplicity, Algorithm\,\Cref{alg:sampling} shows pseudocode for a variant of our sampling algorithm which directly computes a sample without constructing a query that computes the sample.}

\mypara{Complexity}
The runtime of our algorithm
is linear in $\osSize$ and $\card{\db}$ which significantly improves over the naive algorithm which
is in $\oNotation{|\dataDomain|^n}$.

\mypara{Implementation}
Some
DBMS such as Oracle and Postgres support a sample operator out of the box which we can use to implement the $\sampleOp$ operator introduced above. 
However, these
implementations of a sampling operator do not support sampling with replacement
out of the box. We can achieve decent performance for sampling with replacement
using a set-returning function that takes as input the result of applying the
built-in sampling operator to generate a sample of size $\osSize$, caches this
sample, and then samples $\osSize$ times from the cached sample with replacement.
The $\rownumOp_{A}$ operator can be implemented in SQL
using \lstinline!ROW_NUMBER()!.
The expressions $\iftels{\theta}{e_1}{e_2}$ and $\isnull$ can be expressed in SQL using \lstinline!CASE WHEN!  and \lstinline!IS NULL!, respectively.

\subsection{Determining Over-sampling Size}\label{par:samp-agg}
We now discuss how to  choose $\osSize$, the size of the sample $\overSamp$ produced by query $\qBind$, such that the probability that $\overSamp$ contains at least $\sampSize$ derivations from $\whynot(\query,\db,\pt)$ is higher than a threshold $\osThresh$
under the assumption that the sampling method we introduced above samples uniformly random from $\allAnnotRules(\query, \db, \pt)$. We, then, prove that our sampling method returns a uniform random sample.
First, consider the probability $\probProv$ that a derivation chosen uniformly random from $\allAnnotRules(\query, \db, \pt)$  is in $\whynot(\query,\db,\pt)$ which is equal to the faction of derivations from $\allAnnotRules(\query, \db, \pt)$ that are in $\whynot(\query,\db,\pt)$:\\[-3mm]
\begin{minipage}{1\linewidth}
  \centering
  \begin{minipage}{0.05\linewidth}
    \,
  \end{minipage}
  \begin{minipage}{0.9\linewidth}
  $$
\probProv = \frac{\card{\whynot(\query,\db, \pt)}}{\card{\allAnnotRules(\query, \db, \pt)}}
$$
\end{minipage}
\end{minipage}\\

$\card{\allAnnotRules(\query, \db, \pt)}$ can be computed from $\query$, $\pt$, and the attribute domains as explained in \Cref{sec:rule-deri}.
For instance,
consider $\dataDomain = \{1,2,3,4,5,6\}$ as the universal domain and rule
$\rExun$ from our running example, but without the conditional predicate.
Then, there are $\card{\dataDomain}^n = 6^2$ possible derivations of this rule
because neither variable $X$ nor $Z$ are not bound by $\tEx$
and $\dataDomain$ has 6 values.
To determine $\card{\whynot(\query,\db,\pt)}$, we need to know how many derivations in $\allAnnotRules(\query,\db,\pt)$ correspond to missing tuples matching $\pt$. Since in most cases the number of
missing answers is significantly larger than the number of
existing tuples, it is more effective to compute
the number of (successful and/or failed)  derivations of $t \in \query(\db)$ with $t \matches \pt$, i.e., $\card{ \{ t \mid t \in \query(\db) \wedge t \matches \pt \}}$.
This gives us the probability $\probNotProv$ that a derivation is not in
$\whynot(\query,\db,\pt)$ and we get: $\probProv = 1 - \probNotProv$.

Next, consider a random variable $X$ that is the number of derivations from $\whynot(\query, \db, \pt)$ in $\overSamp$. We want to compute the probability $\prob(X \geq \sampSize)$. For that, consider first $\prob(X=i)$, the probability that the sample $\overSamp$ we produce contains exactly $i$ derivations from $\whynot(\query,\db,\pt)$.
We can apply standard results from statistics for computing $\prob(X=i)$, i.e.,  out of a sequence of $\osSize$ picks with probability $\probProv$ we get $i$ successes. The probability to get exactly $i$ successes out of $n$ picks is $\binom{n}{i} \cdot {\probProv}^{i} \cdot (1-\probProv)^{n-i}$ based on the \textit{Binomial Distribution}.
For $i \neq j$, the events $X=i$ and $X=j$ are disjoint (it is impossible to have both exactly $i$ and $j$ derivations from $\whynot(\query,\db,\pt)$ in $\overSamp$). Thus, $\prob(X \geq \sampSize)$ is 
$\sum \prob(X=i)$ for $i \in \{\sampSize, \ldots, \osSize\}$:
$$
\prob(X \geq \sampSize) = \sum_{i = \sampSize}^{\osSize} \binom{\osSize}{i} \cdot {\probProv}^{i}\cdot (1-\probProv)^{\osSize - i}
$$
Given $\probProv$, $\sampSize$, and $\probSucc$, 
we can compute the sample size $\osSize$ such that $\prob(X \geq \sampSize)$ is larger than $\probSucc$ (\cite{AM65,VE01} presents an algorithm for finding the minimum such $\osSize$).
\BG{This is now out of place: remove or merge into beginning
In the following subsections, we explain how to optimize the remaining summarization process
further utilizing a sample $\samp$ generated over our technique,
i.e., using $\samp$ for generating candidate patterns (Section\,\Cref{sec:opt-patgen}) and measuring quality of patterns (Section\,\Cref{sec:approx-summ}).
}

\mypara{Handling Predicates}
Recall that we apply predicates that compare a variable with a constant before creating a sample for this variable. Thus, we do not need to consider these predicates when determining $\osSize$.
  Predicates 
  comparing variables with variables are applied after the step of  creating derivations.
We estimate the selectivity of such predicates using standard techniques~\cite{I03} to estimate how many derivations will be filtered out  and, then,  increase $\osSize$ to compensate for this, e.g.,
for a predicate with $0.5$ selectivity we would double $\osSize$.

\subsection{Analysis of Sampling Bias}\label{sec:approxi-whynot}
We now formally analyze if our approach creates a uniform sample of $\whynot(\query, \db, \pt)$.
We demonstrate this by analyzing the probability $\prob(\fderi \in \samp)$ for an arbitrary derivation $\fderi \in \whynot(\query, \db, \pt)$ to be in the sample $\samp$. If our approach is unbiased, then this probability should be independent of  which $\fderi$ is chosen and for $\fderi' \neq d$ the events $\fderi \in \samp$ and $\fderi' \in \samp$ should be independent.
\begin{theo}\label{theo:unbiasness}
Given derivations $\fderi, \fderi' \in \whynot(\query, \db, \pt)$ and sample sizes $\sampSize$ and $\osSize$, $\prob(\fderi \in \samp) = c$ where $c$ is a constant that is independent of the choice of $\fderi$. Furthermore, the events  $\fderi \in \samp$ and $\fderi' \in \samp$ are independent of each other.
\end{theo}
\ifnottechreport{
\begin{proof}
  We present the proof in~\cite{LL20}.
\end{proof}
}
\iftechreport{
\begin{proof}\label{pf:unbias-proof}
  To prove the theorem, we have to demonstrate that none of the phases of
  sampling introduces bias. In the following, let $\proofA \defas \allAnnotRules(\query,\db, \pt)$,  $\aSize \defas \card{\proofA}$, and $\wnSize \defas \card{\proofWN}$ where $\proofWN \defas \whynot(\query,\db,\pt)$.
  Recall that the first phase of sampling generates
  a sample $\overSamp$ of size  $\osSize$ by independently creating samples for each unbound variable which are combined into a sample
  of $\proofA$. Consider first the case where $\osSize = 1$, i.e., we pick a single value from each domain. Let $\dataDomain_i$ denote the domain for unbound variable $Z_i$ in the single rule $r$ of query $\query$. Since we sample uniform from $\dataDomain_i$, each of the $\card{\dataDomain_i}$ values has a probability of $\frac{1}{\card{\dataDomain_i}}$ to be chosen. Since the sample for $Z_i$ is chosen independently from $Z_j$ for $i \neq j$, any particular derivation
  $d \in \proofA$ to be in $\overSamp$ is $\prob(\fderi \in \overSamp) =  \frac{1}{\card{\dataDomain_1} \times \ldots \times \card{\dataDomain_u}} = \frac{1}{\card{\proofA}}$.
  For $\osSize > 1$, observe that each value in the sample of $\dataDomain_i$ is chosen independently. Thus,  $\prob(\fderi \in \overSamp) = 1 - \prob(\fderi \not\in \overSamp) = 1 - (1 - \frac{\aSize - 1}{\aSize})^{\osSize}$ (the last equivalence is based on $\prob(A \cap B) = \prob(A) \cdot \prob(B)$ when $A$ and $B$ are independent). Furthermore, this implies that $\fderi \in \overSamp$ is independent of $\fderi' \in \overSamp$ for $\fderi \neq \fderi'$.
So far, we have established that $\prob(\fderi \in \overSamp)$ is constant and the events for picking particular derivations are mutually independent. It remains to be shown that the same holds for a derivation $\fderi \in \proofWN$ and the sample $\samp$ we derive from $\overSamp$.
Since $\overSamp$ is sampled from $\proofA$, it may contain derivations $d' \not\in \proofWN$. Our sampling algorithms filters such derivations. Let $\succDeri$ denote the set of all such derivations from $\overSamp$ and $\sSize = \card{\succDeri}$. Observe that for $i \neq j$ the events $\sSize = i$ and $\sSize = j$ are obviously disjoint since $\overSamp$ contains a fixed number of derivations not in $\proofWN$. Furthermore,
$\sum_{i=0}^{\osSize} \prob(\sSize = i) = 1$ since $\overSamp$ has to contain anywhere from zero to $\osSize$ such derivations.
Thus, we can compute the probability $\prob(\fderi \in \samp)$ as the sum over $i=\{0, \ldots, \osSize\}$ of the probability that $\fderi$ is selected to be in $\samp$ conditioned on the probability that $\fderi \in \overSamp$ (otherwise $\fderi$ cannot be in $\samp$) and that $\sSize = i$. 
\begin{align*}
&\prob(\fderi \in \samp) =\\
  &\mathtab\mathtab\sum_{i=0}^{\osSize} \prob(\fderi \in \samp \mid \fderi \in \overSamp \cap \sSize = i) \cdot \prob( \fderi \in \overSamp \cap \sSize = i)
\end{align*}
Now, consider the individual probabilities in this formula. Let $p_1 =  \prob(\fderi \in \samp \mid \fderi \in \overSamp \cap \sSize = i)$. If $\fderi$ is in $\overSamp$ and $\sSize = i$, then there are $\osSize - i - 1$ derivations from $\proofWN$ in $\overSamp$. Our sampling algorithm selects uniformly $\sampSize$ derivation in $\proofWN$ from $\overSamp - \succDeri$ if $\osSize - i > \sampSize$. Thus, the probability for our particular derivation $d$ to be in the final result is:
$$p_1 = \frac{\min(\sampSize, \osSize - i)}{\osSize - i}$$
Next, consider $p_2 = \prob( \fderi \in \overSamp \cap \sSize = i)$. Based on our observation above, any subset of derivations of $\proofA$ has the same probability to be returned as $\overSamp$. Thus, $p_2$ can be computed as the fraction of subsets of $\proofA$ of size $\osSize$ that contain $i$ successful derivation(s), and
$\fderi$ and
$\osSize -i - 1$ other derivations from $\proofWN$. Putting differently, how many of the ${\aSize}^{\osSize}$ possible samples that could be produced by our algorithm contain $\fderi$ and exactly $i$ derivations from $\succDeri$:
\begin{align*}
p_2 = \frac{{\sSize}^{i} \cdot {\wnSize}^{\osSize - i - 1}}{{\aSize}^{\osSize}}
\end{align*}
Observe, that the formulas for $p_1$ and $p_2$ only refer to constants that are independent of the choice of $\fderi$. Thus, $\prob(\fderi \in \samp)$ is independent of the choice of $\fderi$.
\end{proof}
}

\section{Generating Pattern Candidates}\label{sec:generating-pattern-c}
\revm{We now explain the candidate generation step of our summarization approach (phase 2 in \Cref{fig:pug-summ-process}).
  Consider a provenance question (PQ) $\aProvQ = (\pt, \whynot)$ for a query $\query$. For any rule $r$ of $\query$, let $n$ be the number of unbound variables, i.e., $\card{\varsOf{\run}}$ where $\run$ is the unified rule for $r$ and $\pt$, and $m$ be the number of goals in $r$.
  The number of possible patterns for $\run$ is in $\oNotation{(\card{\dataDomain} + n)^n \cdot 2^m}$, because for each variable of $\run$ 
  we can choose either a placeholder or a value from $\dataDomain$ and for each goal we have to pick one of two possible annotations ($\redF$ or $\greenT$). Note that the names of placeholders are irrelevant to the semantics of a pattern, e.g., patterns $p = (A,3)$ and $p' = (B,3)$ are equivalent (matching the same derivations).
}
That is, we only have to decide which arguments of a pattern are placeholders and which arguments share the same placeholder.
Thus, it is sufficient to only consider $n$ distinct placeholders $P_i$ (where $1 \le i \le n$) when creating patterns for a unified rule $\run$ 
  with $n$ variables.
\begin{exam}\label{ex:patt-gen-basic}
  Consider rule $\rExun$ from \Cref{fig:technical-running-example}.
  Let $\dataDomain = \{ 1,2,3,4,5,6 \}$ and $\placeholders = \{ P_1,P_2\}$.
  Let us for now ignore goal annotations. 
  Note that, taking the predicate $X<4$ into account, any pattern where $X \geq 4$ cannot possibly match any derivations for this rules and, thus, we only have to consider patterns where $X$ is bound to a constant less than 4 or a placeholder. The set of viable  patterns 
  is:\\
  \resizebox{1\linewidth}{!}{
  \begin{minipage}{1.0\linewidth}
  \begin{align*}
    &\rExun(P_1,P_2),
      \rExun(P_1,1), 
      \ldots,  \rExun(P_1,6),
    \rExun(1,P_2), \ldots, \rExun(6,P_2),\\
    &\rExun(1,1), \ldots, \rExun(1,6),
      \ldots
    \rExun(3,1), \ldots, \rExun(3,6)
  \end{align*}
\end{minipage}
}\\[2mm]
  This set 
  contains $31$ elements. 
  Considering goal annotations $(\redF,\redF)$, $(\redF,\greenT)$, and $(\greenT, \redF)$, we get $31 \cdot 3 = 93$ patterns. 
\end{exam}

Given the $\oNotation{(\card{\dataDomain} + n)^n \cdot 2^m}$ complexity, it is
not feasible to enumerate all possible patterns. 
Instead, we adapt the \textit{Lowest Common Ancestor}
(LCA) method~\cite{EA14,GF18} for our purpose which generates a number of
pattern candidates from the derivations in a sample $\samp$
that is at most quadratic in $\sampSize$.
Thus, this approach sacrifices
completeness to achieve better performance.  Given a set of derivations (tuples
in the work from~\cite{EA14,GF18}), the LCA method computes the cross-product of
this set with itself and generates candidate explanations by generalizing each
such pair.  The rationale is that each pattern generated in this fashion will at
least match two derivations \revm{(or one derivation for the special case where a derivation is paired with itself).}
In our adaptation, we match derivations on the goal annotations such that only derivations with the same success/failure status of goals are paired.
For each pair of derivations $d_1 = (a_1, \ldots, a_n) - (\bar{g})$ and $d_2 = (b_1, \ldots, b_n) - (\bar{g})$, we generate  a pattern $p = (c_1, \ldots, c_n) - (\bar{g})$.
We determine each element $c_i$ in $p$ as follows.
If $a_i = b_i$ then $c_i = a_i$. That is, constants on which $d_1$ and $d_2$ agree  are retained. Otherwise,  $c_i$ is a fresh
placeholder. 
\begin{exam}\label{ex:lca}
 Reconsider the unified rule $\rExun$ and instance $\rel{R}$ from \Cref{fig:technical-running-example}.
 Two example annotated rule derivations 
 are
 $d_1 = \rExun(2,1) - (\redF,\redF)$ 
 and $d_2 = \rExun(2,5) - (\redF,\redF)$.
 LCA generate a pattern $p = \rExun(2,Z)-(\redF,\redF)$ to  generalize $d_1$ and $d_2$ because $d_1[1] = d_2[1] = 2$  (and, thus, this constant is retained) and $\parg{p}{2} = Z$ since $d_1[2] = 1 \neq 5 = d_2[2]$. 
\end{exam}

We apply $\lca$ to the sample $\samp$ created 
using $\qSamp$ from \Cref{sec:samp-prov}. 
Using $\lca$,
we 
avoid generating exponentially many patterns and, thus, improve the runtime of pattern generation from $\oNotation{\card{\dataDomain}^n}$ to $\oNotation{{\sampSize}^2}$ where typically  $\sampSize \ll |\dataDomain|$. 
Furthermore, this optimization reduces the input size for the final stages of the summarization process leading to additional performance improvements.
Even though LCA is only a heuristic, we demonstrate experimentally in \Cref{sec:experi} that 
it performs well in practice.
\BGDel{We use $\pat_{\lcasm}(\query,\tWithConsVars)$ to denote pattern candidates generated using \lca for a p-tuple $\tWithConsVars$.}{Not used}

\mypara{Implementation}
We implement the \lca method as a query $\qLCA$ joining the query $\qSamp$ (the query producing $\samp$) with itself on a condition $\condLCA \defas \bigwedge_{i=0}^{m} g_i = g_i$ where $m$ is the number of goals of the rule $r$ of $\query$ (recall that we create patterns for each rule of  $\query$ independently  
and merge in the final step). 
Patterns are generated using a projection on an list of expressions $\projLCA$, where the $i^{th}$ argument of a pattern is determined as $\iftels{X_i = X_i}{X_i}{NULL}$.
Note that the LCA method never generates patterns where the same placeholder appears more than once. Thus, it is sufficient to  encode placeholders as \lstinline!NULL!
values.  
$$\qLCA \defas \delta(\projection_{\projLCA}(\qSamp \join_{\condLCA} \qSamp))$$

The query generated for our running example is: \\[-4mm]
\begin{align*}
  \qLCA &\defas \delta(\projection_{e_X \to X, e_Z \to Z}(\qSamp \join_{(g_1=g_1) \wedge (g_2 = g_2)} \qSamp)\\
  e_X &\defas \iftels{X=X}{X}{NULL}\\
  e_Z &\defas \iftels{Z=Z}{Z}{NULL}
\end{align*}

\RevDel{Note that we can replace \lca with any other methods for computing candidate patterns without having to change the other steps of our summarization process. Studying alternative method is an interesting avenue for future work.}

\section{Estimating Completeness}\label{sec:estimating-pattern-c}

\revm{To generate a top-$k$ summary in the next step, we need to calculate the informativeness (\Cref{def:info}) and completeness (\Cref{def:recall}) quality metrics for sets of patterns. 
  Informativeness can be computed from 
  patterns without accessing the data. Recall that completeness is computed as the fraction of 
  provenance matched by a 
  pattern: $\completeness(p) =
 \frac{\card{\allMatches(\query,\db,p,\aProvQ)}}{\card{\explainq(\aProvQ)}}$. Since we can materialize neither $\card{\allMatches(\query,\db,p,\aProvQ)}$ nor $\card{\explainq(\aProvQ)}$ for the why-not provenance, we have to estimate their sizes.
 In this section, we focus on how to estimate the completeness of individual patterns. 
 How to compute the completeness metric for sets of patterns will be discussed in \Cref{sec:computing-top-k-summ}.}

\revm{To determine whether a derivation $d \in \explainq(\aProvQ)$ with goal annotations $\bar{g_1}$ matches a pattern $p$ with goal annotations $\bar{g_2}$ 
  that is in $\allMatches(\query,\db,p,\aProvQ)$,} we have to check  that $\bar{g_1} = \bar{g_2}$ and 
a valuation exists that maps $p$ to $d$.
Then, we count the number of such derivations  
to compute $\card{\allMatches(\query,\db,p,\aProvQ)}$.
The existence of a valuation can be checked in linear time in the number of arguments of $p$ by fixing a placeholder order and, then, assigning to each placeholder in $p$ the corresponding constant in $d$ if a unique such constant exists. 
 The valuation fails if $p$ and $d$ end up having two different constants at the same position.
 \begin{exam}\label{ex:completeness}
   Continuing with \Cref{ex:lca}, we compute completeness of the pattern $p = \rExun(2,Z)-(\redF,\redF)$.
   For sake of the example, assume that $\explainq(\pqEx) = $:\\[-1mm]
   \resizebox{1\linewidth}{!}{
    \begin{minipage}{1.05\linewidth}\centering
   \begin{align*}
     &d_1 = \rExun(2,1) - (\redF,\redF) \hspace{7mm} d_2 = \rExun(2,2) - (\redF,\greenT)\\
     &d_3 = \rExun(2,3) - (\greenT,\redF) \hspace{7mm} d_4 = \rExun(2,4) - (\greenT,\redF)\\
     &d_5 = \rExun(2,5) - (\redF,\redF) \hspace{7mm} d_6 = \rExun(2,6) - (\redF,\redF)
   \end{align*}
   \end{minipage}
   }\\[1mm]
   The completeness of $p$ is $\completeness(p) = \frac{3}{6}$ because $p$ matches all 3 derivations ($d_1$, $d_5$, and $d_6$) for which both goals have failed by assigning $Z$ to 1, 5, and 6.
\end{exam}

\revm{
  To estimate the completeness of a pattern $p$, we compute the number of matches of $p$ with derivations from the sample $\samp$  produced by $\qSamp$ as discussed in \Cref{sec:sampling-provenance}. As long as $\samp$ is an unbiased sample of $\explainq(\aProvQ)$, 
  then the fraction of derivations from $\samp$ matching the pattern is an unbiased estimate of the completeness of the pattern.} 
Continuing with \Cref{ex:completeness}, assume that we created a sample $\samp = \{ d_1,d_3,d_4,d_5 \}$.
    Estimating the completeness of pattern $p$ based on $\samp$, we get $\completeness(p) \simeq \frac{1}{2}$.

\mypara{Implementation}
We generate a query $\qMatch$ which joins the query $\qLCA$ generating pattern candidates with $\qSamp$, the query generating the sample derivations. Let $\run$ be the rule for which we are generating patterns and $A$ be the result attributes of $\qLCA$. We count the number of matches per pattern by grouping on $A$:
$$\qMatch \defas \aggregation_{A,count(*)}(\qLCA \join_{\checkMatchPred} \qSamp)$$
Recall that we  encode placeholders as \lstinline!NULL! values.
Condition $\checkMatchPred$ is a conjunction of conditions, one for each argument $X$ of the pattern/derivation: $X = X \vee \isnull{X}$.
\BG{I THINK THIS WOULD REQUIRE MORE EXPLANATION: based on the observation that the composition of derivations in $\samp$ holds in $\explainq(\aProvQ)$ because each derivation in $\explainq(\aProvQ)$ has same probability of being chosen (uniformly random).}
Since the number of candidates produced by \lca is at most ${\sampSize}^2$, matching is in $\oNotation{{\sampSize}^2 \cdot \sampSize} = \oNotation{{\sampSize}^3}$. 
For our running example, we would create the following query:\\[-5mm]
  \begin{align*}
    \qMatch         & \defas \aggregation_{X,Z,g_1,g_2,count(*)}(\qLCA \join_{\checkMatchPred} \qSamp) \\
    \checkMatchPred & \defas (X=X \vee \isnull{X}) \wedge (Z=Z \vee \isnull{Z})
  \end{align*}\\[-8mm]

\section{Computing Top-k Summaries}\label{sec:computing-top-k-summ}
We now explain how to compute a top-$k$ provenance summary for a provenance question $\aProvQ$ (phase 4 in \Cref{fig:pug-summ-process}). 
  This is the only step that is evaluated on the client-side. Its input is the set of patterns (denoted as $\pat_{\lcasm}$) with completeness estimates returned by evaluating query $\qMatch$ (\Cref{sec:estimating-pattern-c}). 
  We have to find the set  $\sprov \subseteq \pat_{\lcasm}$ of size $k$ that maximizes $\score(\sprov)$. 
  A brute force solution would enumerate all such subsets, compute their scores (which requires us to compute the union of the matches for each pattern in the set to compute completeness), and return the one with the highest score. However, the number of candidates is $\card{\pat_{\lcasm}} \choose k$ and this would require us to evaluate a query to compute matches for each candidate. 
  Our solution  uses lower and upper bounds on the completeness of patterns that can be computed based on the patterns and their completeness alone to avoid running additional queries. Furthermore, we use a heuristic best-first search method
  to incrementally build candidate sets guiding the search using these bounds.

\subsection{Pattern Generalization and Disjointness}\label{sec:p-generalize}
  In general, the exact completeness of a set of patterns cannot be directly computed based on the completeness of the patterns of the set, because the sets of derivations matching two patterns may overlap.
  We present two conditions that allow us to determine in some cases whether the match sets of two patterns are disjoint or one is contained in the other.
  We say a pattern $p_2$ \emph{generalizes} a pattern $p_1$ written as $p_1 \pleq p_2$ if $\forall i: p_1[i] = p_2[i] \vee p_2[i] \in \placeholders$ (infinite set of placeholders),
  and they have the same goal annotations. For instance, $(X,Y,\cnst{a})-(\redF,\redF)$ generalizes  $(X,\cnst{b},\cnst{a})-(\redF,\redF)$.
  From \Cref{def:p-match}, it immediately follows  that if $p_1 \pleq p_2$ then  $\allMatches(\query,\db,p_1,\aProvQ) \subseteq  \allMatches(\query,\db,p_2,\aProvQ)$ since any derivation matching $p_1$ also matches $p_2$ 
  and, thus, $\completeness(\{p_1, p_2\}) = \completeness(p_2)$.
  We say pattern $p_1$ and $p_2$ are \textit{disjoint} written as $p_1 \pind p_2$ if 
  (i) they are from different rules, (ii) they do not share the same goal annotations, or (iii) there exists an  $i$ such that $p_1[i] = c_1 \neq c_2 = p_2[i]$, i.e.,
  the patterns have a different constant at the same position $i$. If $p_1 \pind p_2$, then $\allMatches(\query,\db,p_1,\aProvQ) \cap  \allMatches(\query,\db,p_2,\aProvQ) = \emptyset$ and, thus, we have $\completeness(\{p_1,p_2\}) = \completeness(p_1) + \completeness(p_2)$.
  Note that for any $\sprov$,  $\completeness(\sprov)$ is trivially bound from below by $\max_{p \in \sprov}\completeness(p)$ (making the worst-case assumption that all patterns fully overlap) and by $\min(1,\sum_{p \in \sprov} \completeness(p))$ from above (completeness is maximized if there is no overlap).
  Using generalization and disjointness, we can refine these bounds. Note that generalization is transitive. To use generalization to find tighter upper bounds on completeness for a pattern set $\sprov$, we compute the set $\psub = \{ p \mid p \in \sprov \wedge \neg\exists p' \in \sprov: p \pleq p'\}$. Any pattern not in $\psub$ is generalized by at least one pattern from $\psub$.
  For disjointness, if we have a set of patterns $\sprov$ for which patterns are pairwise disjoint, then $\completeness(\sprov) = \sum_{p \in \sprov} \completeness(p)$. Based on this observation, we find the subset $\pslb$ of  pairwise disjoint patterns  from  $\sprov$ that maximizes completeness, i.e., $\pslb = \argmax_{\sprov' \subseteq \sprov \wedge \forall p \neq p' \in \sprov': p \pind p'} \sum_{p \in \sprov'}\completeness(p)$.\footnote{Note that this is the intractable weighted maximal clique problem. For reasonably small $k$, we can solve the problem exactly and otherwise apply a greedy heuristic.}
  We use $\pslb$ and $\psub$ to define an lower bound $\lbc(\sprov)$ and upper-bound $\ubc(\sprov)$ on the completeness of a pattern set  $\sprov$:
  \begin{align*}
    \lbc(\sprov) &\defas \sum_{p \in \pslb} \completeness(p) &\ubc(\sprov) &\defas \sum_{p \in \psub} \completeness(p)
  \end{align*}\\[-8mm]
  \begin{exam}
    Consider the
    following patterns for $\rExun$: 
    $p = (2,Z)-(\redF,\redF)$,
    $p' = (3,Z)-(\redF,\redF)$,
    $p''=(2,1)-(\redF,\redF)$.
    Assume that $\completeness(p) = 0.44$, $\completeness(p') = 0.55$, and $\completeness(p'') = 0.1$.
    Consider $\sprov = \{p,p',p''\}$ and observe that $p \pind p'$, $p' \pind p''$, and $p'' \pleq p$. Thus, $\psub = \{p, p'\}$ (the pattern $p''$ is generalized by $p$) and $\pslb = \{p, p'\}$ (while also $p' \pind p''$ holds, we have $\completeness(p) + \completeness(p') > \completeness(p') + \completeness(p'')$). We get: $\lbc(\sprov) = \completeness(p) + \completeness(p') = 0.99$ and $\ubc(\sprov) = \completeness(p) + \completeness(p') = 0.99$ from which follows that $\completeness(\sprov) = 0.99$. Note that, without using generalization and disjointness, we would have to settle for a lower bound of $\max_{p \in \sprov} \completeness(p) = 0.55$ and upper bound of $\min(1, \sum_{p \in \sprov} \completeness(p)) = 1$.
  \end{exam}

\subsection{Computing the Top-K Summary}
\label{sec:computing-top-k}

 We apply a best-first search approach to compute a approximate top-$k$ summary given a
 set of patterns $\pat_{\lcasm}$.
 Our approach maintains a priority queue of
  candidate sets sorted on a lower bound $\lbsc$ for the score of candidate sets
  that we compute based on the completeness bound $\lbc$ introduced above. We
  also maintain an upper bound $\ubsc$. For a set $\sprov$ of size $k$, we can
  compute $\informativeness(\sprov)$ exactly.  For incomplete candidates (size
  less than $k$), we bound the informativeness and completeness of any extension
  of the candidate into a set of size $k$ using worst-case/best-case
  assumptions. For example, to bound completeness for an incomplete candidate $\sprov$
  from above, we assume that the remaining patterns will not overlap with any
  pattern from $\sprov$ and have maximal completeness ($\max_{p \in \pat_{\lcasm}} \completeness(p)$). We initialize the priority
  queue with all singleton subset of $\pat_{\lcasm}$ and, then, repeatably take
  the incomplete candidate set with the highest $\lbsc$ and extend it by one
  pattern from $\pat_{\lcasm}$ in all possible ways and insert these new
  candidates into the queue. The algorithm terminates when a complete candidate
  $\sprov_{best}$ is produced for which $\lbsc$ is higher than the highest
  $\ubsc$ value of all candidates we have produced so far (efficiently
  maintained using a max-heap sorted on $\ubsc$). In this case, we return
  $\sprov_{best}$ since it is guaranteed to have the highest score (of course completeness is only an estimation) even though
  we do not know the exact value. The algorithm also terminates when all
  candidates have been produced, but no $\sprov_{best}$ has been found.  In this
  case, we apply the following heuristic: we return the set with the highest
  average ($(\lbsc + \ubsc) / 2$).

\BG{    For instance, to find the top-$2$ summary, we compute the score bounds for each combination. 
    For example, for the set $\sprov_1 = \{ p,p' \}$, we have $\completeness(\sprov_1) = \completeness(p) + \completeness(p') =  0.99$ because $p \pind p'$.
    The score of another set $ps_2 = \{p,p''\}$ where $p \pleq p''$ is $0.44 \le ps_2 \le 0.44$.
    The last set $ps_3$ containing $p'$ and $p''$ are neither of those cases and, thus, we set the score bound as $0.44 \le ps_3 \le 0.54$.
    We, then, return top-$2$ patterns by comparing these scores. 
    In this example, $ps_1$ is returned as a summry (the top-$2$ patterns). 
}

 \section{Experiments}
\label{sec:experi}

We evaluate
(i) the performance of 
computing  summaries and 
(ii) 
the quality of summaries 
produced by our technique.

\begin{figure}[t]
 \centering $\,$\\[-2mm]
 \begin{minipage}{.99\linewidth}
  \centering\scriptsize
  \begin{align*}
    r_1:&\: \rel{InvalidD}(C) \dlImp \rel{LICENSE}(I,B,G,C,T,\textsf{d}), \neg \rel{VALID}(I) \\[1mm]
    \hline \\[-2mm]
    r_2:&\: \rel{Fsenior}(C) \dlImp \rel{LICENSE}(I,B,\textsf{f},C,T,L), \rel{VALID}(I), B < 1953 \\[1mm]
    \hline \\[-2mm]
    r_3:&\: \rel{CasualWatch}(T,E,N) \dlImp \rel{MOVIES}(I,T,Y,R,P,B,V), \rel{GENRES}(I,E),
      \\&\hspace{12mm}\rel{PRODCOMPANY}(I,C), \rel{COMPANY}(C,N), \rel{RATINGS}(U,I,G,S),
    \\&\hspace{7mm}\neg \rel{GENRES}(I,\textsf{thriller}), R < 100, G >= 4\\[1mm]
    \hline \\[-2mm]
    r_4:&\: \rel{Players}(A) \dlImp \rel{MOVIES}(I,T,Y,R,P,B,V), \rel{CASTS}(I,C,H,A,G),
    \\&\hspace{7mm}\rel{GENRES}(I,\textsf{romance}), \rel{RATINGS}(U,I,N,S), Y > 1999, N >= 4\\
    r_4':&\:\rel{Players}(A) \dlImp \rel{MOVIES}(I,T,Y,R,P,B,V), \rel{CASTS}(I,C,H,A,G),
    \\&\hspace{17mm}\rel{GENRES}(I,\textsf{comedy}), \rel{KEYWORDS}(I,\textsf{love}),
    \\&\hspace{16mm}\rel{RATINGS}(U,I,N,S), Y > 1999, N >= 4\\
    r_4'':&\:\rel{Players}(A) \dlImp \rel{MOVIES}(I,T,Y,R,P,B,V), \rel{CASTS}(I,C,H,A,G),
    \\&\hspace{17mm}\rel{GENRES}(I,\textsf{drama}), \rel{KEYWORDS}(I,\textsf{relationship}),
    \\&\hspace{16mm}\rel{RATINGS}(U,I,N,S), Y > 1999, N >= 4\\[1mm]
    \iftechreport{
     \hline \\[-2mm]
      r_{5}:\: &\rel{DirGen}(N) \dlImp \rel{MOVIES}(I,T,Y,R,P,B,V), \\[1mm]
    &\hspace{5mm}\rel{CREWS}(I,W,N,\textsf{director},M), \rel{GENRES}(I,E), B > 20000000\\[1mm]
    \hline \\[-2mm]
      r_{6}:\: &\rel{TomKey}(T,K,E) \dlImp \rel{MOVIES}(I,T,Y,R,P,B,V), \\
      & \hspace{10mm} \rel{CASTS}(I,C,H,\textsf{tom cruise},G), \rel{KEYWORDS}(I,K), \\
      & \hspace{5mm} \rel{GENRES}(I,E), \rel{RATINGS}(U,I,A,S), A \ge 4\\[1mm]
      \hline \\[-2mm]
      r_7:&\: \rel{FavCom}(T) \dlImp \rel{MOVIES}(I,T,Y), \rel{GENRES}(I,\textsf{comedy}),
    \\&\hspace{18mm} \rel{RATES}(U,I,R,M,A), R \ge 4 \\[1mm]
    \hline \\[-2mm]
      r_{8}:&\: \rel{ActMov}(T) \dlImp \rel{MOVIES}(I,T,Y), \rel{GENRES}(I,\textsf{action}), \rel{RATES}(U,I,5,M,A)
    \\[1mm]
    \hline \\[-2mm]
    r_9:&\: \rel{CommCrime}(T) \dlImp \rel{CRIMES}(I,Y,T,L,\textsf{austin}), \neg \rel{ARREST}(I) \\[1mm]
    \hline \\[-2mm]
      r_{10}:&\: \rel{CrimeSince}(T) \dlImp \rel{CRIMES}(I,Y,T,L,C), \neg \rel{ARREST}(I), Y > \textsf{2012} \\[1mm]
       \hline \\[-2mm]
      r_{11}:&\: \rel{Hops}(L) \dlImp \rel{DBLP}(L,R), \rel{DBLP}(R,R1), \rel{DBLP}(R1,R2),
      \\&\hspace{15mm}\rel{DBLP}(R2,R3), \rel{DBLP}(R3,R4), \rel{DBLP}(R4,R5)\\[1mm]
       \hline \\[-2mm]
      r_{12}:&\: \rel{Custs}(CN,NK) \dlImp \rel{CUSTOMER}(CK,CN,C_1,NK,C_2,C_3,C_4,C_5),
          \\&\hspace{19mm}\rel{ORDERS}(OK,CK,O_1,O_2,O_3,O_4,O_5,O_6,O_7),
      \\&\hspace{15mm}\rel{LINEITEM}(OK,L_1,L_2,L_3,
    \cdots ,L_{13},L_{14},L_{15})\\[1mm]
    }
  \end{align*}
 \end{minipage}
 $\,$\\[-5mm]
 \ifnottechreport{\caption{Queries used in the experiments (excerpt)}}
  \iftechreport{\caption{Queries used in the experiments}}
 \label{fig:experi-queries}
\end{figure}

\mypara{Experimental Setup}\label{sec:experi-setup} 
All experiments were executed on a machine with 2 x 3.3Ghz AMD Opteron 
CPUs (12 cores) 
and 128GB RAM running Oracle Linux 6.4.
We use a commercial DBMS (name omitted due to licensing restrictions).

\mypara{Datasets}
We use \iftechreport{TPC-H 
and  several}\ifnottechreport{two} real-world datasets:
(i) the New York State (NYS) license dataset\footnote{\small\url{https://data.ny.gov/Transportation/Driver-License-Permit-and-Non-Driver-Identificatio/a4s2-d9tt}} 
($\sim 16$M tuples), and 
(ii) a movie dataset\footnote{\small\url{https://www.kaggle.com/rounakbanik/the-movies-dataset}} ($\sim 26$M tuples)\ifnottechreport{.}\iftechreport{, (iii) a Chicago crime dataset\footnote{\small\url{https://data.cityofchicago.org/Public-Safety/Crimes-2001-to-present/ijzp-q8t2}} ($\sim 6$M tuples),
and
(iv) a co-author graph relation extracted from DBLP\footnote{\small\url{http://www.dblp.org}}. 
}
For each dataset, we created several subsets; 
$\rel{R_{x}}$  denotes a subset of $\rel{R}$ with x rows. 

\mypara{Queries}
\Cref{fig:experi-queries} shows the queries used in the experiments. 
\iftechreport{In \Cref{tab:numofvalues}, we provide the number of distinct values from the largest datasets of (i), (ii), and (iii) for the attributes of these queries.}
For the license dataset, 
we use $\rel{InvalidD}$ ($r_1$) which returns cities with invalid driver's licenses and $\rel{Fsenior}$ ($r_2$) which returns cities with valid licenses held by female seniors.
For the movie dataset,  $\rel{CasualWatch}$ ($r_3$) returns movies with their genres and production companies if their runtime is less than 100 minutes and they have received high ratings ($G \ge 4$).
  $\rel{Players}$ ($r_4$) computes actresses/actors who have been successful (rating higher than 4) in a romantic comedy after 1999. \ifnottechreport{We report additional experiments in~\cite{LL20}.}
  \iftechreport{In addition, $\rel{DirGen}$ ($r_{5}$) computes name of person who has directed a movie 
    with over 2M dollars budget, and
  $\rel{TomKey}$ ($r_{6}$) returns movie title with its keyword and genre that Tom Cruise has played. 
  $\rel{FavCom}$ ($r_7$) computes popular movies
  (
  if it has received ratings over 3) 
  in comedy genre 
  and $\rel{ActMov}$ ($r_8$) 
  returns titles of action movies that have been rated with the highest score. 
  For the crime dataset, $\rel{CommCrime}$ ($r_9$) and $\rel{CrimeSince}$ ($r_{10}$) return types of unarrested crimes in the community Austin and anywhere since $2012$, respectively.
For the DBLP dataset, $\rel{Hops}$ ($r_{11}$) returns authors that are connected to each other by a path of length $6$ in the co-author graph. 
Over TPC-H, $\rel{Custs}$ ($r_{12}$) returns ids and the nations of customers who have at least one order. 
}

\iftechreport{
\begin{figure}
  \begin{minipage}{1.0\linewidth}
    \centering \scriptsize 
    \hspace{-5mm}
    \begin{minipage}{0.51\linewidth}
      \centering
    \begin{tabular}{|c|c|c|} \hline
	\thead {Q} & \thead {Why} & \thead {Why-not} 
		\\ \hline
		$r_1$ & \textsf{new york} & \textsf{swanton} 
    		\\ \hline
      $r_2$ & \textsf{brooklyn} & \textsf{delaware} 
                \\ \hline
      $r_3$ & \textsf{drama ($E$)} & \textsf{family ($E$)} 
               \\ \hline
      $r_4$ & \textsf{jack black} & \textsf{tom ford} 
                                                                                                                      \\ \hline
                $r_{5}$ & \textsf{steven} & \textsf{robert}\\
                & \textsf{spielberg} & \textsf{altman}
                \\ \hline
               $r_{6}$ & \textsf{mission ($K$)} & \textsf{spying ($K$)}
		\\ \hline
    \end{tabular}
  \end{minipage} 
    \begin{minipage}{0.45\linewidth}
      \centering
    \begin{tabular}{|c|c|c|} \hline
      \thead {Q} & \thead {Why} & \thead {Why-not}
		\\ \hline
 	         $r_7$ & \textsf{forrest gump} & \textsf{babysitting}
    		\\ \hline
      $r_8$ & \textsf{fight club} & \textsf{avalanche}
                                        		\\ \hline
      $r_9$ & \textsf{battery} & \textsf{domestic} 
     \\
      & & \textsf{violence}                                                                                                                \\ \hline
               $r_{10}$ & \textsf{theft} & \textsf{ritualism}

		\\ \hline
                $r_{11}$ & - & \textsf{xueni pan} 
                \\ \hline
      $r_{12}$ & - & various
		\\ \hline
    \end{tabular}
    \end{minipage}
  \end{minipage}
  $\,$\\[-3mm]
  \caption{Bindings for 
 why and why-not provenance questions used in the experiments.} 
  \label{tab:bindings}
\end{figure}

\begin{figure} $\,$\\[-1mm]
  \begin{minipage}{1\linewidth}
    \begin{minipage}{0.5\linewidth}
      \centering
      \scriptsize 
   \resizebox{1\textwidth}{!}{
    \begin{tabular}{c|c|c|c|c}
      \textbf{Dataset} & \multicolumn{4}{c}{$\rel{LICENSE_{16M}}$} \\ \hline
      \textbf{Attribute} & $I$ & $B$ & $G$ & $T$\\ \hline
      \textbf{\#Dist.} & 16M & 118 & 2 & 64 \\ 
    \end{tabular}
   }
 \end{minipage} 
 \begin{minipage}{0.49\linewidth}
      \centering
      \scriptsize 
   \resizebox{1\textwidth}{!}{
    \begin{tabular}{c|c|c|c|c}
      \textbf{Dataset} & \multicolumn{4}{c}{$\rel{CRIME_{6M}}$} \\ \hline
      \textbf{Attribute} & $I$ & $Y$ & $L$ & $C$ \\ \hline
      \textbf{\#Dist.} & 6M & 19 & 181 & 105 \\ 
    \end{tabular}
   }
 \end{minipage}
  \end{minipage}
\\[2mm]
  \begin{minipage}{1\linewidth}
    \centering \scriptsize 
   \resizebox{1\textwidth}{!}{
    \begin{tabular}{c|c|c|c|c|c|c|c|c}
      \textbf{Dataset} & \multicolumn{8}{c}{$\rel{MOVIES_{26M}}$} \\ \hline
      \textbf{Attribute} & $I$ & $T$ & $Y$ & $R$ & $P$ & $B$ & $V$ & $E$ \\ \hline
      \textbf{\#Dist.} & 45K & 42K & 135 & 350 & 44K & 1K & 7K & 20 \\ \hline \hline
      \textbf{Attribute} & $C$ & $C_{\rel{CASTS}}$ & $H$ & $G_{\rel{CASTS}}$ & $U$ & $G$ & $S$ & \\ \hline
      \textbf{\#Dist.} & 61K & 24K & 170K & 3 & 270K & 10 & 21M \\ 
    \end{tabular}
   }
  \end{minipage}
  $\,$\\[-3mm]
  \caption{Number of distinct values from the largest datasets for attributes of the experimental queries}
  \label{tab:numofvalues}
\end{figure}
}

\ifnottechreport{
\begin{figure}[t]
\begin{minipage}{1\linewidth}
 \centering$\,$\\[-2mm]
\begin{minipage}{1\linewidth}
  \centering \hspace{5mm}
  \begin{minipage}{.47\linewidth}
    \subfloat[\scriptsize\rel{InvalidD}-Why]{
      \includegraphics[width=0.76\columnwidth,trim=100 20 0 0]{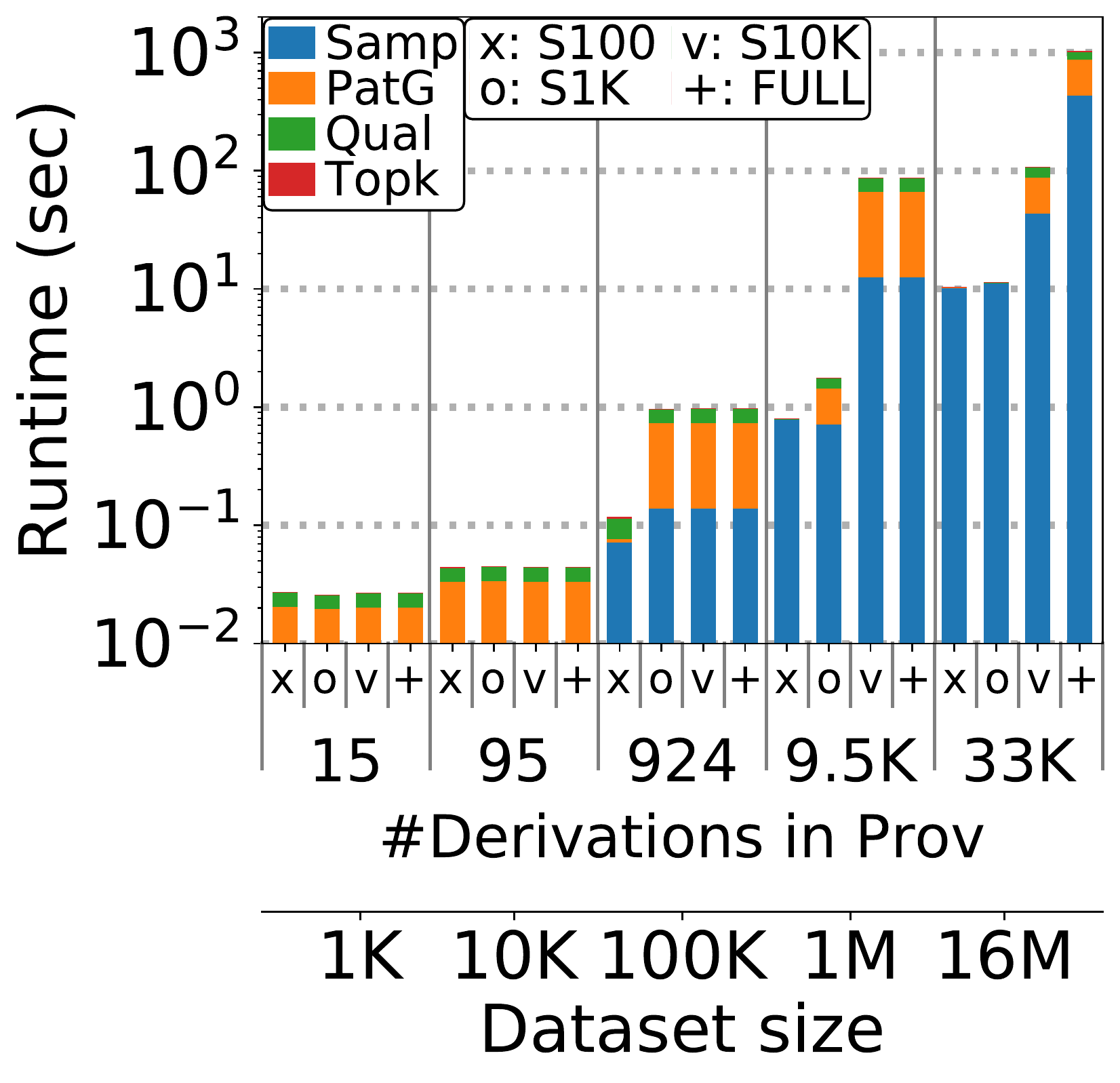}
    \label{fig:perf-r1-why}
    }
  \end{minipage}\hspace{-2mm}
  \begin{minipage}{.47\linewidth}
   \centering
   \subfloat[\scriptsize\rel{InvalidD}-Whynot]{
     \includegraphics[width=0.76\columnwidth,trim=110 20 10 10]{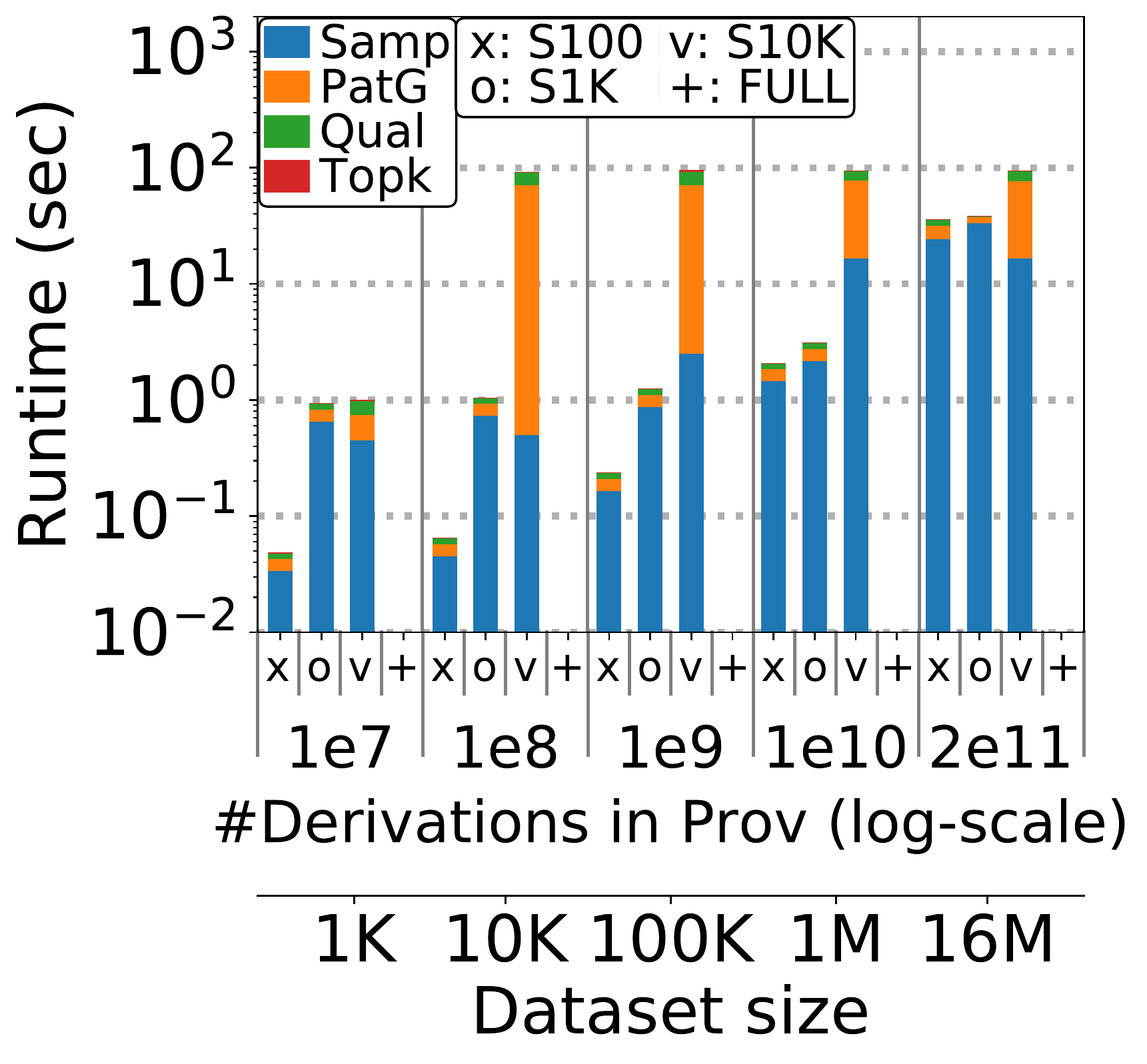}
    \label{fig:perf-r1-whynot}
   }
 \end{minipage}
\end{minipage}\\
\begin{minipage}{1\linewidth}
  \centering \hspace{3mm}
 \begin{minipage}{.47\linewidth}
  \centering \vspace{2mm}
  \subfloat[\scriptsize\rel{CasualWatch}-Whynot]{
    \includegraphics[width=0.78\columnwidth, trim=110 20 10 10]{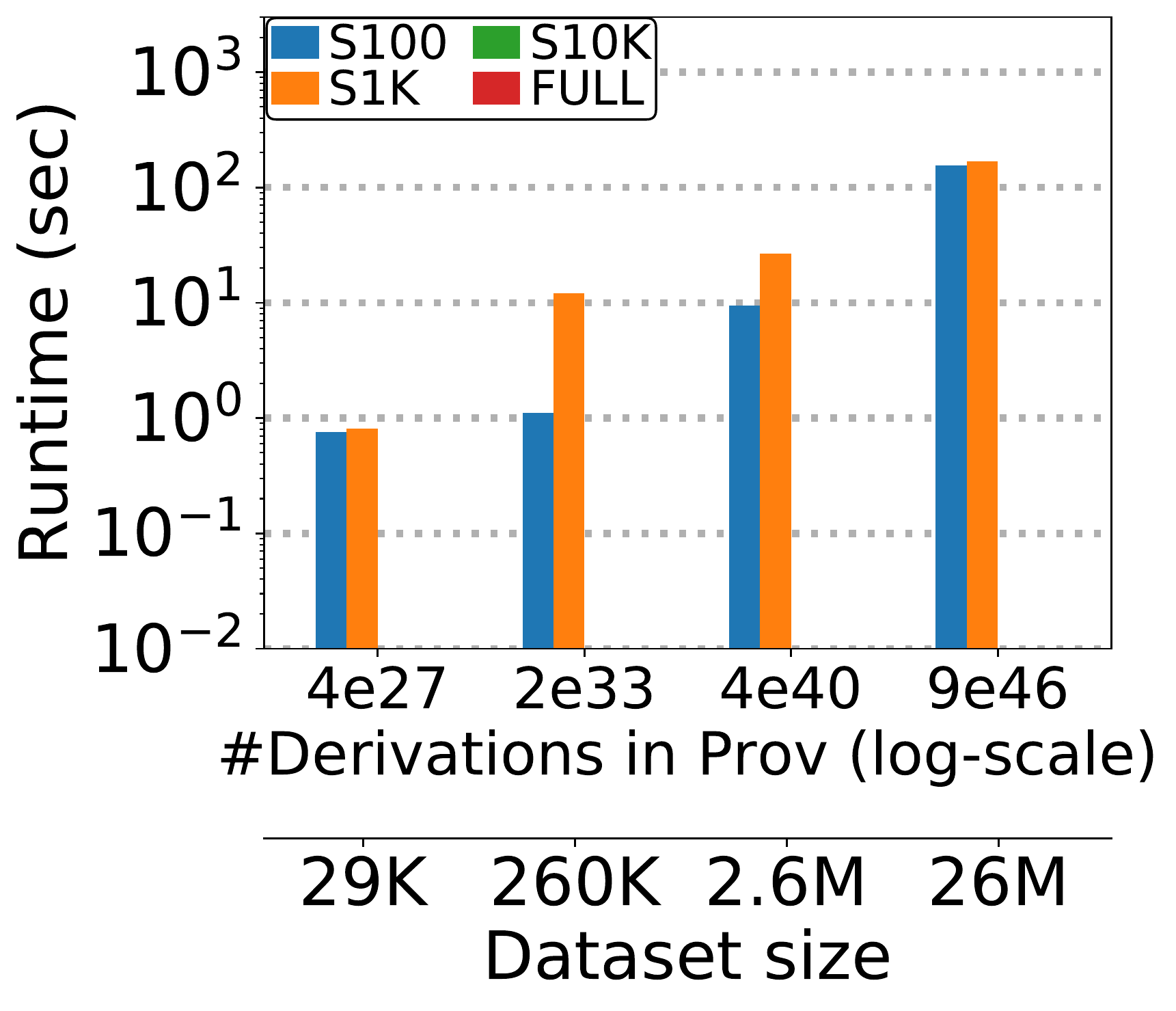}
   \label{fig:perf-cw-whynot}
  }
\end{minipage}
 \begin{minipage}{.47\linewidth}
   \centering \vspace{2mm}
   \subfloat[\scriptsize\rel{Players}-Whynot]{
     \includegraphics[width=0.78\columnwidth, trim=110 20 10 10]{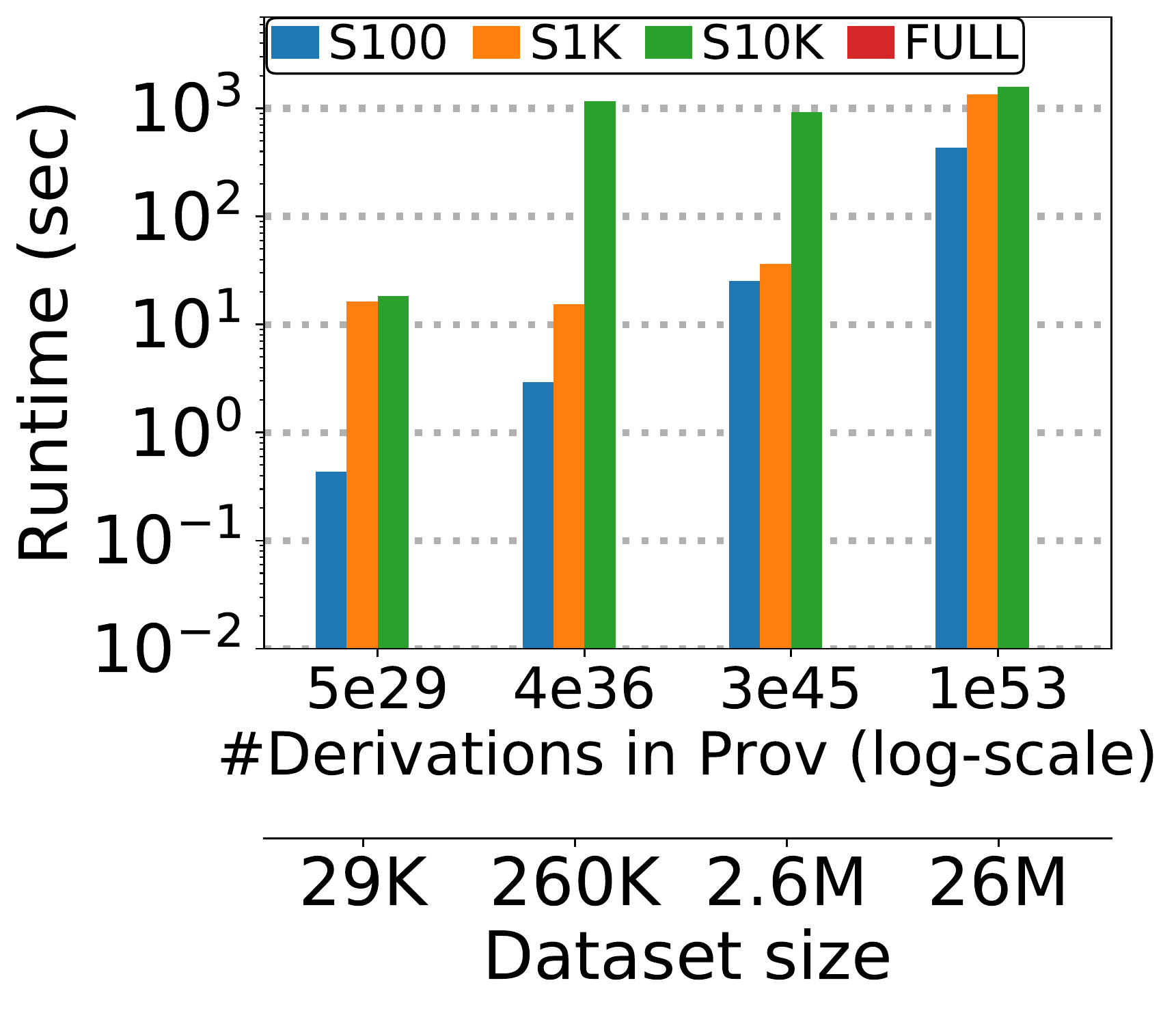}
    \label{fig:perf-pl-whynot}
   }
 \end{minipage}
\end{minipage}\\[0.5mm]
\caption{Measuring performance of generating summaries.} 
\label{fig:vary-prov}
\end{minipage}
\end{figure}
}

\iftechreport{
\begin{figure*}[t]
\begin{minipage}{1\linewidth}
 \centering 
 \begin{minipage}{0.48\linewidth}
   \centering \hspace{4mm}
  \begin{minipage}{.47\linewidth}
    \subfloat[\scriptsize\rel{InvalidD}-Why]{
      \includegraphics[width=0.76\columnwidth,trim=100 20 0 0]{perf-prov-q1-break-new.pdf}
    \label{fig:perf-r1-why}
    }
  \end{minipage}\hspace{-3mm}
  \begin{minipage}{.47\linewidth}
   \centering
   \subfloat[\scriptsize\rel{InvalidD}-Whynot]{
     \includegraphics[width=0.76\columnwidth,trim=110 20 10 10]{perf-whynot-q1-break-new.pdf}
    \label{fig:perf-r1-whynot}
   }
 \end{minipage}
  \centering \hspace{3mm}
 \begin{minipage}{.47\linewidth}
  \centering \vspace{2mm}
  \subfloat[\scriptsize\rel{CasualWatch}-Why]{
    \includegraphics[width=0.78\columnwidth, trim=100 20 0 0]{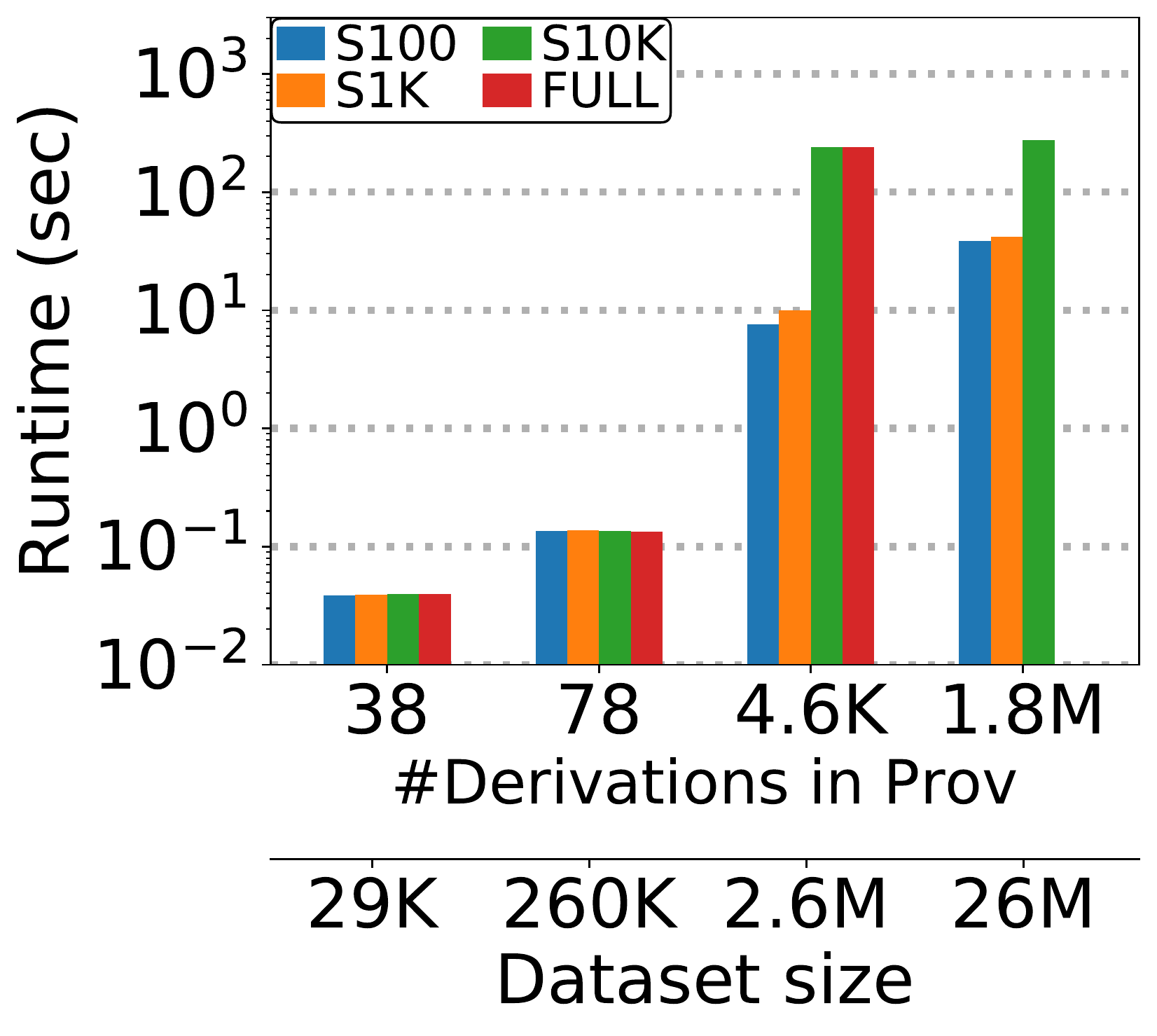}
   \label{fig:perf-cw-why}
  }
\end{minipage}
 \begin{minipage}{.47\linewidth}
   \centering \vspace{2mm}
   \subfloat[\scriptsize\rel{CasualWatch}-Whynot]{
     \includegraphics[width=0.78\columnwidth, trim=110 20 10 10]{casualwatch-whynot.pdf}
    \label{fig:perf-cw-whynot}
   }
 \end{minipage}
  \centering \hspace{3mm}
 \begin{minipage}{.47\linewidth}
  \centering \vspace{2mm}
  \subfloat[\scriptsize\rel{Players}-Why]{
    \includegraphics[width=0.78\columnwidth, trim=100 20 0 0]{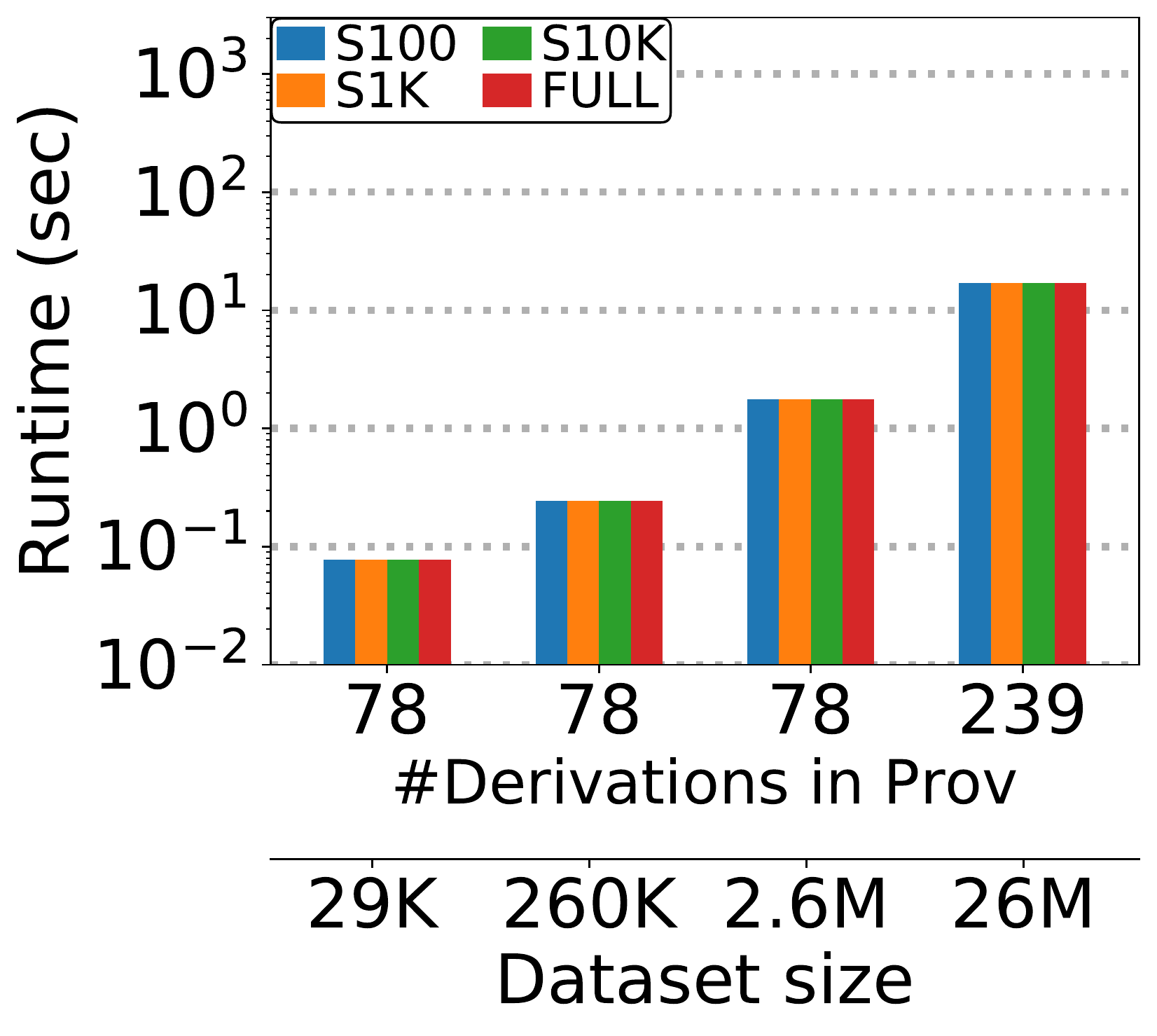}
   \label{fig:perf-pl-why}
  }
\end{minipage}
 \begin{minipage}{.47\linewidth}
   \centering \vspace{2mm}
   \subfloat[\scriptsize\rel{Players}-Whynot]{
     \includegraphics[width=0.78\columnwidth, trim=110 20 10 10]{players-whynot.pdf}
    \label{fig:perf-pl-whynot}
   }
 \end{minipage}
\end{minipage}
\begin{minipage}{0.48\linewidth}
  \centering \hspace{1mm}
 \begin{minipage}{.47\linewidth}
  \centering 
  \subfloat[\scriptsize\rel{DirGen}-Why]{
    \includegraphics[width=0.78\columnwidth, trim=100 20 0 0]{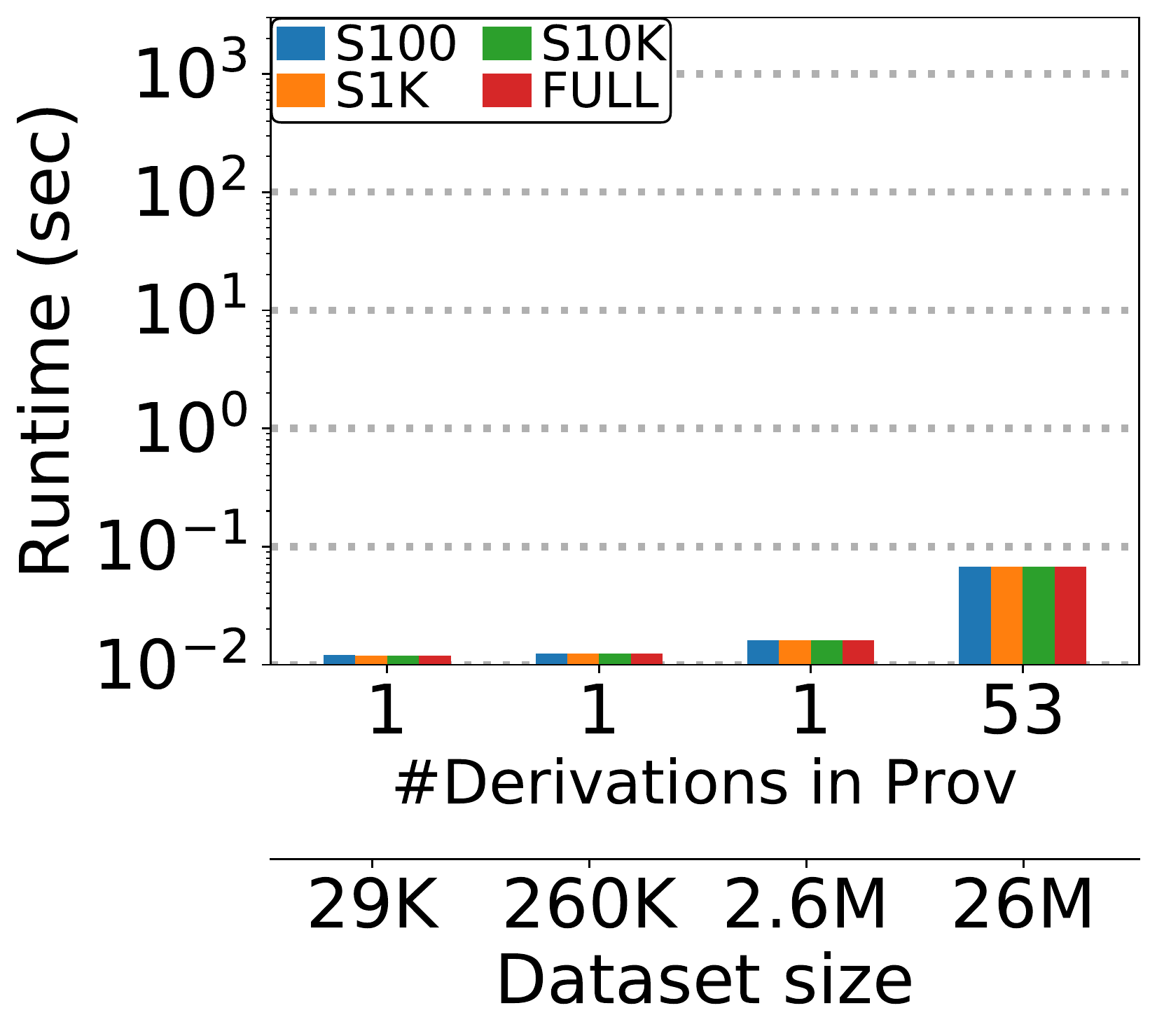}
   \label{fig:perf-dg-why}
  }
\end{minipage}
 \begin{minipage}{.47\linewidth}
   \centering 
   \subfloat[\scriptsize\rel{DirGen}-Whynot]{
     \includegraphics[width=0.78\columnwidth, trim=110 20 10 10]{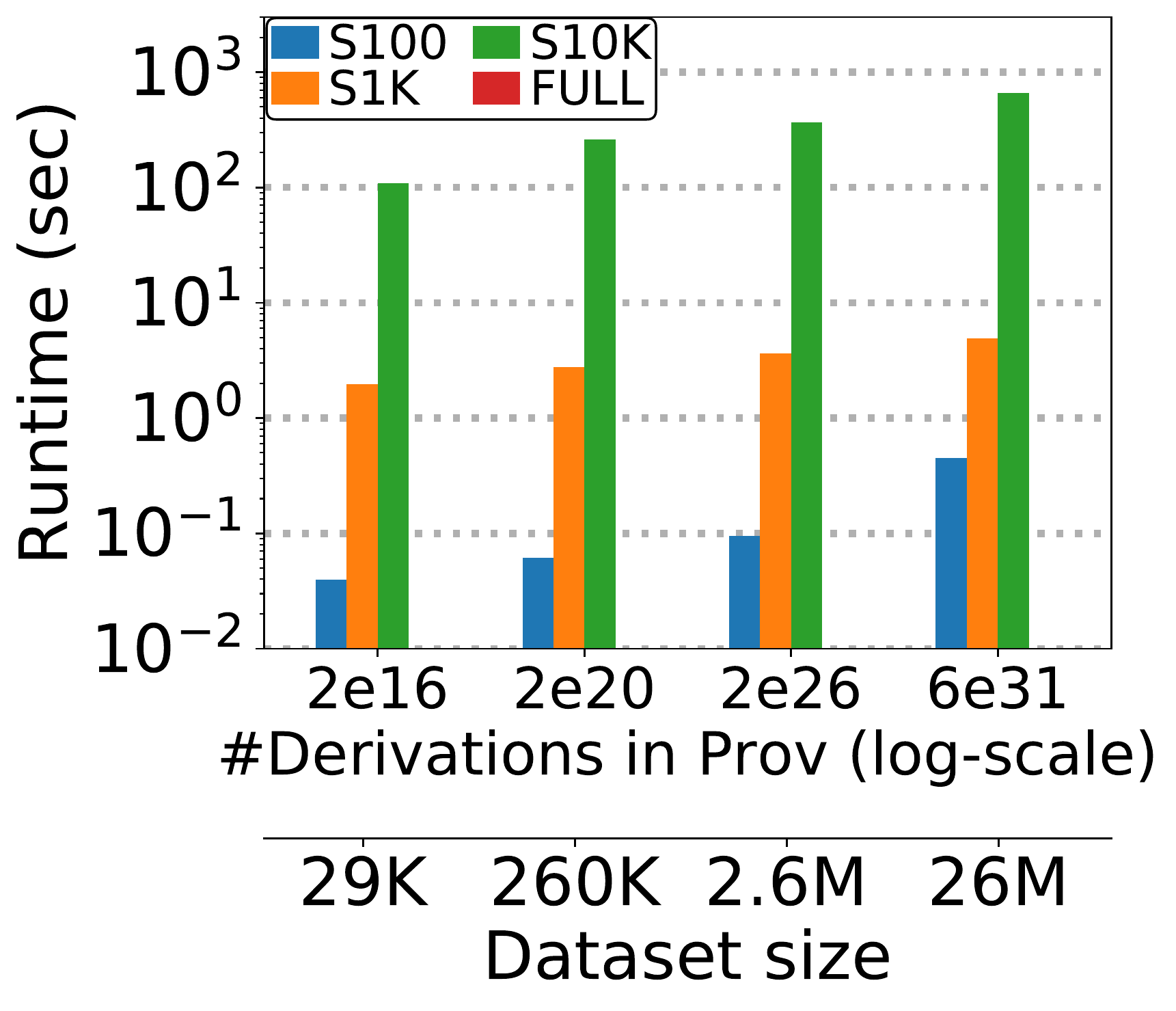}
    \label{fig:perf-dg-whynot}
   }
 \end{minipage}
\\
 \begin{minipage}{.47\linewidth}
  \centering \vspace{2mm}
  \subfloat[\scriptsize\rel{TomKey}-Why]{
    \includegraphics[width=0.78\columnwidth, trim=100 20 0 0]{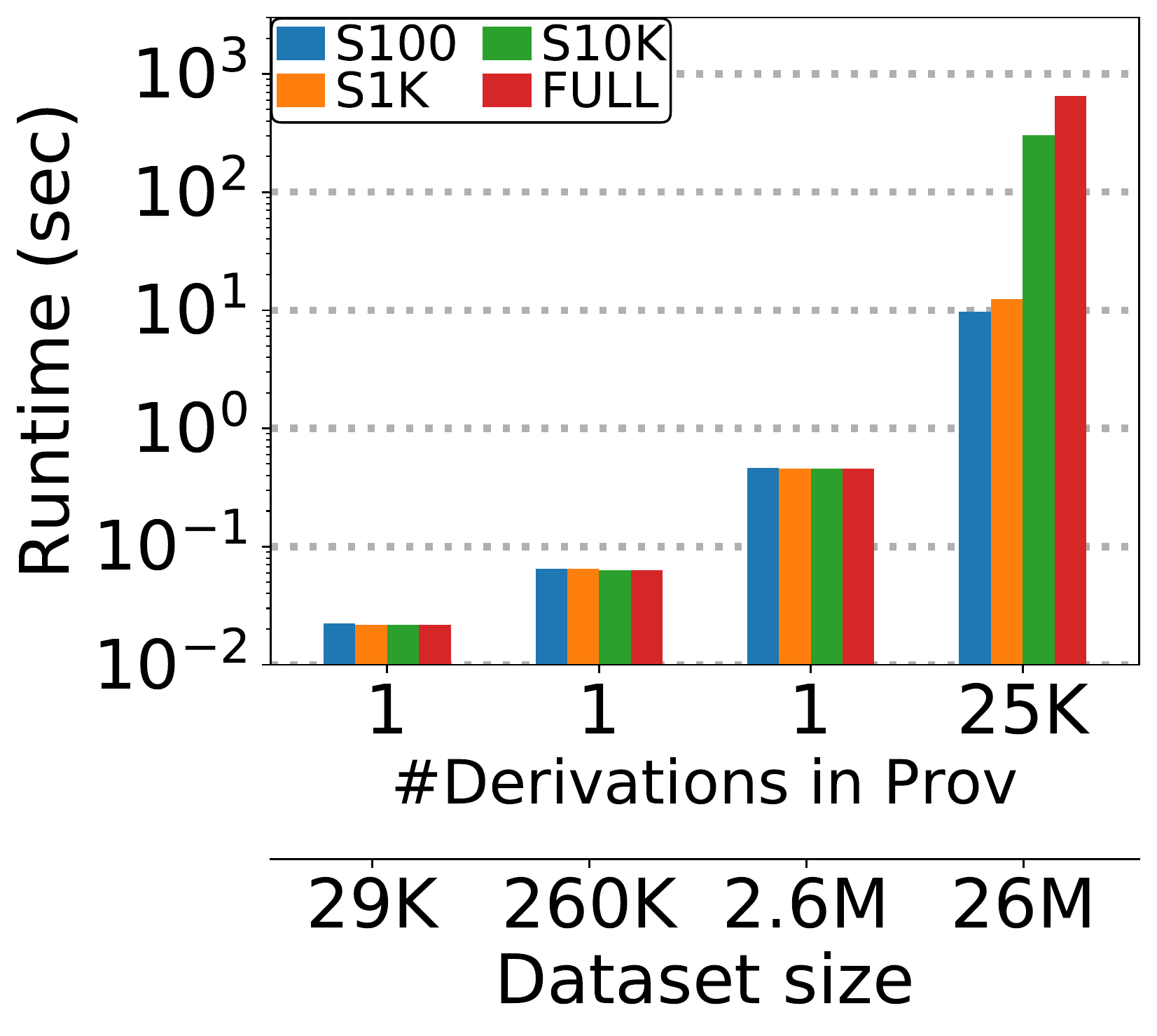}
   \label{fig:perf-tk-why}
  }
\end{minipage}
 \begin{minipage}{.47\linewidth}
   \centering \vspace{2mm}
   \subfloat[\scriptsize\rel{TomKey}-Whynot]{
     \includegraphics[width=0.78\columnwidth, trim=110 20 10 10]{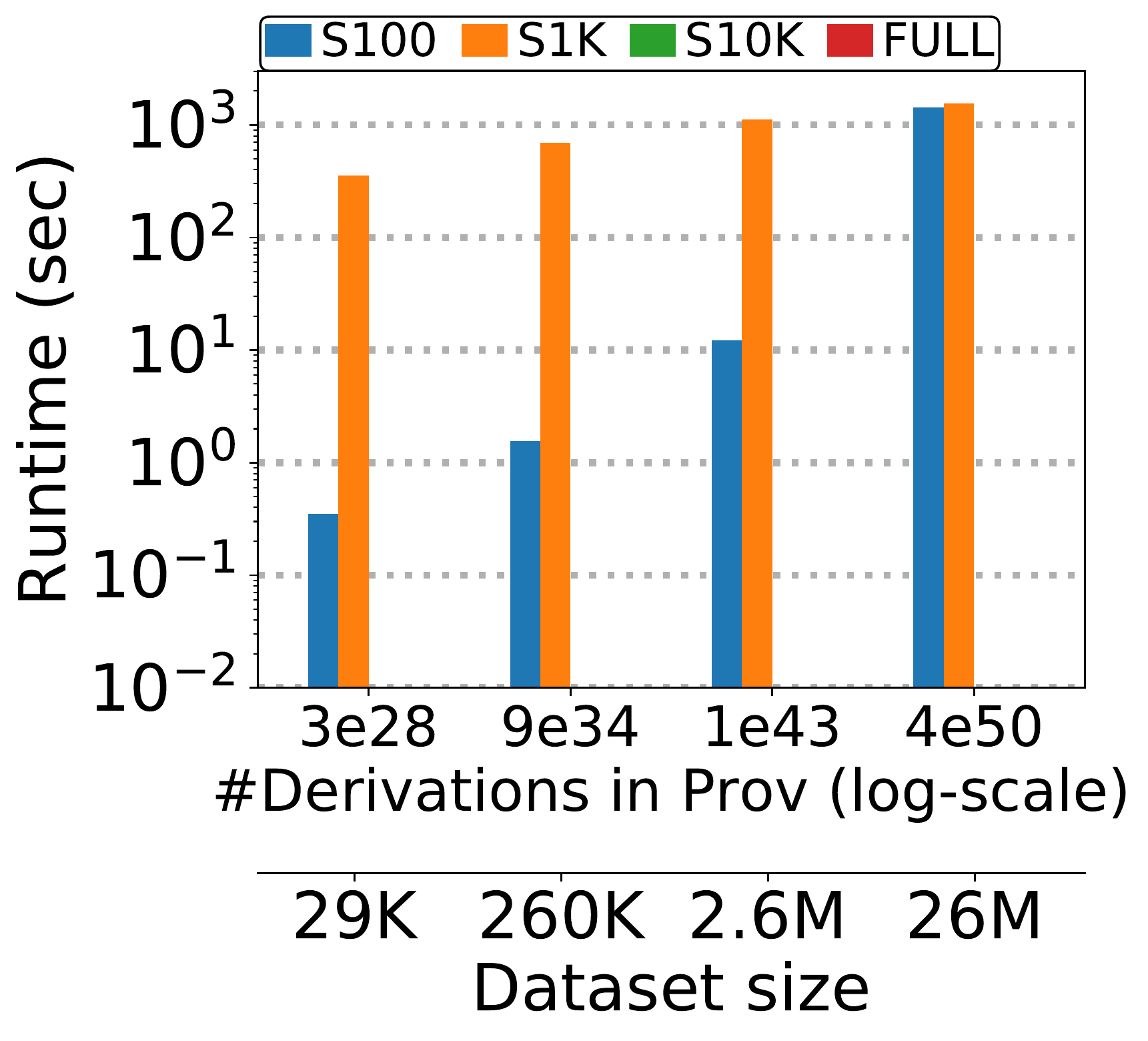}
    \label{fig:perf-tk-whynot}
   }
 \end{minipage}
 \\
  \begin{minipage}{.47\linewidth}
  \centering \vspace{2mm}
  \subfloat[\scriptsize\rel{FavCom}-Why]{
    \includegraphics[width=0.78\columnwidth, trim=110 20 0 0]{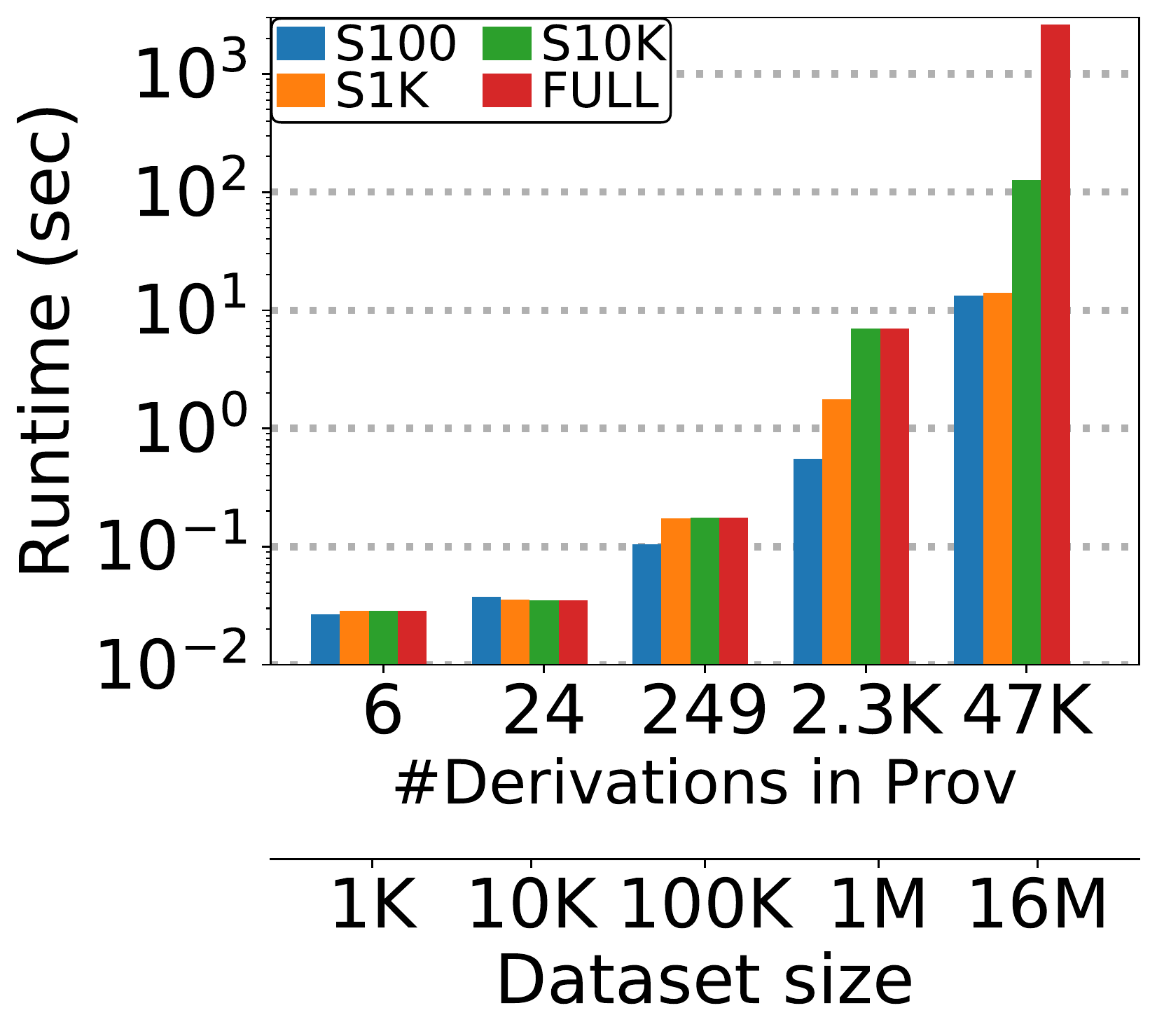}
   \label{fig:perf-q5-why}
  }
\end{minipage}
 \begin{minipage}{.47\linewidth}
   \centering \vspace{2mm}
   \subfloat[\scriptsize\rel{FavCom}-Whynot]{
     \includegraphics[width=0.78\columnwidth, trim=110 20 10 10]{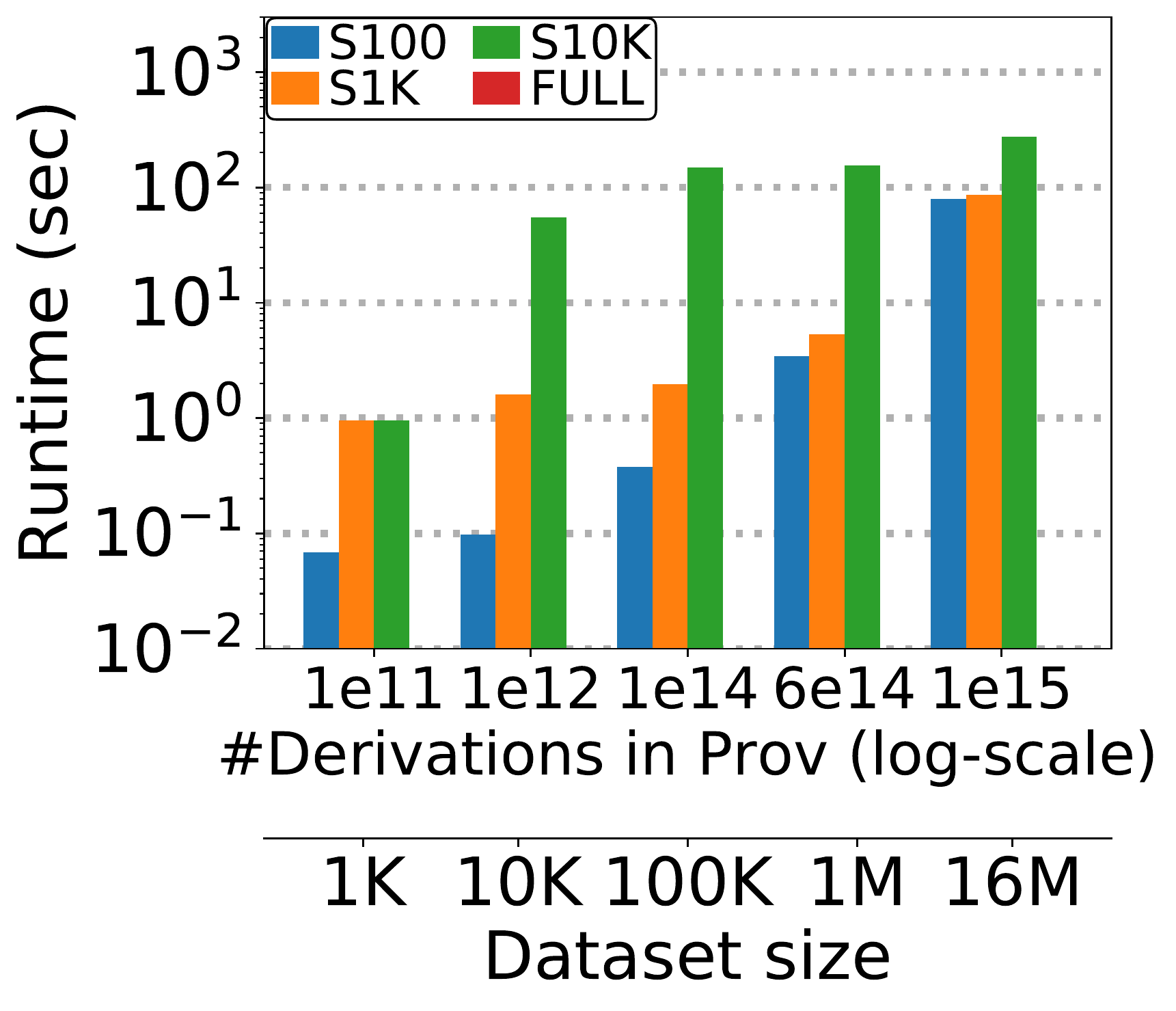}
    \label{fig:perf-q5-whynot}
   }
 \end{minipage}
\end{minipage}
\\
\caption{Measuring performance of generating summaries for why and why-not provenance for queries
}
\label{fig:vary-prov}
\end{minipage}
\end{figure*}
}

\subsection{Performance} \label{sec:perf-eval}
We consider samples of varying size  (\texttt{Sx} denotes a sample with $x$ rows). Furthermore, \texttt{Full} denotes using the full provenance as input for summarization.
Missing bars indicate timed-out experiments ($30$ minute default timeout).

\mypara{Dataset Size}
We measure the runtime of our approach for computing top-$3$ summaries varying dataset and sample size over the queries from \Cref{fig:experi-queries}. 
On the x-axis of plots, we show both the provenance size (\#derivations) and dataset size (\#rows). 
In \Cref{fig:perf-r1-why,fig:perf-r1-whynot}, we show the runtimes of the individual  steps of our algorithm (sampling, pattern generation, computation of quality metrics, and computing the top-3 summary) for query $r_1$ when using a provenance question that binds $C$ to  \textsf{new york} (why) and \textsf{swanton} (why-not), respectively\iftechreport{ (the bindings for provenance questions are shown in \Cref{tab:bindings})}.
Observe that, even for the largest dataset, we are able to generate summaries within reasonable time if using sampling. Overall, pattern generation dominates the runtime 
for why provenance.
\ifnottechreport{From now on, we focus on why-not provenance.  
    For queries $r_3$ (many joins with a negation) and the union of $r_4$, $r_4'$, and $r_4''$, the runtimes are shown in \Cref{fig:perf-cw-whynot,fig:perf-pl-whynot}, respectively.}
For why-not provenance, sampling dominates the runtime for smaller sample sizes while pattern generation is dominant for \texttt{S10K}. 
\texttt{FULL} does not finish even for the smallest license dataset (1K).
\iftechreport{ For queries $r_3$ (many joins with a negation) and $r_4$ (union of $r_4'$ and $r_4''$), 
  \Cref{fig:perf-cw-why,fig:perf-pl-why} show the runtimes for why and \Cref{fig:perf-cw-whynot,fig:perf-pl-whynot} show the runtimes for why-not.
  }
We observe the same trend as for $r_1$ even though why-not provenance is significantly larger (up to $10^{52}$ derivations).
\iftechreport{
  This trend is preserve over the runtimes of other queries $r_5$, $r_6$, and $r_7$ (\Cref{fig:perf-q5-why,fig:perf-dg-why,fig:perf-tk-why} for why and \Cref{fig:perf-q5-whynot,fig:perf-dg-whynot,fig:perf-tk-whynot} for why-not, respectively).
}

\iftechreport{
 \mypara{Performance Comparison with Naive Approach}
 We also compare the performance of our sample-based summaries to the summary 
 over full provenance (\texttt{FULL}). 
 The 
 \texttt{FULL} is shown as 
 `+' or red bars in \Cref{fig:vary-prov}. 
 The results of \texttt{FULL} for why provenance 
 are quadratic increase over the size of successful derivations while 
 summaries result in almost linear increase. Computing summaries over \texttt{FULL} why-not provenance 
 are not feasible within the allocated time slot for any size 
 (
 bars are omitted in \Cref{fig:vary-prov}). 
}

\begin{figure}[t]
\begin{minipage}{1\linewidth}
 \centering$\,$\\[-3mm]
 \begin{minipage}{1\linewidth}
  \centering
  \begin{minipage}{.9\linewidth}
   \centering
     \includegraphics[width=0.83\columnwidth, trim=80 20 0 0]{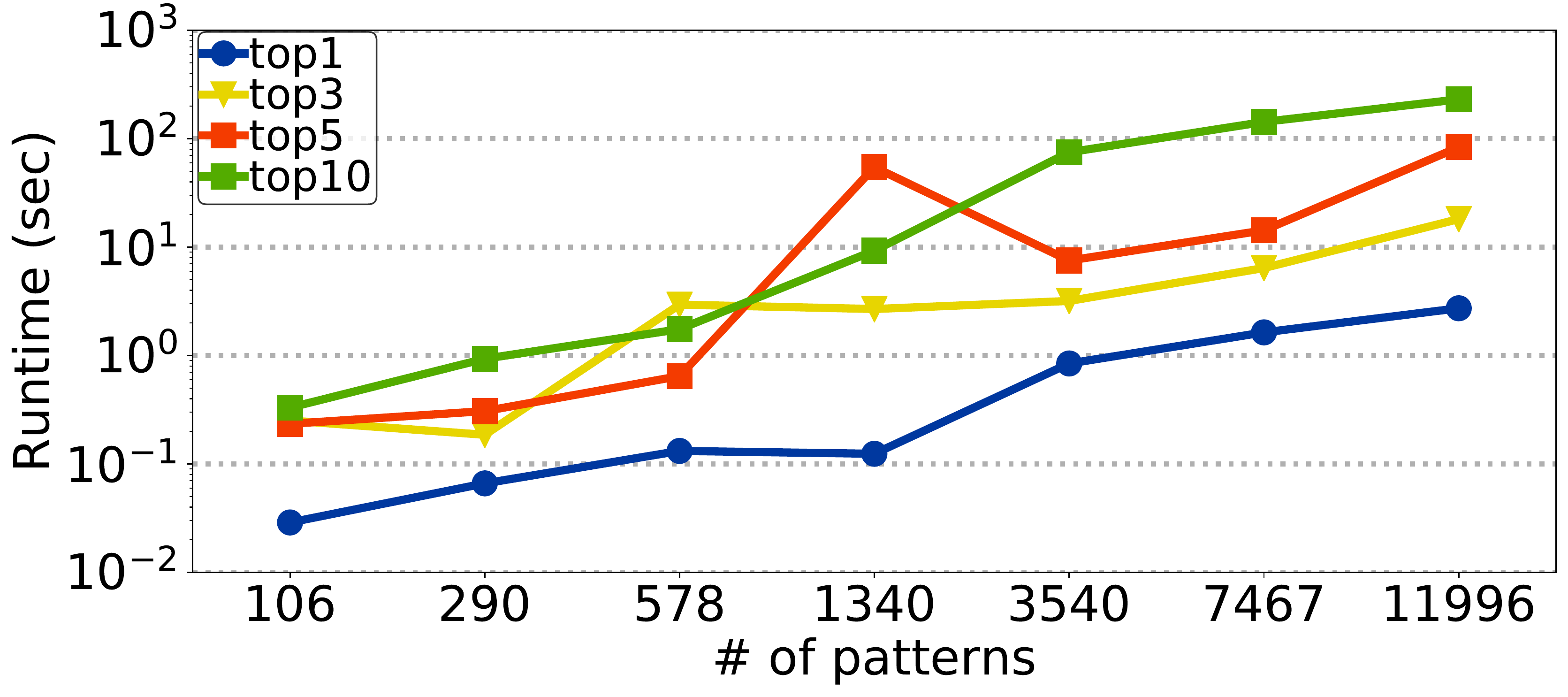}
     \label{fig:topk-why}
\end{minipage}
\end{minipage}
\\[1mm]
\caption{Runtime for computing top-$k$ summaries when patterns are provided as input. 
}
\label{fig:perf-topk}
\end{minipage}
\end{figure}

\mypara{Generating Top-$k$ summaries}
\Cref{fig:perf-topk} shows the runtime of computing the top-$k$ summary over the patterns produced by the first three steps (\Cref{sec:sampling-provenance} through \Cref{sec:estimating-pattern-c}).
We selected sets of patterns from different queries and sample sizes to get a roughly exponential progression of the number of patterns.
We vary $k$ 
from $1$ to $10$.
The runtime is roughly linear in $k$ and in the number of patterns.
Note that this is significantly better than the theoretical worst-case of our algorithm ($\oNotation{n^k}$ where $n$ is the number of patterns). The reason is that typically a large number of patterns is clearly inferior and will be pruned by our algorithm early on.

\begin{figure}[t]
  \vspace{-2mm}
\begin{minipage}{1\linewidth}
 \centering 
 \begin{minipage}{.47\linewidth}
   \centering
      \subfloat[\scriptsize Chain query]{
     \includegraphics[width=0.77\columnwidth,trim=80 20 40 0]{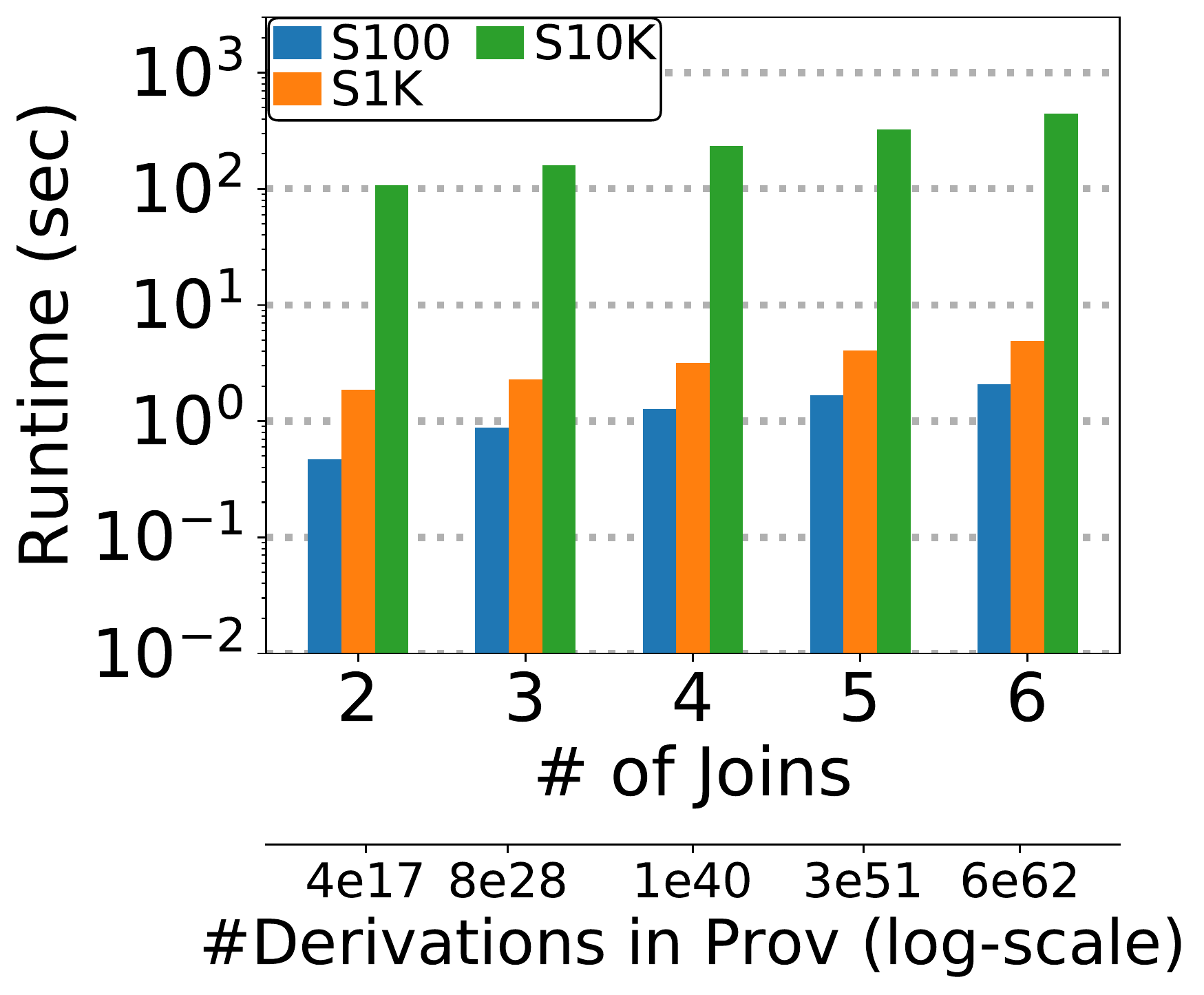}
   \label{fig:perf-joinsc}}
  \end{minipage}\hspace{0.5mm}
  \begin{minipage}{.47\linewidth}
    \centering
      \subfloat[\scriptsize Star query]{
     \includegraphics[width=0.77\columnwidth,trim=80 20 40 0]{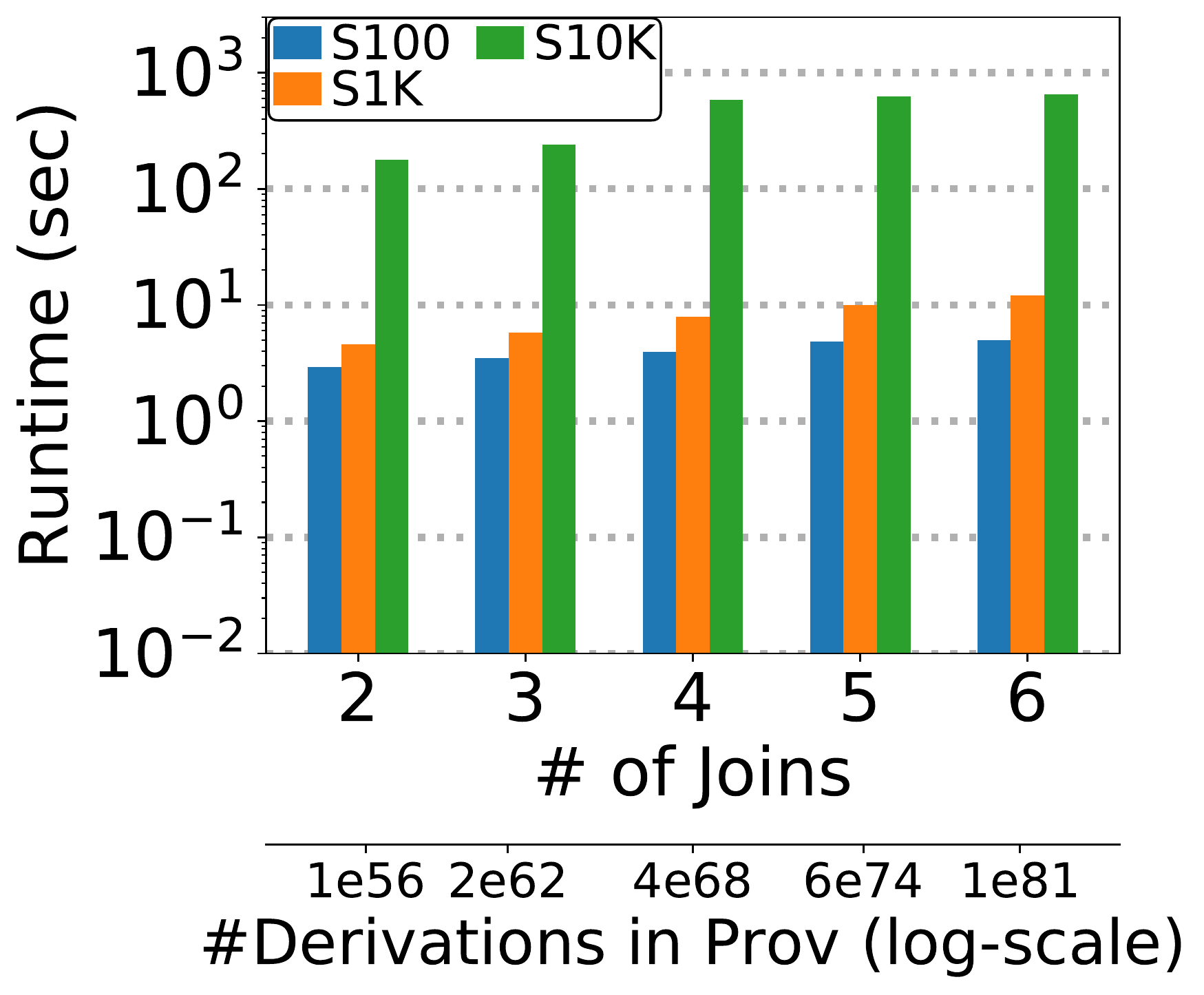}
   \label{fig:perf-joinss}}
  \end{minipage}
\\
  \begin{minipage}{.47\linewidth}
   \centering
    \subfloat[\scriptsize Chain query]{
      \includegraphics[width=0.77\columnwidth,trim=80 20 40 0]{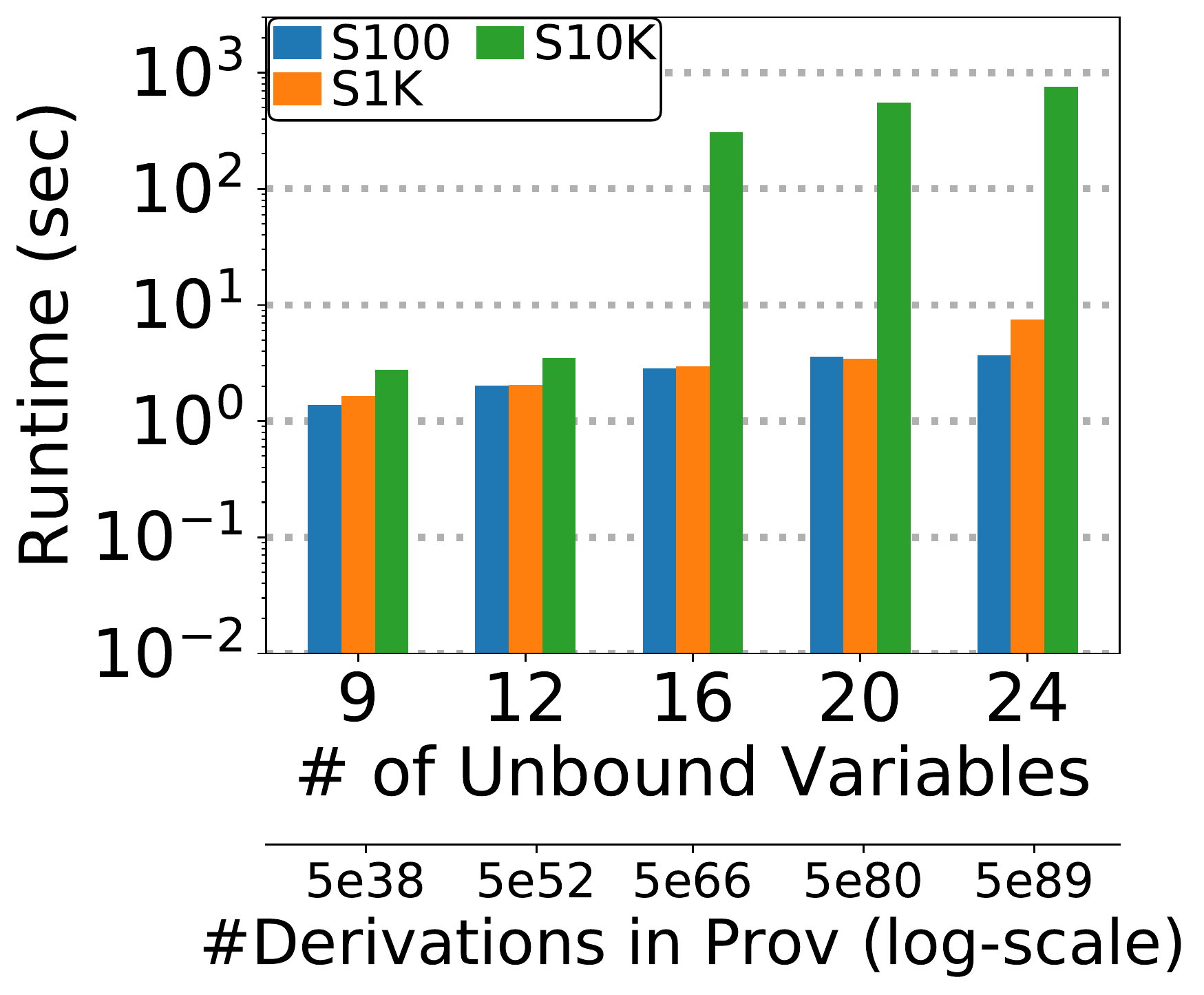}
      \label{fig:perf-varsc}}
 \end{minipage}
  \begin{minipage}{.47\linewidth}
   \centering
    \subfloat[\scriptsize Star query]{
      \includegraphics[width=0.77\columnwidth,trim=80 20 50 00]{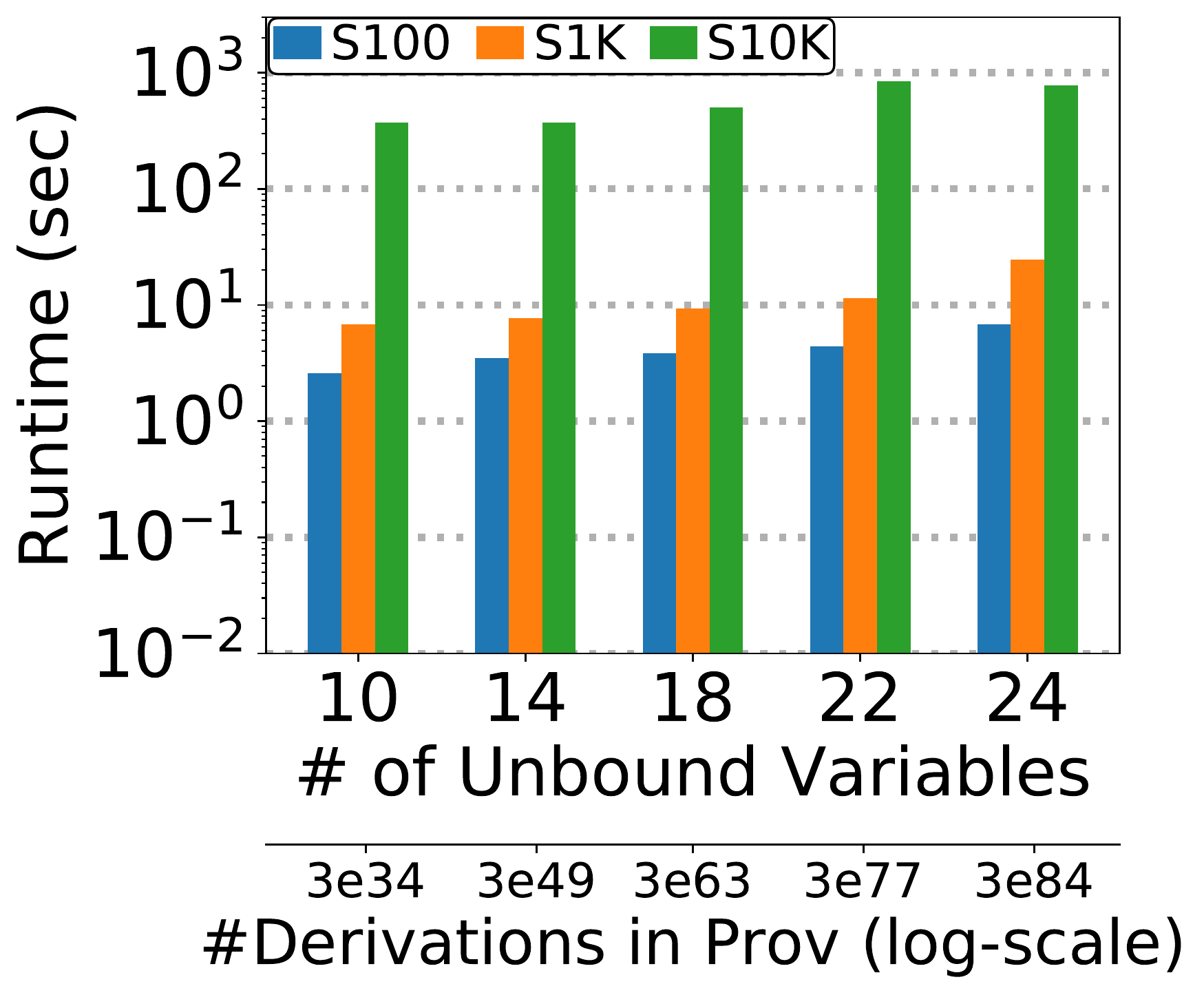}
   \label{fig:perf-varss}}
\end{minipage}
\iftechreport{
    \\
  \begin{minipage}{.47\linewidth}
   \centering
      \subfloat[\scriptsize  $r_{11}$ over $\text{DBLP}_{100\text{K}}$]{
     \includegraphics[width=0.77\columnwidth,trim=80 20 55 0]{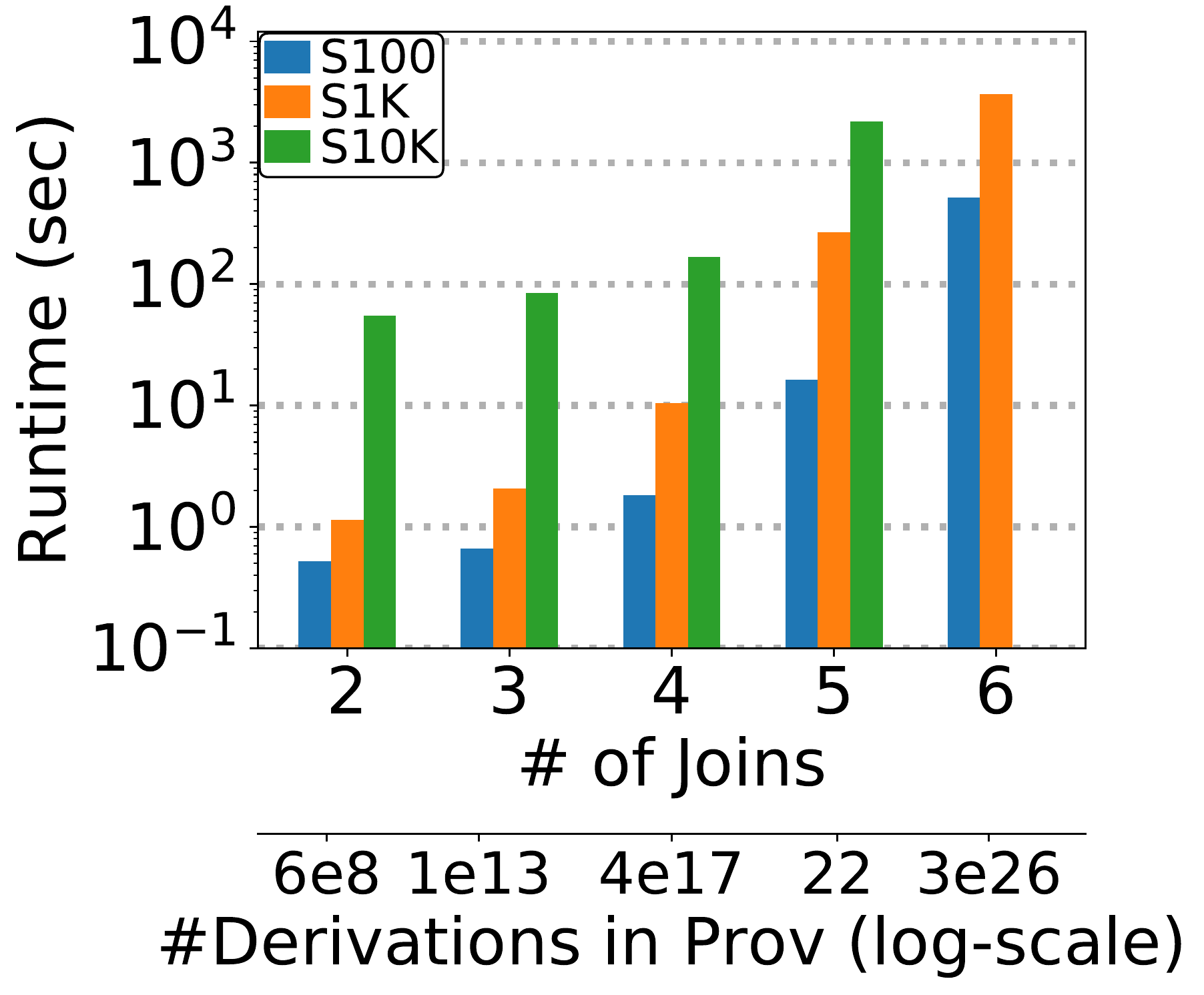}
   \label{fig:perf-dblp}}
  \end{minipage}\hspace{0.5mm}
  \begin{minipage}{.47\linewidth}
    \centering
      \subfloat[\scriptsize $r_{12}$ over $\text{TPC-H}_{150\text{K}}$]{
     \includegraphics[width=0.77\columnwidth,trim=80 20 60 0]{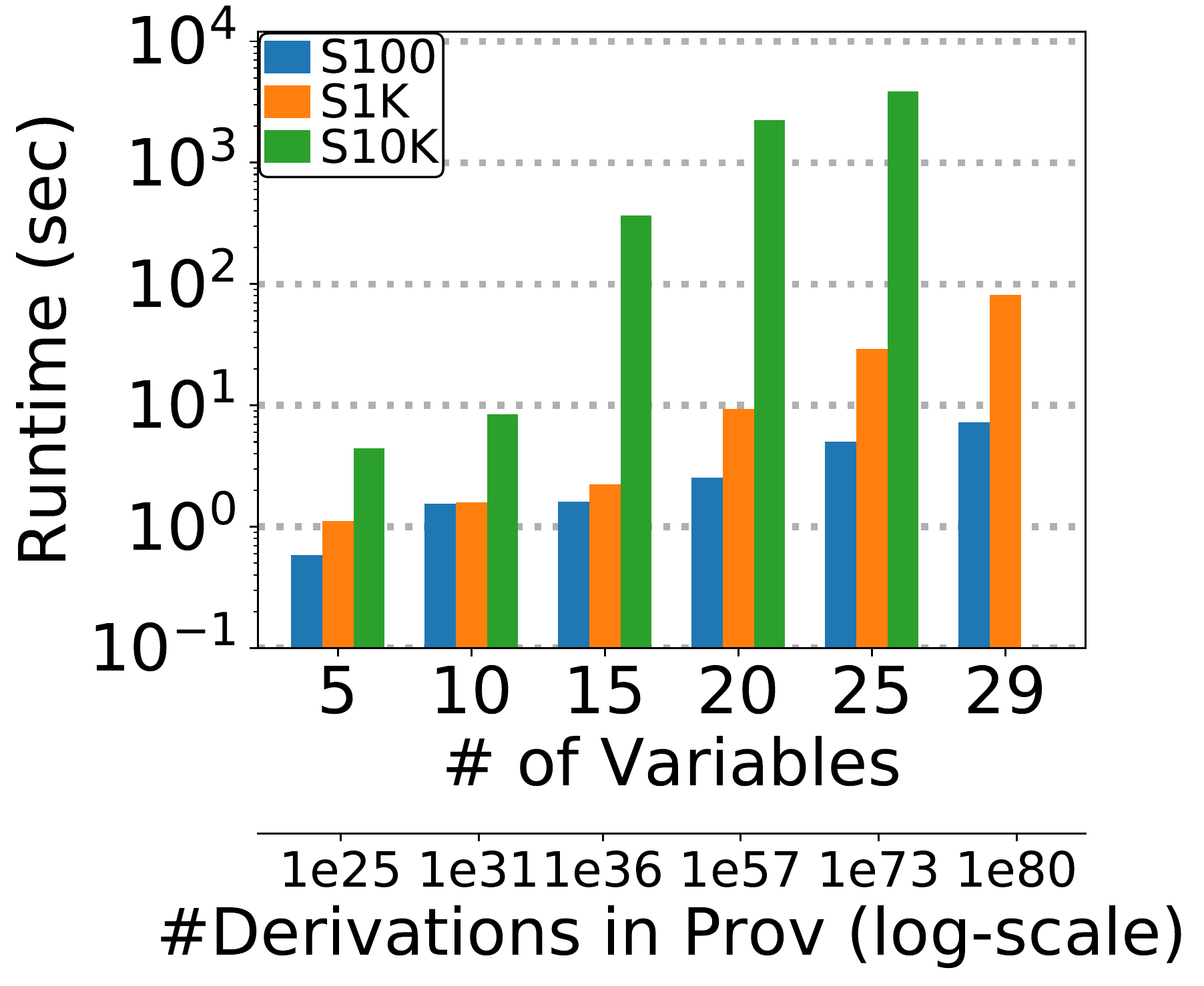}
   \label{fig:perf-tpch}}
\end{minipage}
}
\\[0mm]
 \caption{Varying \#joins and \#variables  for why-not}
\label{fig:joins-vars}
\end{minipage}
\end{figure}

\mypara{Query Complexity and Structure}
In this experiment, we vary the query complexity in terms of number of joins and variables.
We randomly generated synthetic queries whose join graph is either a star or a chain.
We  
compute the top-3 patterns for why-not provenance. 
In \Cref{fig:perf-joinsc,fig:perf-joinss} we vary the number of joins. 
The results confirm that our approach scales to very large provenance sizes (more than $10^{60}$ derivations) regardless of join types.
To evaluate the impact of the number of variables on performance, we use chain queries with $8$-way joins and star queries with  $5$-way joins. We vary the number of variables bound to constants by the query from $1$ to $16$ (out of $25$ variables).
The head 
and join variables 
are never bound.
The results  shown in \Cref{fig:perf-varsc,fig:perf-varss} 
 confirm that our approach works well, 
 even for queries with up to $24$ unbound variables 
 (provenance sizes of up to $\sim 10^{80}$).

\iftechreport{
  We now extend the evaluation with other queries and datasets.
  For the extension of variable impacts, we use query $r_{12}$ (the join size is fixed to $3$) over $\text{TPC-H}_{\text{150K}}$ and compute summaries of why-not provenance. 
By binding an increasing number of variables from $r_{12}$ to constants, we generate 6 rules that contain between $5$ and $29$ existential variables.
The result shown in \Cref{fig:perf-tpch} confirms that our approach scales well over TPC-H dataset. 
Using the $\text{DBLP}_{\text{100K}}$ dataset, 
we vary the number of joins (path length) of query $r_{11}$.
For example, $\rel{2Hop}(L) \dlImp \rel{DBLP}(L,R),$ $\rel{DBLP}(R,R1)$ is the query we use for a  $2$-way join. 
We use a p-tuple that binds $L=\textsf{xueni pan}$ (\Cref{tab:bindings}). 
\Cref{fig:perf-dblp} shows that even for real-world dataset (with a 6-way join where the provenance contains $3 \cdot 10^{26}$ derivations), we produce a 
summary for sample sizes \texttt{S100} and \texttt{S1K}. 
}

\begin{figure}[t]
\begin{minipage}{1.0\linewidth}$\,$\\[-4mm]
  \begin{minipage}{.47\linewidth}
   \centering
   \subfloat[\scriptsize Why\iftechreport{ (\rel{InvalidD})}]{
     \includegraphics[width=0.74\columnwidth,trim=60 20 40 0]{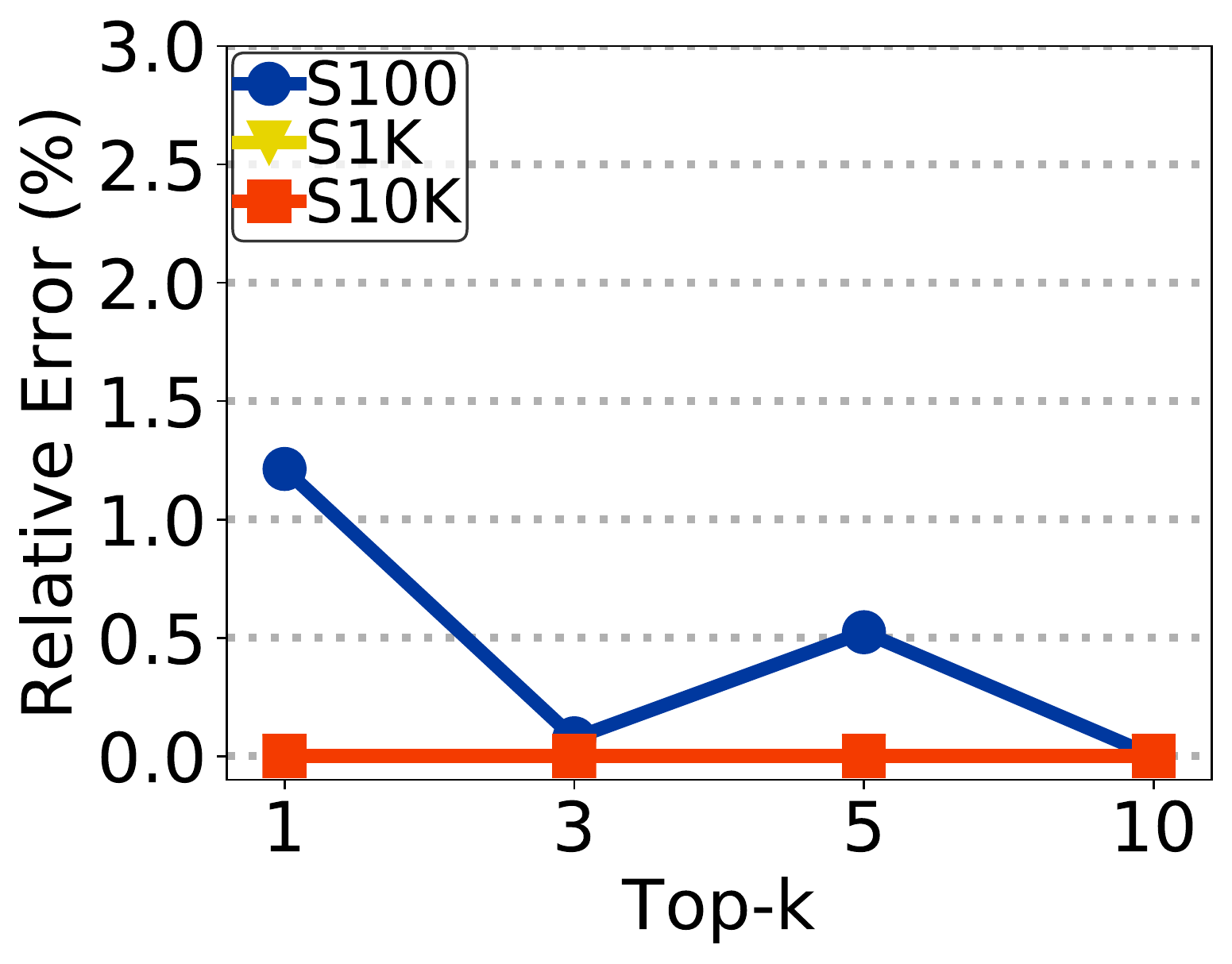}
   \label{fig:err-r1-why}
   }
   \end{minipage}
  \begin{minipage}{.47\linewidth}
   \centering
   \subfloat[\scriptsize Why-not\iftechreport{ (\rel{InvalidD})}]{
     \includegraphics[width=0.74\columnwidth,trim=60 20 40 0]{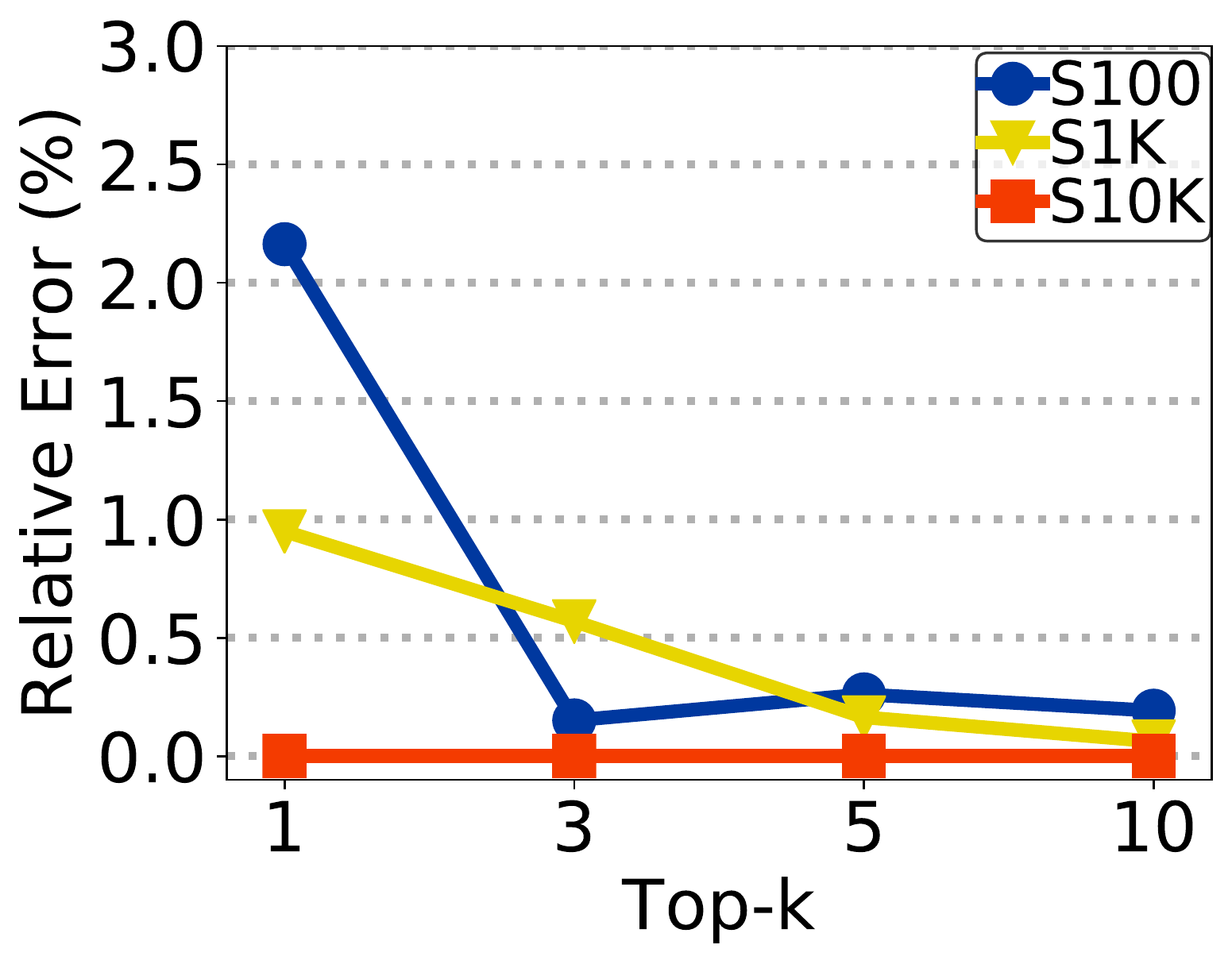}
   \label{fig:err-r1-whynot}
   }
 \end{minipage}\\
 \iftechreport{
  \begin{minipage}{.47\linewidth}
   \centering
   \subfloat[\scriptsize Why (\rel{CrimeSince})]{
      \includegraphics[width=0.74\columnwidth,trim=60 10 40 0]{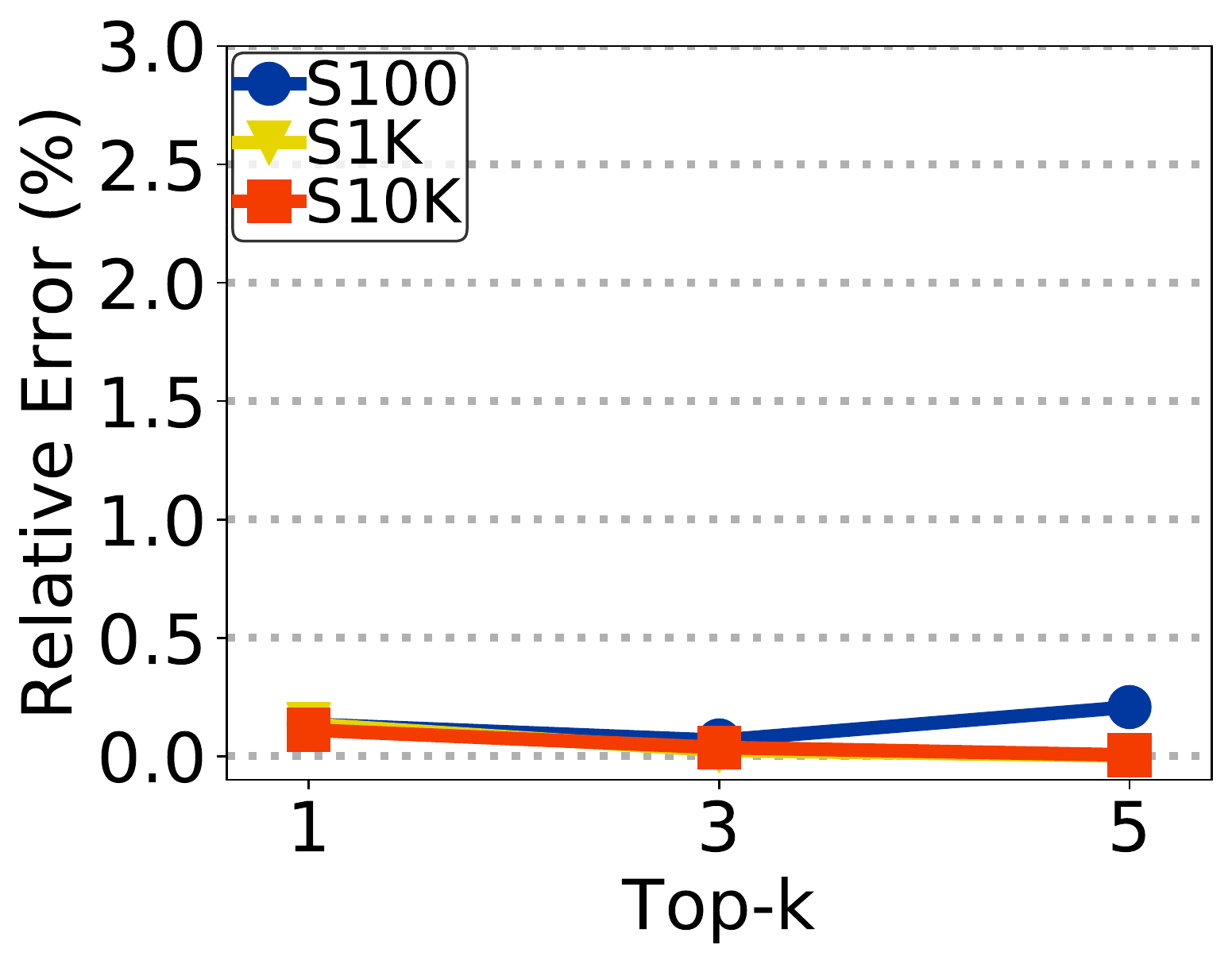}
   \label{fig:err-r4-why}
   }
   \end{minipage} \hspace{-2mm}
  \begin{minipage}{.52\linewidth}
   \centering
   \subfloat[\scriptsize Why-not (\rel{CrimeSince})]{
     \includegraphics[width=0.75\linewidth,trim=50 10 0 0]{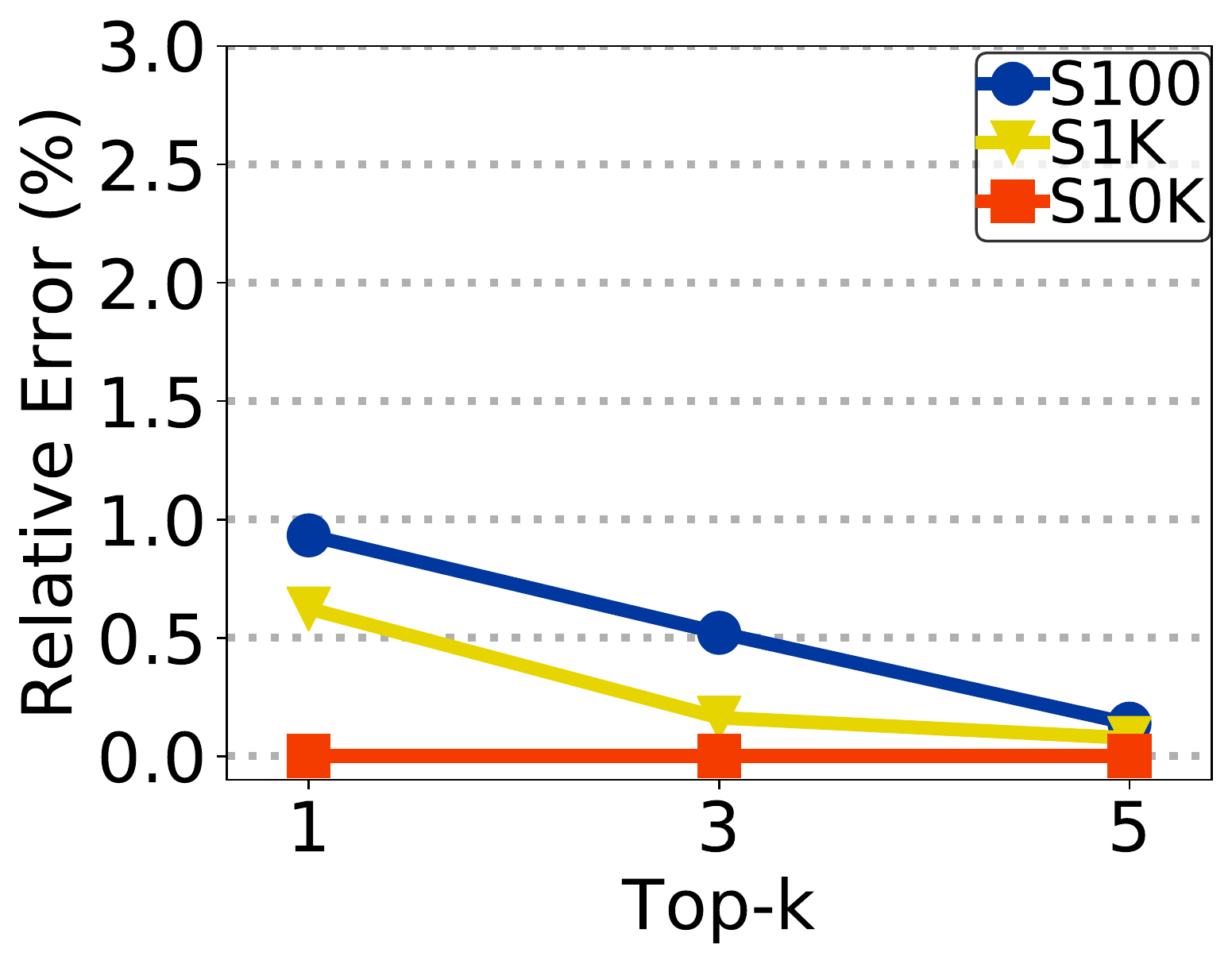}
   \label{fig:err-r4-whynot}
   }
 \end{minipage}\\
}
   \caption{
     Quality metric error caused by sampling.}
\label{fig:exp-error}
 \end{minipage}
\end{figure}

\begin{figure}[t]
\begin{minipage}{1.0\linewidth}
  \begin{minipage}{.47\linewidth}
   \centering
   \subfloat[\scriptsize Why]{
     \includegraphics[width=0.75\linewidth, trim=60 20 40 0]{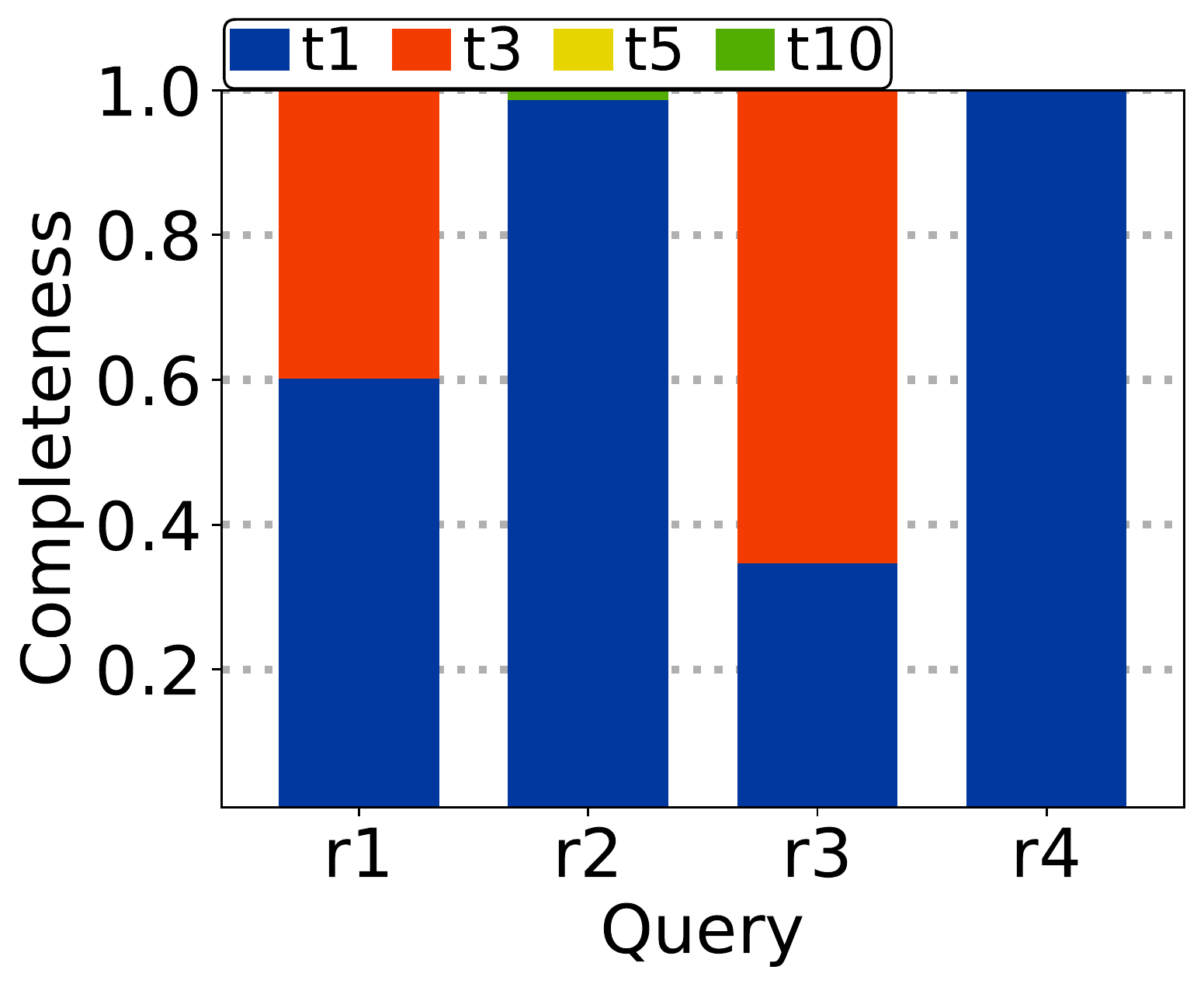}
   \label{fig:qual-comp-why}
   }
   \end{minipage}\ifnottechreport{\hspace{3mm}}
  \begin{minipage}{.47\linewidth}
   \centering
   \subfloat[\scriptsize Why-not]{
     \includegraphics[width=0.75\linewidth, trim=60 20 40 0]{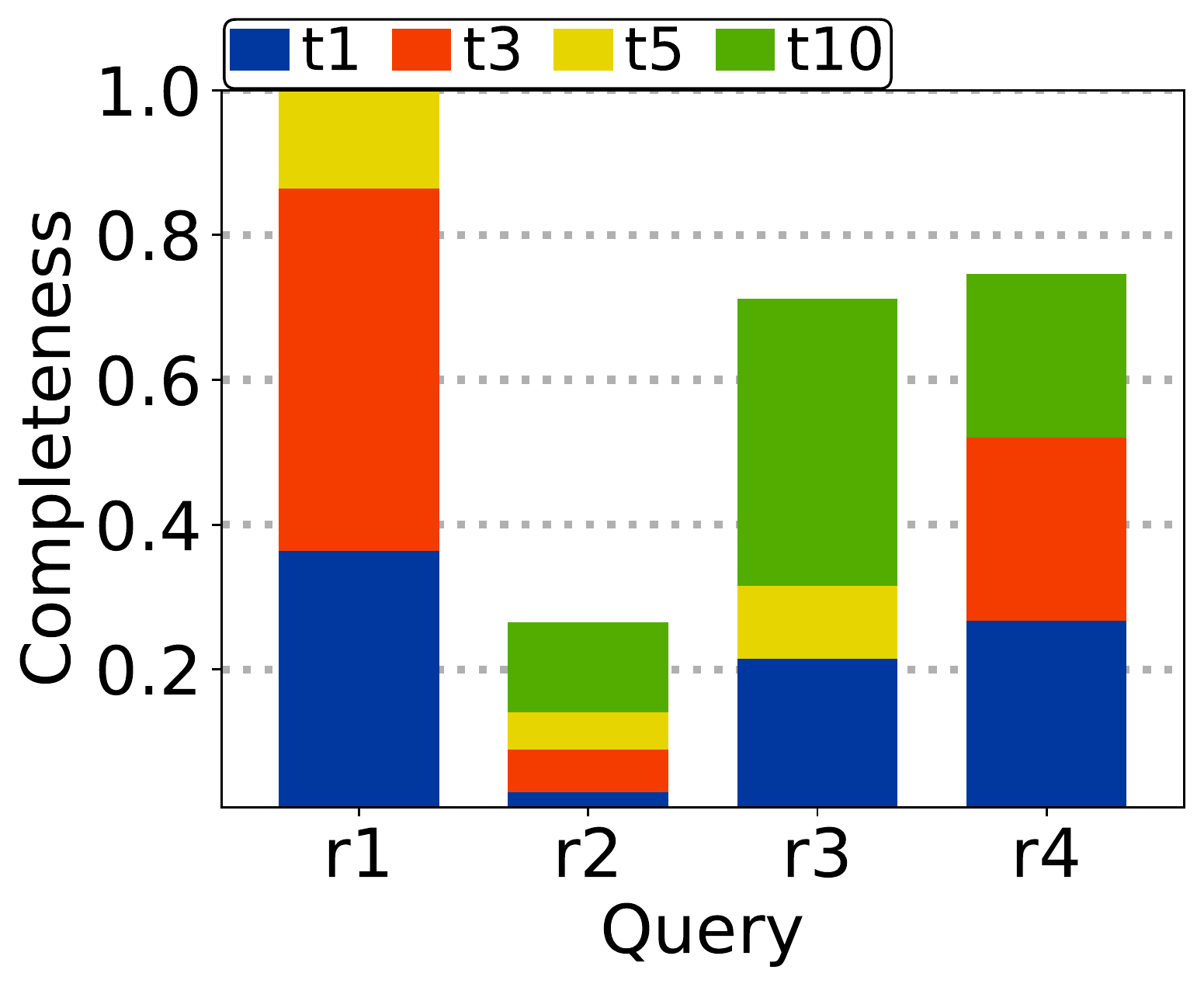}
   \label{fig:qual-comp-whynot}
   }
 \end{minipage} \\
\caption{Completeness - varying $k$.} 
\label{fig:exp-completeness}
 \end{minipage}
\end{figure}

\subsection{Pattern Quality} \label{sec:qual-eval}
We now measure the difference between the quality metrics approximated using sampling and the  exact values when using  full provenance.
For why-not provenance where it is not feasible to compute full provenance, we compare against the largest sample size (\texttt{S10k}) instead.

\mypara{Quality Metric Error}
\Cref{fig:err-r1-why,fig:err-r1-whynot} show the relative quality metric error for query $r_1$ over $\rel{InvalidD}_{\text{100K}}$ varying sample size and $k$.
The error is at most $\sim$ 2\% and typically decreases in $k$.
\ifnottechreport{The results for other queries are presented in~\cite{LL20}.}
\iftechreport{
 For query $r_{10}$ over $\rel{Crimes}_{\text{1M}}$, 
\Cref{fig:err-r4-why,fig:err-r4-whynot} show the results for why and why-not, respectively. 
Similarly, the overall relative error 
caused by sampling is quite low (below 1\%) and descreases in $k$ and sample size.
}

\mypara{Summary Completeness}
\Cref{fig:exp-completeness} shows the completeness scores of summaries returned by our approach for queries from \Cref{fig:experi-queries}.
We measure this by calculating the upper bound of completeness of the set of top-$k$ patterns  for each query as described in \Cref{sec:computing-top-k-summ}.
For $k=10$, we achieve $100\%$ completeness 
for why provenance and $\sim 75\%$ completeness 
for why-not 
except 
for $r_2$  (\Cref{fig:qual-comp-whynot})
for which the relatively large number of distinct values for the domains of unbound variables
prevents us from achieving better completeness scores.

\subsection{Comparisons with other systems}\label{sec:comparison}

\begin{figure}[t]
\ifnottechreport{\vspace{-3mm}}
\begin{minipage}{1\linewidth}
 \centering 
  \begin{minipage}{.47\linewidth}
   \subfloat[\scriptsize Artemis]{
     \includegraphics[width=0.77\columnwidth,trim=60 20 50 0]{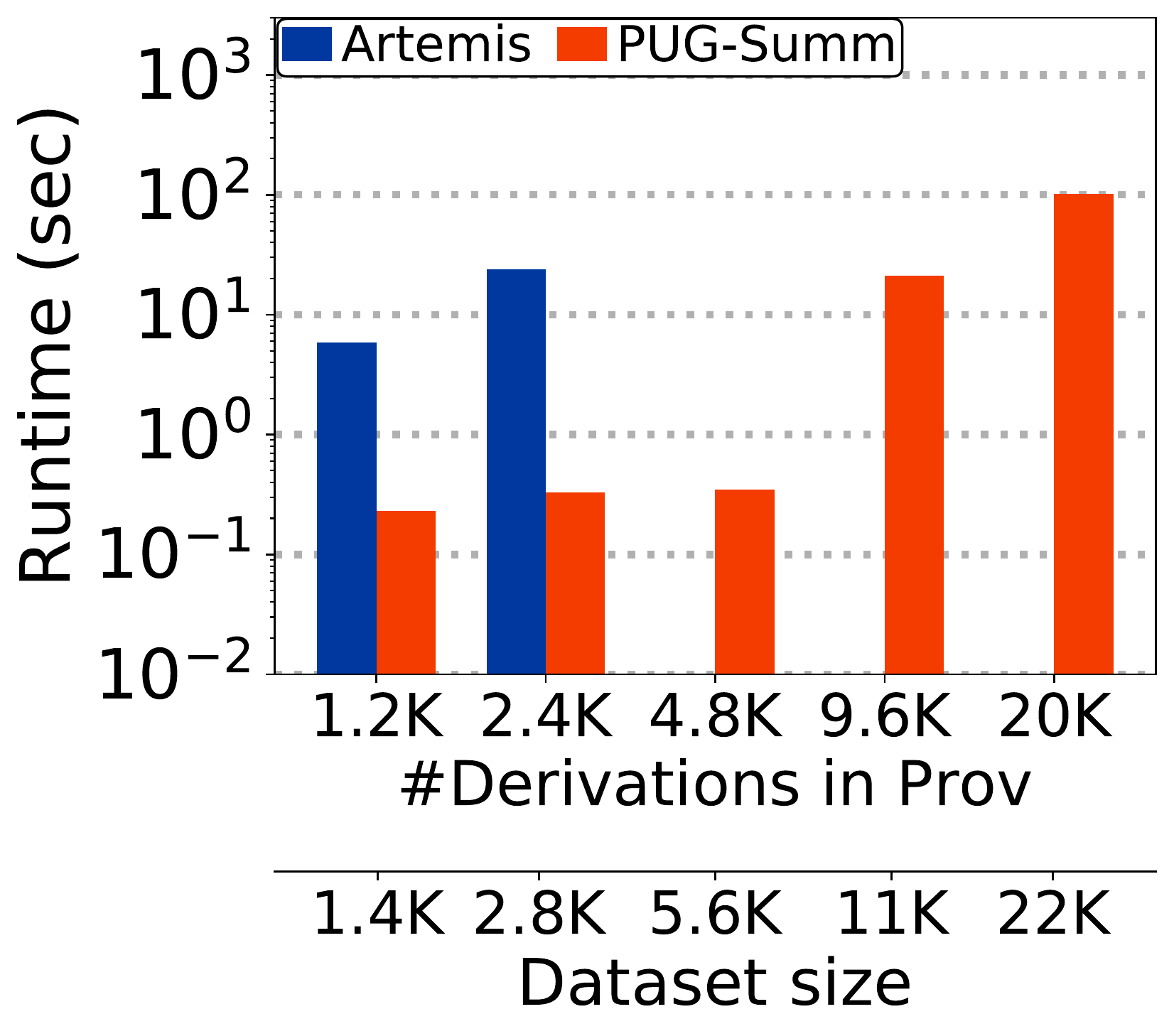}
   \label{fig:perf-artemis}}
  \end{minipage}
  \begin{minipage}{.47\linewidth}
   \centering
    \subfloat[\scriptsize \SingleDer]{
      \includegraphics[width=0.77\columnwidth,trim=60 20 50 0]{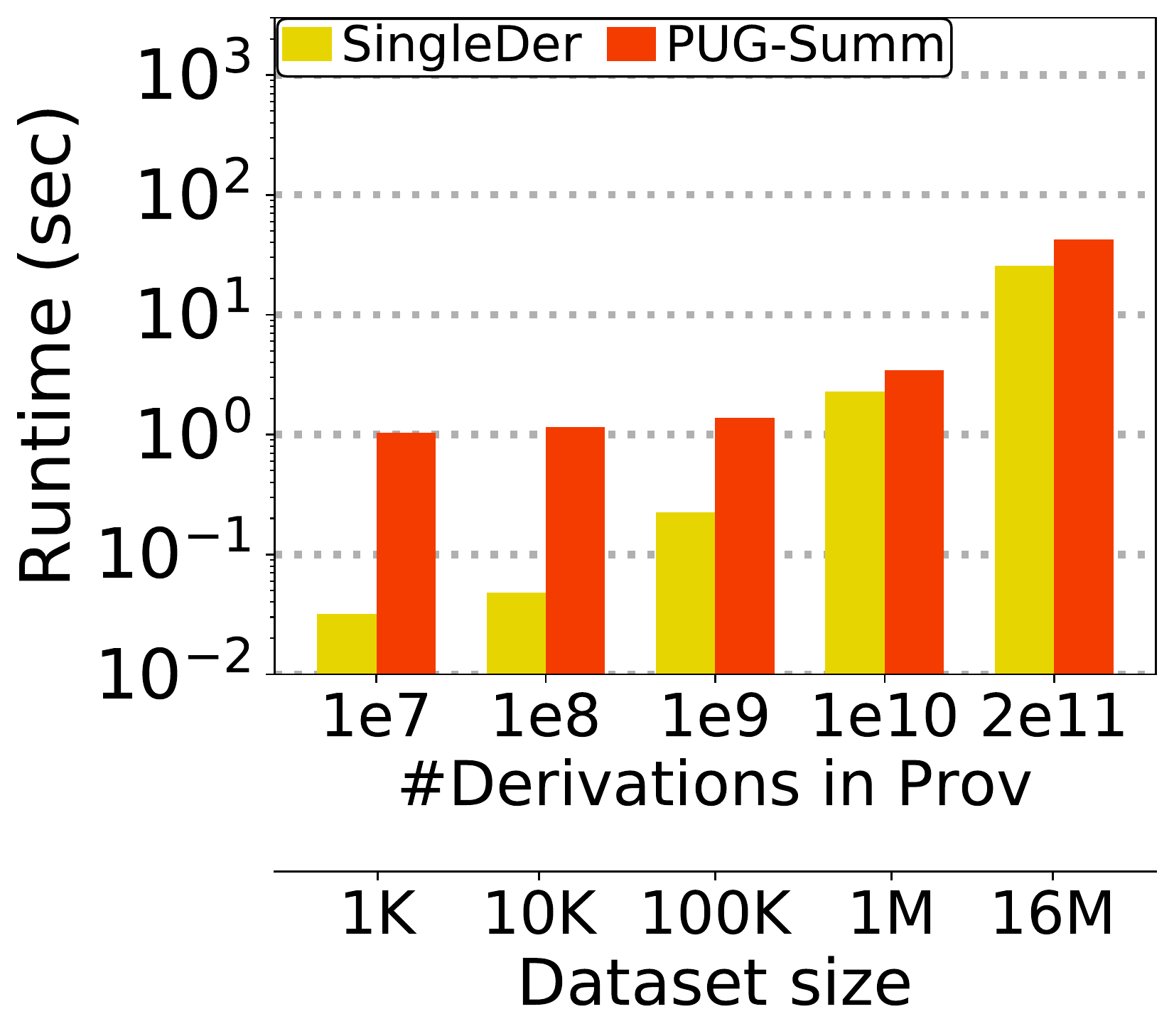}
    \label{fig:perf-ynot}}
 \end{minipage}
\\[0mm]
 \caption{Performance comparisons for why-not}
\label{fig:perf-comparison}
\end{minipage}
\end{figure}

We now compare our system against~\cite{HH10} (\allDer{}) and a \singleDer{} approach implemented in our system.

\mypara{Artemis}
The authors of~\cite{HH10} made their system \texttt{Artemis} available as a virtual machine (VM). We ran both systems in this VM (4GB memory) and used Postgres as a backend since it is supported by both systems. We used a query from the VM installation that computes the names of witnesses ($N$) who saw a person with a particular cloth and hair color ($C$ and $H$) perpetrating a crime of a particular type $T$.
\\[-3mm]
\begin{align*}
\rel{CrimeDesc}(&T,N,C,H) \dlImp \rel{CRIME}(T,S), \rel{WITNESS}(N,S),\\
&\rel{SAWPERSON}(N,H,C), \rel{PERSON}(M,H,C), S > 97
\end{align*}
\ifnottechreport{We use the provenance question  provided by Artemis which bounds all head variables.} \iftechreport{We use the provenance question  provided by Artemis which bounds all head variables: $T$ = `trespassing', $N$ = `aarongolden', $C$ = `midnightblue', and $H$ = `lavender'.} The original dataset is $\rel{CRIME_{\text{1.4K}}}$  which we scaled up to $\rel{CRIME_{\text{22K}}}$. We use $\sim10\%$ as the sample size (e.g., \texttt{S2K} for $\rel{CRIME_{\text{22K}}}$) and compute top-5 summaries. The result of this comparison is shown in \Cref{fig:perf-artemis}.
Our system (\texttt{PUG-Summ}) outperforms Artemis by $\sim 2$ orders of magnitude for the two smallest datasets for which Artemis did not time out.
Artemis returned the most general pattern (all placeholders) as the top-1 explanation:\\[-3mm]
\begin{align*}
  p &= (\textsf{tresp.},\textsf{aarongolden},\textsf{midnightblue},\textsf{lavender},S,M), S > 97 \end{align*}
 Unlike Artemis, 
 PUG returned a summary that contains a pattern which covers $\sim 50\%$ of the provenance:\\[-5mm]
 $$p' = (\textsf{tresp.},\textsf{aarongolden},\textsf{midnightblue},\textsf{lavender},\textsf{98},M)$$
 \\[-8mm]

\mypara{Single Derivation Approach}
We implemented a simple \singleDer{} approach (\texttt{SingleDer}) in our system by applying $\sampSize = 1$.  
  That is, the explanation is computed based on only one value from $\dataDomain$ for each unbound variable.
   We use query $r_1$ from \Cref{fig:experi-queries}, sample size  \texttt{S1K}, and compute a top-3 summary.
As shown in \Cref{fig:perf-ynot}, 
   \texttt{SingleDer} outperforms 
   \texttt{PUG-Summ} about an order of magnitude for small datasets. The gap between the two approaches is less significant for larger datasets (more than 1M tuples).

 \section{Related Work}
\label{sec:rel-work}

\RevDel{Our work is closely related to  approaches for compactly representing provenance
and 
explaining missing answers. 
}

\mypara{Compact Representation of Provenance}
The need for compressing provenance to reduce its size has been recognized early-on, e.g., \cite{AB09,CJ08a,OZ11}. However, the compressed representations produced by these approaches are often not semantically meaningful to users.
More closely related to our work are techniques for generating higher-level explanations for binary outcomes~\cite{EA14,WD15}, missing answers~\cite{CC15}, or query results~\cite{RS14, WM13,AB15a} as well as methods for summarizing data or general annotations which may or may not encode provenance information~\cite{XE}. Specifically, like ~\cite{EA14,WD15,RS14,WM13} we use patterns with placeholders.
Some approaches use ontologies~\cite{CC15,WD15}  or logical constraints~\cite{RS14,EA14,WM13} to derive semantically meaningful and compact representations of a set of tuples. 
The use of constraints to compactly represent large or even infinite database instances has a long tradition~\cite{IL84a,KL13a} and these techniques have been adopted to compactly explain missing answers~\cite{HH10,RK14}. However, the compactness of these representations comes at the cost of computational intractability.

\mypara{Missing Answers}
The missing answer problem was first stated for query-based explanations (which parts of the query 
are responsible for the failure to derive the missing answer) in the seminal paper by Chapman et al.~\cite{CJ09}. Most follow-up work~\cite{BH14a,BH14,CJ09,TC10} is based on this notion. Huang et al.~\cite{huang2008provenance} first introduced an instance-based approach, i.e., which existing and missing input tuples
caused the missing answer~\cite{HH10,huang2008provenance,LS17,LL18j}). Since then, several techniques have been developed to exclude spurious explanations and to support larger classes of queries~\cite{HH10}.  
As mentioned before, approaches for instance-based explanations use  either the \allDer{} (giving up performance) or the \singleDer{} approach (giving up completeness). 
In contrast, 
using summarizes we guarantee performance by compactly representing large amounts of provenance without forsaking completeness.
Artemis~\cite{HH10} uses c-tables to compactly represent sets of missing answers. 
However, this comes at the cost of additional computational complexity.

 \section{Conclusions}
\label{sec:conclusions}

We have presented an approach for efficiently computing summaries 
of why and why-not provenance. 
Our approach uses sampling to generate summaries that are guaranteed to be concise while balancing completeness (the fraction of provenance covered) and informativeness (new information provided by the summary).  Thus, we overcome a severe limitation of prior work which sacrifices either completeness or performance.  \RevDel{A core idea of our method is to integrate sampling 
  into provenance capture and summarization.}  We demonstrate experimentally that our approach efficiently produces meaningful summaries of provenance graphs with up to $10^{80}$ derivations.  \RevDel{Our experimental evaluation confirms that this is quite effective. We produce meaningful summaries over very large provenance (up to $10^{80}$ derivations) in only a few minutes.}
In future work, we plan to investigate 
how to utilize additional information, e.g., integrity constraints, in the summarization process.

{
\bibliographystyle{abbrv}

}

\end{document}